\RequirePackage{fix-cm}
\documentclass[smallextended]{svjour3}       
\smartqed  
\usepackage{graphicx}
\immediate\write18{makeindex \jobname.nlo -s nomencl.ist -o \jobname.nls}
\usepackage{url}
\usepackage{longtable}
\usepackage{multirow}
\usepackage{subcaption}
\usepackage{hyperref}
\usepackage{listings}
\usepackage{xcolor}
\definecolor{codegreen}{rgb}{0,0.6,0}
\definecolor{codegray}{rgb}{0.5,0.5,0.5}
\definecolor{codepurple}{rgb}{0.58,0,0.82}
\definecolor{backcolour}{rgb}{0.95,0.95,0.92}
\lstdefinestyle{mystyle}{
    backgroundcolor=\color{backcolour},   
    commentstyle=\color{codegreen},
    keywordstyle=\color{magenta},
    numberstyle=\tiny\color{codegray},
    stringstyle=\color{codepurple},
    basicstyle=\ttfamily\footnotesize,
    breakatwhitespace=false,         
    breaklines=true,                 
    captionpos=b,                    
    keepspaces=true,                 
    numbers=left,                    
    numbersep=5pt,                  
    showspaces=false,                
    showstringspaces=false,
    showtabs=false,                  
    tabsize=2
}
\usepackage[section]{placeins}

\usepackage{graphicx}
\usepackage{listings}
\definecolor{lightgray}{rgb}{9, 9, .9}
\definecolor{darkgray}{rgb}{.4, .4, .4}
\definecolor{purple}{rgb}{0.65, 0.12, 0.82}
\definecolor{gray}{rgb}{0.4,0.4,0.4}
\definecolor{line-numbers}{rgb}{0.4,0.4,0.4}
\definecolor{tags}{rgb}{1, 0, 0}
\definecolor{darkblue}{rgb}{0.0,0.0,0.6}
\definecolor{cyan}{rgb}{0.0,0.6,0.6}
\definecolor{highlight}{HTML}{ffffff}

\usepackage{soul}

\newcommand{\hlc}[2][yellow]{{%
    \colorlet{foo}{#1}%
    \sethlcolor{foo}\hl{#2}}%
}

\usepackage[]{algorithm2e}
\SetKwRepeat{Do}{do}{while}

\usepackage{amsfonts}
\usepackage{amsmath}
\usepackage{verbatim}
\usepackage{longtable}
\usepackage{listings}
\usepackage[titletoc]{appendix}
\usepackage{appendix}
\usepackage{xcolor}
\definecolor{codegreen}{rgb}{0,0.6,0}
\definecolor{codegray}{rgb}{0.5,0.5,0.5}
\definecolor{codepurple}{rgb}{0.58,0,0.82}
\definecolor{backcolour}{rgb}{0.95,0.95,0.92}
\lstdefinestyle{mystyle}{
    backgroundcolor=\color{backcolour},   
    commentstyle=\color{codegreen},
    keywordstyle=\color{magenta},
    numberstyle=\tiny\color{codegray},
    stringstyle=\color{codepurple},
    basicstyle=\ttfamily\footnotesize,
    breakatwhitespace=false,         
    breaklines=true,                 
    captionpos=b,                    
    keepspaces=true,                 
    numbers=left,                    
    numbersep=5pt,                  
    showspaces=false,                
    showstringspaces=false,
    showtabs=false,                  
    tabsize=2
}
\begin{document}

\title{Predicting sensitive information leakage in IoT applications using flows-aware machine learning approach
}


\author{Hajra Naeem         \and
        Manar H. Alalfi 
}


\institute{F. Author \at
              first address \\
              Tel.: +123-45-678910\\
              Fax: +123-45-678910\\
              \email{fauthor@example.com}           
           \and
           S. Author \at
              second address\\
              Dataset Link: https://github.com/HajraNaeem/Dataset
}

\date{Received: date / Accepted: date}

\maketitle

\begin{abstract}
\hlc[highlight]{This paper presents an approach for identification of vulnerable IoT applications. The approach focuses on a category of vulnerabilities that leads to sensitive information leakage which can be identified by using taint flow analysis. Tainted flows vulnerability is very much impacted by the structure of the program and the order of the statements in the code, designing an approach to detect such vulnerability needs to take into consideration such information in order to provide precise results. In this paper, we propose and develop an approach, FlowsMiner, that mines features from the code related to program structure such as control statements and methods, in addition to program's statement order. FlowsMiner, generates features in the form of tainted flows. We developed, Flows2Vec, a tool that transform the features recovered by FlowsMiner into vectors, which are then used to aid the process of machine learning by providing a flow's aware model building process. The resulting model is capable of accurately classify applications as vulnerable if the vulnerability is exhibited by changes in the order of statements in source code. When compared to a base Bag of Words (BoW) approach, the experiments show that the proposed approach has improved the AUC of the prediction models for all algorithms and the best case for $Corpus1$ dataset is improved from 0.91 to 0.94 and for $Corpus2$ from 0.56 to 0.96}.
\end{abstract}

\section{Introduction}
\label{intro}

The Internet of Things (IoT) is the interconnection of uniquely identifiable, ubiquitous, sensing/actuating capable, programmable and self configurable things over the internet. These things share data and services without the need for human involvement \cite{Ref56}.
IoT applications are becoming popular in many fields including: smart homes, automobiles, health care and agriculture. Along the benefits of these applications, there are also some serious risks associated with their usage. The interconnection of multiple devices, sharing data using the internet and the insecure development make these applications vulnerable to serious attacks as explained by the OWASP IoT top Ten vulnerabilities \cite{Ref54}. 

Software vulnerability is a weakness in a software that can be exploited to harm a system or steal sensitive data. If an IoT application has access to the private data then misuse of such data can create problems for the application users. Techniques to identify vulnerabilities in such applications are much needed.
Repeated incidents of IoT security exploitation show an increasing trend of IoT attacks. According to Avast Smart Home security report published in 2019, 40.8\% of the smart homes have at least one vulnerable device \cite{Ref13} which makes the whole network exploitable to attackers. 

Unlike traditional less connected software applications, the nature of IoT applications makes addressing these vulnerabilities more crucial, since these applications have access to private data like when one comes home, what programs they watch on TV, who visits them, when are they alone. The state of devices connected to the SmartApps is itself a sensitive information. Thus, requiring to have privacy risks properly analysed for these applications. But, such tools for SmartApps are not widely available. There are some solutions that have used static analysis to identify vulnerable IoT applications \cite{Ref8,Ref16,Ref39}. 

In this paper, we aim at identifying vulnerable IoT applications using machine learning algorithms. The approach focuses on a category of vulnerabilities that leads to sensitive information leakage which can be identified using taint analysis. Recovering features related to information leakage requires an analysis that can identify tainted flows in the applications under analysis. Taint analysis required to identify tainted flows by tracking information flows from sensitive sources to sensitive sinks. Previous work use machine learning or deep learning algorithms for identification of similar category of vulnerabilities in platforms other than IoT applications, such as android and web applications. \cite{Ref1,Ref2,Ref11,Ref4,Ref7,Ref40,Ref41,Ref42,Ref43,Ref44,Ref45}. 
This paper is an extension of the research paper published at SANER2020 in the early research achievement track where we proposed to use the tainted flows along source code token frequencies to detect vulnerable SmartApps \cite{Ref20}. For that research, only flow sensitivity was considered to identify the tainted flows. We improve the proposed technique by the following research contributions for more accurate and efficient automatic identification of vulnerable and non-vulnerable SmartApps:

\begin{itemize}
    
    \item  We have used an improved method for extraction of tainted flows. In our short paper, we extract tainted flows ignoring control structures like if statements, loops and functions \cite{Ref20}. This paper extends the approach to consider path and context sensitivity analysis. 
    Our approach to perform taint analysis is based on source code mining which we demonstrate as more efficient than taint analysis done with static analysis. We have implemented our approach to develop a tool called FlowsMiner. 
    
    \item We have conducted comprehensive experiments by using a more complex dataset and a larger set of machine learning algorithms. The mutated dataset named as $Corpus2$ used for this research has larger number of SmartApps and these SmartApps are prepared by injecting vulnerabilities. Identification of these vulnerabilities require deeper analysis as compared to dataset used for our previously published research.
    
    \item The gain with performance measures for the comprehensive experiments is better than previously published paper. This is mainly because of the refined technique for identification of tainted flows. For training machine learning algorithms, we have combined two categories of features that are source code token frequencies and tainted flows recovered from SmartApps. 
    
    
    \item We have added a new step in the taint flows identification process. This step formats the source code. We call it preprocessing of source code of SmartApps. This reduces the number of cases to be handled for taint analysis and optimises the taint flows identification process.
\end{itemize}

\hlc[highlight]{The remainder of this paper is organised as follows. Section 2 presents the background work. Section 3 provides an overview of the proposed approach and presents all of its steps in detail including features extraction, identification of tainted flows by using the text mining and using them to build models with machine learning algorithms. 
All conducted experiments are presented in section 4. In section 5, we discuss our findings and related work is presented in section 6. Section 7 concludes the paper.}

\section{Background}
\label{sec:background}
The research presented in this paper addresses a category of IoT applications, Smart-Home applications. There are many frameworks that can be used to develop Smart-Home IoT applications including
SmartThings,
OpenHAB,
Google Home,
Nest Mobile,
Android Things,
ThingSpeak and
DeviceHive. In this paper, we specifically focus on SmartThings, a popular, open source development platform. This platform uses Groovy programming language \cite{Ref14}.
The applications developed in this platform are called SmartApps. SmartApps developed in this platform allow their users to connect with external services and devices to perform actions which makes their homes intelligent by allowing them to control the hardware like lights, fancy door locks and thermostats. SmartApps can perform actions like: turn off lights when no motion is detected, notify a user if the user is not at home and a door opens, turn off the heating system when a user leaves home. SmartApps can communicate with web services to perform even more sophisticated tasks. Intentional or unintentional vulnerabilities are prevalent in the SmartApps. \hlc[highlight]{If a SmartApp has access to the private data then misuse of such data can create problems to application users}. 


The basic skeleton of a SmartApp contains a method call and four method definitions: definition, preferences, installed, updated and initialize. 

\begin{lstlisting}[language=Python, caption={Groovy Application Template},label = {lst:template}]
definition(
    name: "Template",
    category: "Convenience")
preferences {
	section("Change to this mode") {
		input "newMode", "mode", title: "Mode?"
	}
}
def installed() {
	log.debug "Current mode = ${location.mode}"
	createSubscriptions()
}
def updated() {
	log.debug "Current mode = ${location.mode}"
	unsubscribe()
	createSubscriptions()
}
def createSubscriptions()
{
	subscribe(switches,"switch.off", switchOffHandler)
}
def switchOffHandler(evt) {
	takeActions()
}
private takeActions() {
	def message= newMode
	sendNotificationToContacts(newMode)
	send(message)
	setLocationMode(newMode)
	log.debug message
}
private send(msg) {
        sendSms(msg)
        log.debug msg
}
\end{lstlisting}

The definition method call provides the metadata of an application and expects parameters in the form of a map from the developer. The information that an application requires from a user is defined in the preferences. A user has to input this information at the installation of the application. The installed method is called when an application is installed by a user and the updated method is called when a user updates the installation e.g. if a user selects that a different switch should be turned on when the application is installed, the updated method is called at such modifications. The initialize method is called from both installed and updated methods. 

A smart application subscribes to various events to detect occurrence of events, which is done with the usage of the subscribe method. The subscribe method is called from the initialize method. All the event handlers and helper methods are registered to these methods. An example of a Groovy application is provided in the Listing $\ref{lst:template}$. The example in the listing does not contain complete code of the application, we have kept important statements from different methods required to demonstrate presented concepts. To detect vulnerabilities in IoT applications, researchers have used static and dynamic analysis to find the flows of paths that leak the sensitive information \cite{Ref10,Ref3}.

Our research focuses on devising machine learning techniques that help to detect vulnerable SmartApps using automatically extracted features that include tainted flows or in other words the SmartApps include information leakage vulnerability. In the next subsection we present some concepts required for the tainted flows. 

\subsection{Information Leakage vulnerability}
\label{subsec:informationLeakageVulnerability}
\hlc[highlight]{Information leakage caused by tainted flows is always attributed to the execution order of the program statements. For instance, if a sensitive value is passed to a sensitization function before it reaches sensitive sinks then the flow is not tainted and the SmartApp is not vulnerable. Tools developed to identify tainted flows can provide three levels of analysis depending on the capability/sensitivity of the analysis to different language constructs. 

Tools that are sensitive to the statement execution order in the program as well as content change in variables are said to provide flow sensitive analysis. For instance, the code in listing} \ref {lst:flow} \hlc[highlight]{defines a variable \textit{var1} and assigns to it a non-vulnerable string value. The value for \textit{var1} was changed to include sensitive information on line 3, then the value of that sensitive variable is sent to a sensitive sink \textit{sendSms} which would potentially leak the sensitive information. This example needs a flow sensitive analysis to flag it as vulnerable. If statments 2 and 3 are of opposite order, a sensitive flow analysis tool would declare the program as non-vulnerable.}
\begin{lstlisting}[language=Java, caption={Flow Sensitivity Example },label = {lst:flow}]

def var1= "safe input";
var1 = "This variable contains '${ sensitive source}'";
sendSms(var1); // sendSms is a sensitive sink
\end{lstlisting}

\hlc[highlight]{A path sensitive analysis takes the execution path into account. This is exemplified in how it deals with conditional statements; a path sensitive analysis would treat each conditional block as a separate path. Listing} \ref{lst:path} \hlc[highlight]{presents an example where a path sensitive analysis would only detect \textit{sendSms} on line 7 as a tainted sink and that if the condition is evaluated to True, while an insensitive analysis would mark the code vulnerable, regardless of the condition result.}

\begin{lstlisting}[language=Java, caption={Path Sensitivity Example },label = {lst:path}]
def var2="111"
def var3="Hello"
if(condition)
    var3="${source}"
else
	var3="Hello"
sendSms(var3,var2)
\end{lstlisting}
  
\hlc[highlight]{An insensitive path analysis will flag Listing }\ref{lst:nonvul} \hlc[highlight]{as vulnerable even though, the value of \textit{msg} was changed from including sensitive content into a regular string in both blocks of the if statement.}

\begin{lstlisting}[language=Java, caption={Non-vulnerable in $Corpus2$ Example },label = {lst:nonvul}]
def phone="11111"
def msg="${source}"
	if(condition)
		msg="Hello"
	else
		msg="Hi"
	sendSms(phone,msg)
\end{lstlisting}
   
\hlc[highlight]{Context sensitivity mainly deals with function calls and callbacks within the program. A context sensitive analysis identifies each call as its own and can track back to the context of the call.
Listing} \ref{lst:context} \hlc[highlight]{presents a case where a method is called twice. In a context insensitive analysis, both calls on lines 4 and 5 might be conflated, so the tainted return in \textit{firstCall} will also be considered in \textit{secondCall}, marking the flow tainted. A context sensitive approach on the other hand, doesn't confuse method calls and distinguishes each call site, so it won't mark the flow to \textit{sendSms} tainted.}
 
\begin{lstlisting}[language=Java, caption={Context Sensitivity Example },label = {lst:context}]
//$sensitiveData defined
    def takeAction() {
        def message = "This contains $sensitiveData";
        def firstCall = returnMessage(message);
        def secondCall = returnMessage("no sensitive data");
        sendSms(secondCall); }
    private returnMessage(message) {
        return message;}
\end{lstlisting}
  

\hlc[highlight]{Looking at the above code listings, we can note that the difference between vulnerable and non-vulnerable version is mostly the execution order of statements, the location of the sensitive information within a conditional statement or the context of the method call which involves sensitive information as parameter. In the following section, we present architecture of the proposed approach along a detailed description of each step including the tool FlowsMiner that can be used to recover features from the source code related to tainted flows.}
\label{sec:tfId}
\begin{figure}
\centerline{\includegraphics[width=1\textwidth]{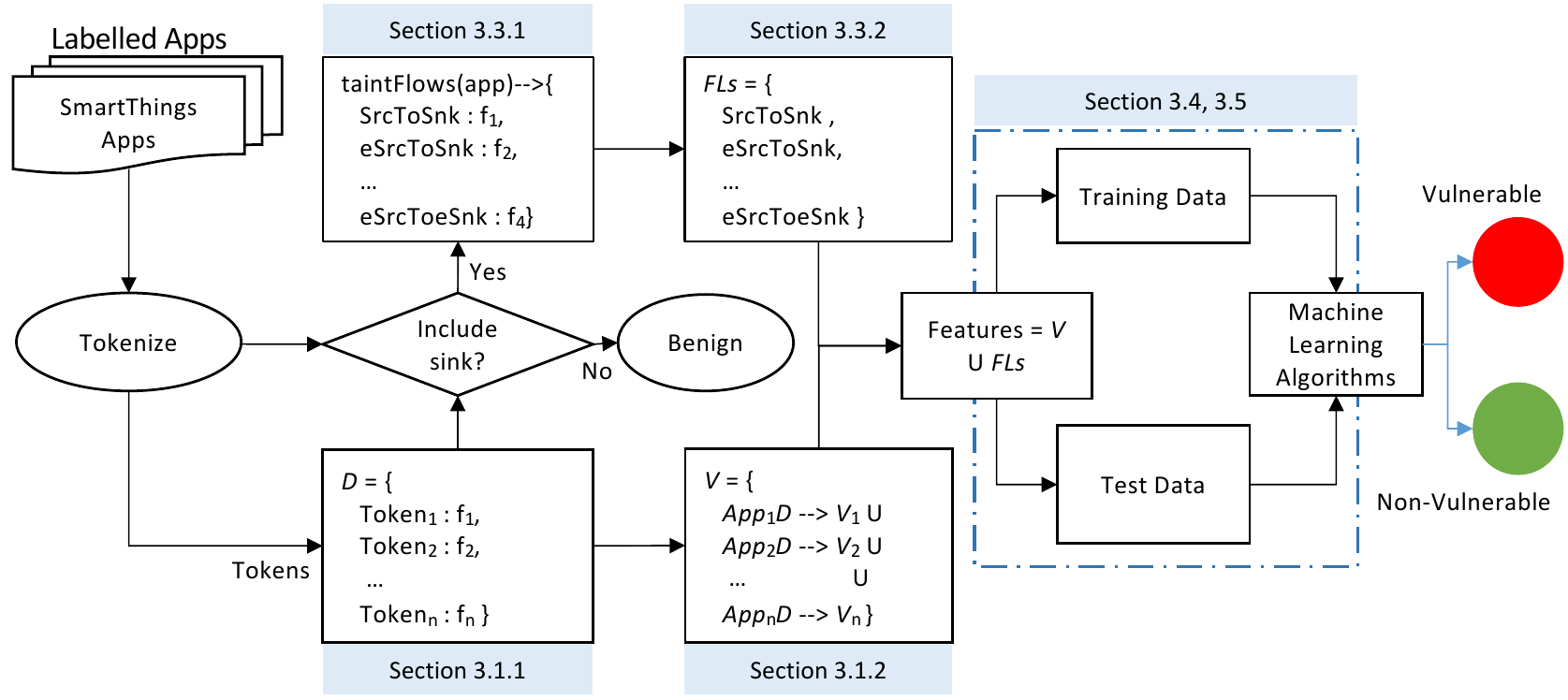}}
\caption{Flows-aware Machine Learning Approach}
\label{fig:approachBOW_FLOW}
\end{figure}
\section{Flows-aware Machine Learning}\label{sec:MB_Flows}
In this section, we present an approach that uses a combined bag of words of source code and tainted flows information recovered from source code and that to build models capable of accurately classifying applications as vulnerable if they contain information leakage vulnerability, such vulnerability is mostly exhibited by changes in the order of statements in source code. 

Figure \ref{fig:approachBOW_FLOW} \hlc[highlight]{presents the architecture of the proposed approach, highlighted labels in the figure mark the corresponding sections describing the details of each step.} 

Machine learning algorithms require some inputs from data to build behavioral models. These inputs are called features of the data. Most of machine learning algorithms are trained on numeric features to build the models. Our research focuses on identifying vulnerable SmartApps using machine learning algorithms. The data for us is the source code of these applications. The source code is raw data in terms of features. 
To build a model for detection of vulnerable SmartApps, we need numeric features from the raw data. We have used text mining to extract and tune numeric features from the source code. 

The features used for training a machine learning algorithm to build a model have a significant impact on the performance of the algorithm. The relevant features have a positive impact and are important to build an exact behavior of an application. The irrelevant or partially relevant features have a negative impact on the performance of the machine learning algorithms. This makes selection of relevant features important. It is hard to decide which of the features are relevant to build appropriate models, especially considering that for almost every dataset the set of relevant features changes \cite{Ref50,Ref51}.

\hlc[highlight]{To build accurate models, every dataset has a different set of relevant features which depends on the information in the dataset and the research questions. A different research question may require a different set of features.}
Considering all these aspects we have decided to prepare a feature set 
according to the following two categories
of features:
\begin{itemize}
\item Initially, we build a bag of words (BoW) representation of the source code and
then we analyse it to select relevant tokens as features. Details about this step are presented in section \ref{sec:featuePrep}. \hlc[highlight]{In the evaluation section, we attempt to evaluate the effectiveness of the BoW approach alone in predicting information leakage vulnerabilities in SmartApps and to investigate on how does automatically extracted features from the BoW of source code perform when the BoW representations of vulnerable and non-vulnerable applications are almost similar? (RQ1, }\ref{sec:EvaluationBOW}).

\item Then, using a text mining approach, we extract tainted flows from the source code and add them to the features set we gathered from the BoW. The proposed text mining technique is implemented in a tool called FlowsMiner, presented in section \ref{sec:FlowsMiner}. The tainted flows are converted to feature vectors using a proposed tool called Flow2Vec, presented in section \ref{sec:flow2Vec}. Details of this step are presented in section \ref{sec:featuePrep2}. \hlc[highlight]{In the evaluation section, we investigate whether a text mining approach is capable of extracting tainted flows features from SmartApps? and how such approach performs in comparison to static analysis techniques? (RQ2, }\ref{sec:TaintedFlowsEvaluation}\hlc[highlight]{). Then we analyze the impact of combining the tainted flows features extracted from text mining with the features extracted from the BoW (RQ3,} \ref{sec:evaluationF2V}).
\end{itemize}

\subsection{Features Extraction and Preparation- Bag of Words}\label{sec:featuePrep}
We build a BOW representation of the source code and analyse it to select features from relevant tokens as presented in section \ref{tokenFrequencies}. \hlc[highlight]{The Token2Vec presented in section} \ref{subsec:token2Vec} \hlc[highlight]{prepares the features set from these BoW which is then fed to machine learning algorithms to build the behavioral models. When we apply the Algorithm} \ref{algo:tokenFrequencies} \hlc[highlight]{ to the SmartApp of Listing } \ref{lst:template}, \hlc[highlight]{the output BOW are presented in the Listing} \ref{lst:frequencies}. 

The BoW feature set for a SmartApp is generated by crawling each line of its source code. The source code is read line by line and it's tokens are generated. The frequencies of tokens are calculated to prepare BoW. The definition method call from source code of a SmartApp is excluded from BoW because the method call only contains metadata about the SmartApp. The BoW is then converted to feature vectors and used to train the machine learning algorithms to build the models.
\subsubsection{Token Frequencies} \label{tokenFrequencies}
Tokens in methods of an application like $installed$, $updated$ and $initialize$ are crawled to form a vocabulary of tokens $V_i$ for the application. The $preferences$ method is excluded from this step. The vocabulary $V_i$ contains distinct set of tokens for the application. The frequencies of the tokens are recorded against the tokens. The process of identifying vocabulary set $V_i$ and calculating the frequencies is described in function $calTokenFrequencies$ presented in the Algorithm $\ref{algo:tokenFrequencies}$. 

\begin{algorithm}
\KwIn{Tokens for all code lines of methods}
\KwOut{$D$ containing tokens $\textit{t}$ and frequencies $f$}
\BlankLine
\SetKwFunction{FMain}{calTokenFrequencies}
\SetKwProg{Pn}{Function}{:}{\KwRet ($D$)}
\Pn{\FMain{tknsCodeLines}}{
	$D$ $\leftarrow$ empty dictionary\\
	\While{a line of code in tknsCodeLines} {				
		\For{every token $t$ in a line} {
		    \eIf{$t$ $\in$ $D$}
		        {increment by 1 $f$ of $t$ in $D$}
			    {add $t$ to $D$ with $f$ as 1}
		}					
	}
}
\caption{Calculate Token Frequencies}
\label{algo:tokenFrequencies}
\end{algorithm}

The algorithm starts by defining an empty dictionary $D$, the dictionaries are [key, value] data structures. The tokens of vocabulary $V_i$ are recorded as keys $t$ and their respective frequencies $f$ are recorded as values in the dictionary $D$. While crawling the lines of tokens, when a new element for $V_i$ is identified, it is added to the $D$ along value 1, and for the existing elements of $V_i$, the value of element in $D$ is incremented by 1. All string tokens are counted as $aStr$ tokens and all numeric tokens are counted as $aNum$ tokens. Once tokens of all code lines are crawled, the algorithm returns $D$.

When the $calTokenFrequencies$ function is applied to the Groovy file of the Listing $\ref{lst:template}$, the tokens identified by the function and their frequencies are shown in the Listing $\ref{lst:frequencies}$. We have implemented the function in Python 3. The Listing $\ref{lst:frequencies}$ shows all tokens $t$ and their frequencies $f$ returned as a dictionary $D$ from the implementation. The function implementation takes name of $D$ from the name of the application which is $Template$ in this case.

\begin{lstlisting}[language=Python, caption={Tokens and Frequencies for Listing $\ref{lst:template}$},label = {lst:frequencies}]
'Template': {'def': 5, 'installed': 1, '(': 15, ')': 15, '{': 6, 'log.debug': 4, 'aStr': 3, 'createSubscriptions': 3, '}': 6, 'updated': 1, 'unsubscribe': 1, 'subscribe': 1, 'switches': 1, ',': 2, 'switchOffHandler': 2, 'evt': 1, 'takeActions': 2, 'private': 2, 'message': 3, '=': 1, 'newMode': 3, 'sendNotificationToContacts': 1, 'send': 2, 'setLocationMode': 1, 'msg': 3, 'sendSms': 1}
\end{lstlisting}
\subsubsection{Token2Vec}\label{subsec:token2Vec}
In the literature, different techniques are used to prepare vectors to train machine learning algorithms. One such technique is Word2Vec \cite{Ref33,Ref34} which converts words to vectors that can be used to build models. We don't need a vector for every source code token but a vector for source code of an application.
Therefore, we have developed a tool Token2Vec, which transforms the source code BoW into numeric feature vectors. To prepare the numeric feature vectors, we use token frequencies from SmartApps under analysis computed by Algorithm \ref{algo:tokenFrequencies}. 
The process to identify distinct tokens $V_i$ for an application is
described in the \hlc[highlight]{section} $\ref{tokenFrequencies}$. Algorithm \ref{algo:tokenFrequencies} calculates the frequencies of distinct tokens $V_i$ for an application. 
We have implemented the technique named as Token2Vec using Python 3 which identifies distinct tokens for a complete dataset of SmartApps. It prepares a set of distinct tokens $\mathbb{V}$ for an entire dataset by combining the set of distinct tokens identified for each SmartApp i.e.\ $\mathbb{V} \leftarrow V_1 \cup V_2 \cup \ldots \cup V_n$. The set of distinct tokens $\mathbb{V}$ is used to create a dataFrame object of the pandas library \cite{Ref19}
with its column names initialised to $\mathbb{V}$. A vector for each SmartApp is prepared against $\mathbb{V}$ and it is appended to the dataFrame. Once this process is completed for all SmartApps, the dataFrame serves as a training and testing set for the machine learning algorithms.

For $Corpus1$ dataset presented in \hlc[highlight]{section} \ref{subsubsec:dataset}, the tool finds 3834 distinct tokens and then keeps only those tokens that are at least present in 5 applications and it finds 515 such tokens. For the $Corpus2$ dataset it finds 1603 distinct tokens and selects 1592 tokens that are at least present in 5 applications. In $Corpus2$ there are around 18 versions of every application (the applications are prepared by injecting mutations in an application), which is the reason to have fewer distinct but more frequent tokens in $Corpus2$ as compared to $Corpus1$.

\hlc[highlight]{On finding that models built from features extracted from BOW representation of SmartApps does not work in all cases, we have proposed to extract and use the tainted flows as features for building the models with classifiers. In the next section, we present extracting the flows from the SmartApps and using them as features.}

\subsection{ Features Extraction- Tainted Flows traces}
\label{subsubsec:FE_Flows}

\hlc[highlight]{This section focuses on extracting flows from the SmartApps which track the sensitive information}. Our approach uses text mining to extract the flow of sensitive information. The task is accomplished in four steps. First step includes preparing BoW by splitting the source code into tokens and then calculating their frequencies. The second step is to define a list of sensitive methods called sinks. We have used the list of sinks identified by Celik et al. \cite{Ref8}.
Third, a list of sensitive information called sources is identified and the last step is to use all of these to identify the flows that can track a potential leakage of the sensitive information.

To demonstrate the process, we have used the source code of a SmartApp provided in the Listing $\ref{lst:template}$ as an example. And the process needs to be applied to all SmartApps of the dataset under test. All single-line and multi-line comments are removed from the SmartApps to put the source code in a so called standard format through a preprocessing stage which is described in detail in \hlc[highlight]{section} \ref{sec:preProcessing}. To identify flows, each line of the source code is crawled. The source code files are read line by line, and their tokens are generated. The definition method call tokens are not included in this analysis because the method call only contains metadata about a SmartApp. This step ends up in creating a BoW of the source code as presented in the \hlc[highlight]{section} \ref{tokenFrequencies}.
\subsubsection{Preprocessing the Source Code}\label{sec:preProcessing}
The goal of this stage is to put the source code in a standard format. This will reduce the number of cases to be handled by our implementation, and thus reduces the complexity of our implementation for sensitive flows identification. We name this process as preprocessing the source code. The process of preprocessing consists of the following steps.

\begin{lstlisting}[language=Java, caption={Removing comments},label = {lst:formatComments}]
/*if evt.value is "active"
invoke sendPush method*/
if (evt.value == "active") {//the if satement
    sendPush("Alarm motion detected")
}// else {
   //sendPush("Motion stopped")
//}

/*is modified to the following*/

if (evt.value == "active") {
    sendPush("Alarm motion detected")
}
\end{lstlisting}
  
\begin{enumerate}
  \item All single line and multi line comments are removed. 
 
  This step is important because it allows to remove the text which is not source code. Making comments part of the features will be incorrect because they are actually not part of behavior of a SmartApp. An example for removing comments is presented in Listing \ref{lst:formatComments}.
  
  \item Get multi line $input$ method calls on one line.
  
  The information required by a SmartApp from a user  is defined in the $preferences$ method. The SmartApp needs to know the devices that it will work with. The arguments of the $input$ method call specify what a device will be referred as in a SmartApp. The $input$ method calls are grouped in a $section$ call which are statements in the $preferences$ method definition.
\begin{lstlisting}[language=Java, caption={Preprocessing input method calls},label = {lst:formatInput}]
input(
name:"serverIp",
title:"In order for SmartThings to send commands back to the SmartBlocks on your Minecraft server,you will have to enter your Server Address",
type:"text",
required:false
)

/*is modified to the following*/

input(name:"serverIp", title:"In order for SmartThings to send commands back to the SmartBlocks on your Minecraft server,you will have to enter your Server Address", type:"text", required:false)
\end{lstlisting}
  The first positional argument in the $input$ method call is an identifier for a device in a SmartApp. This identifier is used as a reference to the device in all methods of the SmartApp. If it is a keyword argument, it can be at any position in the $input$ method call. Having all arguments of the $input$ method call on one line makes it possible to just analyse one line and identify an identifier which is used as a reference to a device. We call these identifiers as sources of a SmartApp. Getting all parts of the $input$ method call on one line enables us to process the positional argument as a source by extracting it from first position of the arguments and in case of keywords arguments, we prepare a Python dictionary for all the keyword arguments and read the value of $name$ key. An example of this step is presented in Listing \ref{lst:formatInput}.
  
  \item Get multi line map definitions on one line. 
  
  A map definition may contain an item which contains a source. We process all items of a map to locate any sensitive information. Having all items on one line makes it easier to analyse map items for sensitive information by processing just one line of code. Otherwise, we will have to process multiple lines to detect beginning and end of the source code of a map. If some sensitive information is part of an item then we will have to go back to the map identifier which was processed on a previous line and mark it as sensitive. An example of this step is presented in Listing \ref{lst:formatMap}. 
\begin{lstlisting}[language=Java, caption={Preprocessing map definitions},label = {lst:formatMap}]
takeParams=[    
uri:"attacker.com",
path:"",    
requestContentType:"app/x-w...",    
body:[    
"message":"${switches}"   
]   
]

/*is modified to the following*/

takeParams=[uri:"attacker.com",path:"",requestContentType:"app/x-w...",body:["message":"${switches}"]]
\end{lstlisting}  
  
  \item If there are multiple nested conditional statements on one line, split them to have at most one conditional on one line.
  
  To track the sensitive information flows, considering that values of some identifiers might be switching between sensitive to non sensitive information. We need to identify the beginning and end of conditional blocks. 
  On identification of beginning of a conditional, all source code after left curly brace is shifted to next line. An example for preprocessing multiple conditionals on one line is presented in Listing \ref{lst:formatIfCur}.
\vspace{1.2cm}
\begin{lstlisting}[language=Java, caption={Preprocessing multiple conditionals on one line},label = {lst:formatIfCur}]
if(sendSms2(switches,1)){if(sendSms2(switches,1)){sendPush(switches,1)

/*is  modified  to the  following*/

if(sendSms2(switches,1)){
if(sendSms2(switches,1)){
sendPush(switches,1)
\end{lstlisting}
  \item If after the beginning of a conditional, a \textit{left curly brace} on next line, get it on the same line as presented in Listing \ref{lst:formatIfNoCur}. 
  
  We have used a stack to identify beginning and end of the conditionals. If a conditional is found, we push a left curly brace into the stack. On finding a right curly brace we pop from the stack. When the stack is empty that means the body block of a conditional ends. If the first line does not have a left curly brace that means processing of next line will get an empty stack. For simplicity, we ensure that every conditional line contains a left curly brace.
\begin{lstlisting}[language=Java, caption={Preprocessing to get left curly brace on the same line},label = {lst:formatIfNoCur}]
if(motion&&minutes)
{

/*is  modified  to the  following*/

if(motion&&minutes){
\end{lstlisting}

 \item If there is a right curly brace outside of a function invocation, move all the code after right curly brace to the next line as presented in Listing \ref{lst:formatElse}.
\begin{lstlisting}[language=Java, caption={Preprocessing to get code after right curly brace to next line},label = {lst:formatElse}]
} else {

/*is modified to the following*/

}
else {
\end{lstlisting}
  This step enables to easily identify the end of a block that appears before a conditional by processing a right curly brace to pop an item from a stack and beginning of the next block by pushing conditional along the right curly into the stack. The stack enables to detect beginning and ending of the conditional blocks.
  
  \item If there are multiple statements on one line. Split on semi colon to have at most one statement on one line, as presented in Listing \ref{lst:formatMultiStatements}.
  
  The SmartApps are written in the groovy language. It is optional to end a groovy statement on a semicolon. But, if we want to have multiple statements on one line then we need to end all statements on semicolon with the exception of the last statement on the line. 
  \vspace{1.2cm}
\begin{lstlisting}[language=Java, caption={Preprocessing to keep at most one statement on a line},label = {lst:formatMultiStatements}]
log.debug("sending text message");sendSms(phoneNumber, msg)

/*is modified to the following*/

log.debug("sending text message");
sendSms(phoneNumber, msg)
\end{lstlisting}
  We process the lines with multiple statements to have at most one statement on a line. Without this step we will have to process all statements on the line at once. Identify the start and end a statement which will complicate processing of the statements.
\end{enumerate}
We have done few similar processing for $do$, $try$, $catch$ and $finally$ blocks in the source code. The processing of these blocks is similar to above mentioned points. Preprocessing of SmartApps source code makes feature and flow mining simpler. In the next section, we describe preparing the initial set of features that we use for building the behavioral models of the SmartApps.

\subsubsection{Sinks identification} \label{sinks}
 Our approach requires the identifications of the presence of tainted flows. Some tokens of the source code indicate presence of a potential tainted flow, for example, if an application contains sink or a token from sinks set, that application may potentially contain a tainted flow. If we find a token which confirms the presence of a sink in an application, we investigate that application further to identify tainted flows. The set of sinks $SN$ that we have used for this study is provided in the Listing \ref{lst:sinks}.

\begin{lstlisting}[language=Python, caption={Set of Sinks $SN$},label = {lst:sinks}]
Sinks = {'sendNotification', 'httpHead', 'sendPush', 'sendPushMessage', 'sendSms', 'sendSmsMessage', 'sendText', 'httpGet', 'httpPost', 'httpPut', 'httpPostJson', 'sendPushAndFeeds', 'httpDelete', 'sendNotificationEvent', 'contactBookEnabled', 'httpPutJson', 'sendNotificationToContacts', 'inline_html'}
\end{lstlisting}

If we check the tokens of $D$ from the Listing \ref{lst:frequencies}. We can note that a sink $sendSms$ is present in the tokens. This means that there might be a tainted flow in the application and for that we need to investigate further. To identify the tainted flows, we prepare the set of sources as presented in \hlc[highlight]{section} $\ref{sources}$.

\subsubsection{Sources identification} \label{sources}
The information required from users by a SmartApp is defined in the $preferences$ method. On installation of a SmartApp, a user configures the devices a SmartApp will be operating on. The SmartApp specifies references of the devices in arguments of the $input$ method. These references are called $sources$ by us. The $section$ call allows us to group the relevant inputs and give them a text description. The section calls are located in the preferences method definition.

The preferences method tokens are crawled to identify the $sources$. If the first argument of the input method call is a positional argument; it is a reference that is used in a SmartApp to refer to an input device. If the arguments of the input method call are keyword arguments then the identifier of a device is specified in the keyword argument named as $name$ which can be at any index in the method call. The value of $name$ is an identifier for the input device in a SmartApp. The input device identifiers from the tokens of preferences method are selected to form a set named as $sources$. The process of identifying set of sources $S$ from the preferences method definition is described in the function $findSources$ presented in Algorithm $\ref{algo:sources}$. The algorithm initializes the set of sources $S \leftarrow \emptyset$ and crawls the tokens of the preferences method by going through each line and adds a source found to $S$. Once all lines are crawled, it returns a set of sources $S$ for that application. The set of sources $S$ for the application in Listing $\ref{lst:template}$ is $\{newMode\}$.

\begin{algorithm}
\KwIn{Tokenized code lines of a method}
\KwOut{Sources $S$ in a method}
\BlankLine
\SetKwFunction{FMain}{findSources}
\SetKwProg{Pn}{Function}{:}{\KwRet ($S$)}
\Pn{\FMain{tokensMethodCodeLines}}{
	$S \leftarrow \emptyset$\\
	\For{a line of code in tokensMethodCodeLines} {				
		\If{first token ft of line is `input'}
		{
    		\If{all arguments are keyword arguments}
    		{
    		    find value $s$ of $name$ argument\\
    		    add $s$ to $S$
    		}
    		\Else{
    		    find first positional argument $s$\\
    		    add $s$ to $S$
    		}
		}
	}
}
\caption{Identify sources from preferences method}
\label{algo:sources}
\end{algorithm}

\subsubsection{Method Invocations and Extended Sources Identification}
\label{extSourcesMethodInvocations}

If a source is assigned to an identifier that identifier also becomes a source. We name such identifiers as extended sources $extS$. For identification of tainted flows, we also need to search for the extended sources. The set of extended sources is maintained separately for every method a SmartApp because of the scope of extended source identifiers. In the same method, an identifier may later get a value from a variable which is neither a source nor an extended source, if this happens then the identifier is removed from the extended sources set. 

\begin{algorithm}
\KwIn{Tokenized code lines of a method, Sources S}
\KwOut{$funcSignature$, extended sources set $extS$ and method invocations $mthdInvocations$}
\BlankLine
\SetKwFunction{FMain}{findExtS\_mtdInv}
\SetKwProg{Pn}{Function}{:}{\KwRet ($mthdSignature$,$extS$,$mthdInvocations$)}
\Pn{\FMain{tokensMtdCodLines,S}}{
	$extS \leftarrow \emptyset$\\
	$mthdInvocations \leftarrow$ empty list\\
	find $mthdSignature$ from first line of code\\
	\For{a line of code in $tokensMtdCodLines$ } {				
		\If{line is assignment statement}
		{
		    find identifier $eS$ and assigned expression $val$\\
    		\eIf{$\exists s \in S \cup extS : s=val \lor s\in val$} 
    		{add $eS$ to $extS$}
    		{if $eS \in extS$
    		remove $eS$ from $extS$}
		}
		\If{line contains a method invocation}
		{add line to $mthdInvocations$}
	}
}
\caption{Identify extended sources and method invocations for every method definition}
\label{algo:extSourcesMethodInvocs}
\end{algorithm}

All method invocations in the definition of a method of a SmartApp are located despite of the locations of the invocations. A method can be invoked directly from a method or a method invocation could be wrapped inside a conditional. A method can be invoked from another method invocation or it can also be a default parameter value in a method definition. The ternary statements may also include a method invocation. All method invocations $mthdInvocations$ that are present in a method definition of an application are saved against the name of that method. The function $findExtS\_mtdInv$ finds elements of $mthdInvocations$, $extS$ and returns along the name of the method as described in Algorithm $\ref{algo:extSourcesMethodInvocs}$.

The set of extended sources $extS$ identified by function $findExtS\_mtdInv$ for the application in the Listing $\ref{lst:template}$ is $\{takeActions:[(message,newMode)]\}$. Which means that if $message$ is passed as a parameter to any method invocation in the method $takeActions$, it is treated as if a source is passed to that method. Lists of method invocations for all method definitions in the Listing $\ref{lst:template}$ identified by $findExtS\_mtdInv$ are shown in Listing $\ref{lst:methodInvocations}$. 

\begin{lstlisting}[language=Python, caption={Method invocations list for every method definition in Listing $\ref{lst:template}$},label = {lst:methodInvocations}]
{def createSubscriptions(): [subscribe(switches, "switch.off ", switchOffHandler)],
 def installed(): [createSubscriptions()],
 private send(msg): [sendSms(msg)],
 def switchOffHandler(evt): [takeActions()],
 private takeActions(): [sendNotificationToContacts(newMode), send(message), setLocationMode(newMode)],
 def updated(): [unsubscribe(), createSubscriptions()]}
\end{lstlisting}
\subsubsection{Extended Sinks Identification}
\label{extSinks}
Our technique locates all method invocations which are a member of the Sink $SN$ and if an element from the set of sources $S \cup extS$ is passed as a parameter to that method invocation then it reports it as a tainted flow. But, a parameter $p_1$ from signature of a method $eSN$ defined in an application could also be passed to a method invocation which is a sink $S$. If the invocation of the method $eSN$ in definition of another method accepts an element from $S \cup extS$ as a parameter at the index of $p_1$ then the method $eSN$ is identified as an extended sink method.

\begin{algorithm}
\KwIn{List of parameters in method definition, method invocations from the method}
\KwOut{data for extended sinks}
\BlankLine
\SetKwFunction{FMain}{findExtSink}
\SetKwProg{Pn}{Function}{:}{\KwRet ($extendedSinkData$)}
\Pn{\FMain{params,mthdInvocations}}{
	$extendedSinkData \leftarrow$ empty dictionary\\
	$SN \leftarrow$ $Sinks$\\
	\For{$p$ in $params$} {				
		\For{$mtdIn$ in $mthdInvocations$}
    		{\If{$mtdIn \in SN \wedge p \in$ parameters of $mtdIn$}
    		    {add the following to $extendedSinkData$\\
    		    $p$, index of $p$ in $params$
    		    }
    		}
	}
}
\caption{collect data for extended sinks}
\label{algo:extSinks}
\end{algorithm}

To identify elements of extended sink methods set, the signature tokens of every method definition are crawled to locate parameters of that method. Suppose $M$ is name of a method. For every parameter $p$ of $M$ check all the method invocations identified by the function $findExtS\_mtdInv$ described in Algorithm $\ref{algo:extSourcesMethodInvocs}$. If any $p$ of $M$ is passed in the method definition to a method invocation $mtdIn$ and that invocation is for a sink $SN$ then save $M$, $p$ and index of $p$ in the extended Sinks set. Algorithm $\ref{algo:extSinks}$ returns this information in $extendedSinkData$. When the invocation of method $M$ in the definition of another method accepts a source at the index of $p$ then a tainted flow is reported. The set of extended sinks identified by $findExtSink$ for the application in the Listing $\ref{lst:template}$ is $\{send: [((0, msg), sendSms(msg))]\}$, the parameter $msg$ of the method $send$ which is at the index 0 makes it extended sink because in the definition of the $send$ method, $msg$ is passed as a parameter to the sink method $sendSms$. If method $send$ is called from another method and an element from $S \cup extS$ is passed as a parameter at index 0, It results in passing a source to a sink.
\subsubsection{Tainted Flows Identification}
\label{tainted_Flows}
Algorithm \ref{algo:taintedFlows} \hlc[highlight]{focuses on identifying tainted flows}. It initializes $S$ with empty set, $allMthdsData$, $extendedSinks$ and $taintedFlows$ with empty dictionary. All information extracted from each method definition like method invocations and extended sources, is saved in $allMthdsData$. The identified extended sinks are saved in $extendedSinks$ and set of tainted flows in $taintedFlows$.

\begin{algorithm}
\KwIn{Source code of an application, set of Sinks $SN$}
\KwOut{Tainted Flows}
\BlankLine
\Begin{
$S \leftarrow \emptyset$, $allMthdsData$ $\leftarrow$ empty dictionary\\
$extendedSinks$ $\leftarrow$ empty dictionary\\
$taintedFlows$ $\leftarrow$ empty dictionary\\
create tokens for each line of preprocessed source code\\
create $tknsCodeLines$ by ignoring the following\\
$\hspace*{3mm}$ lines for $definition$ method call\\
save $preferences$ code lines in $tokensPCodeLines$\\
$D$ $\leftarrow$ calTokenFrequencies($tknsCodeLines$)\\
$setTokens\leftarrow$ get all tokens from $D$\\
\If{$SN \cap setTokens \ne \emptyset$}
{   $S \leftarrow$ findSources$(tokensPCodeLines)$\\
    \For{tMtdCodLines in tknsCodeLines}
        {$x \leftarrow$ FindExtS\_MthdInv($tMtdCodLines$)\\
        add $x$ to $allMthdsData$\\
        }
    \For{aMethodData in allMthdsData}
    { $ms$,$es$, $mInvs$ $\leftarrow aMethodData$\\
    // $ms$ is method Signature,\\
    // $es$ is set of extended sources,\\ 
    // $mInvs$ is method Invocations\\ 
    get parameters $params$ from $ms$\\    
     $p, ip\leftarrow$ findExtSink($params$,$mInvs$)\\
     add $ms, p, ip$ to $extendedSinks$
    }
    $eSN\leftarrow$ all method names from $extendedSinks$
    \For{aMethodData in allMthdsData}
    {$ms$, $es$, $mInvs$ $\leftarrow aMethodData$\\
    $\forall inv : inv \in (mInvs \cap SN)$ \\
    $\hspace*{3mm} \wedge \exists p : p \in (parameters(inv) \cap (S \cup es))$\\
    add ($ms \rightarrow inv \rightarrow p$) to $taintedFlows$\\
    $\forall inv : inv \in (mInvs \cap eSN)$
    $\hspace*{1mm} \wedge \exists p : p \in (parameters(inv) \cap (S \cup es))$
    $\hspace*{1mm} \wedge index\_p(inv) \in extSinkIndices(inv)$\\
    add ($ms \rightarrow inv \rightarrow p$) to $taintedFlows$\\
    }
    \Return $taintedFlows$
}
}
\caption{Tainted Flows Identification}
\label{algo:taintedFlows}
\end{algorithm}

The algorithm starts by creating tokens of the preprocessed source code. The preprocessing is described in the \hlc[highlight]{section} \ref{sec:preProcessing}. The $definition$ method call tokens are not included in the analysis because it is only metadata of a SmartApp. The tokens are created for each line and the lines of tokens are analyzed at a method level because of the scope of local variables of the method. The tokens for the code lines of the $preferences$ method are saved in $tokensPCodeLines$, whereas rest of the tokens are saved in $tknsCodeLines$. The frequencies of $tknsCodeLines$ tokens are calculated by using the function $calTokenFrequencies$ and are saved in $D$. The algorithm checks that if a token from set of sinks $SN$ is present in the tokens of $D$ then, the algorithm proceeds to search for the tainted flows, otherwise, the algorithm terminates. 

If a sink is found in the tokens of $D$, the algorithm proceeds to the next step and crawls the tokens of $tokensPCodeLines$ line by line to identify set of sources $S$ by calling the $findSources$ method. Tokens for lines of codes of every method $tMtdCodLines$ are taken from $tknsCodeLines$ and are passed step by step to the function $FindExtS MthdInv$ to identify extended sinks and method invocations for every method, which are saved in $allMthdsData$. 

Once this process is complete for all methods, the algorithm processes data for each method from $allMthdsData$ by using the function $findExtSink$ to identify set of extended sinks $extendedSinks$ and the parameter index of the parameters that make a method extended sink. The algorithm again analyses the method invocations from $allMthdsData$ for each method and compares those invocations with the set of sinks $SN$ and extended sinks $extendedSinks$ to identify any matches. If an invocation $inv$ belongs to a sink method $SN$, the algorithm checks if an element $p$ from $S \cup es$ is passed as a parameter to $inv$. If the algorithm is successful, it adds method name, $inv$ and $p$ to tainted flow $taintedFlows$. Here, $es$ is the set of extended set of sources which is different for each method. In parallel, the algorithm also checks if a method invocation $inv$ matches to an extended sink $eSN$ from $extendedSinks$, if the algorithm is successful then it goes on to check if a parameter $p$ from $S \cup es$ is passed as a parameter to invocation $inv$ and index of $p$ in $inv$ is same as the index of a parameter which makes the method $eSN$ an extended sink. If all constraints are met, the algorithm adds the method name, $inv$ and $p$ to tainted flow $taintedFlows$. 
The tainted flows identified by the Algorithm \ref{algo:taintedFlows} from Listing $\ref{lst:template}$ are presented in the Listing $\ref{lst:taintedFlow}$.
\vspace{1.5cm}
\begin{lstlisting}[language=Python, caption={Tainted sink flows identified for code in Listing $\ref{lst:template}$},label = {lst:taintedFlow}]
Tainted Sink Flow: Method: takeActions --> sendNotificationToContacts(newMode) --> newMode
Tainted ExtSink Flow: Method: takeActions --> send(message) --> message --> newMode
\end{lstlisting}
\vspace{1.5cm}
\subsection{Features Preparation-Tainted Flows}\label{sec:featuePrep2}

\subsubsection{FlowsMiner}\label{sec:FlowsMiner}
The Python 3 implementation of Algorithm \ref{algo:taintedFlows} is called FlowsMiner. The implementation processes the source code of a SmartApp as described in the \hlc[highlight]{section} \ref{sec:preProcessing} before tokenizing the source code. This reduces the number of cases to be handled at each step while identifying the flows. The first step involves identifying sources of a SmartApp as described in \hlc[highlight]{section} \ref{sources}. If the arguments of the input method call are positional arguments then the first argument is selected as a source, otherwise, the implementation prepares a Python 3 dictionary object for the keyword arguments of the input method, and then reads the value of $name$ key. Dictionary is a suitable data structure for keyword arguments because the keyword arguments are separated with colons. Once the sources set is prepared, the sinks are read from a file named as $Sinks.txt$. For our research, we have used a fixed set of sinks. The FlowsMiner provides us flexibility to provide any set of sinks and the tool is capable of handling the change. In case we want to use a different set of sinks, all we have to do is change the set of sinks in $Sinks.txt$. Changing them in the file will have impact on the whole analysis and the updated sinks set will be used for identification of tainted flows. After identifying the extended sources and extended sinks the tainted flows are identified as described in Algorithm \ref{algo:taintedFlows}.

The usage of FlowsMiner is fairly straightforward, one needs to provide source code and the tool outputs the tainted flows if there are any. A sample output of FlowsMiner for the source code presented in Listing \ref{lst:template} is provided in the Listing \ref{lst:taintedFlow}. 

\subsubsection{Flow2Vec}\label{sec:flow2Vec}
Our Python 3 implementation Flow2Vec maps the tainted flows identified by our text mining taint analysis approach described in Algorithm \ref{algo:taintedFlows} to the numeric feature vectors. 
 To map the tainted flows to vectors, Flow2Vec calculates occurrences for each type of flow and records them as frequencies for the flows.
The Flow2Vec analyses the flows identified by Algorithm \ref{algo:taintedFlows} and categorises them into six types of tainted flows named as $FLs$. The six categories of $FLs$ are: from a source to a sink, from a source to an extended sink, from an extended source to a sink, from an extended source to an extended sink, a sink in body of a tainted conditional, an extended Sink in body of a tainted conditional. The frequencies for these flows are denoted by $f_1$, $f_2$, $f_3$, $f_4$, $f_5$ and $f_6$, respectively. 

The set $FLs$ is mapped to distinct numbers by mapping first element to 0, second element to 1, third element to 2 and repeating the process for all elements. These numbers are the indices of these elements in the columns of the feature vectors table and the elements are the names of the columns. The rows are labelled with the names of the SmartApps. 
 
 For each SmartApp, we loop through the elements of the $FLs$ set, if an element matches to a token in $FLs$, it's frequency is selected accordingly.
 We populate all flows frequencies of an application in the corresponding columns of the table and if the corresponding tokens are not found from $FLs$, zeros are filled. This process is repeated for all SmartApps to prepare a numeric vector of flows for each application.

\subsection{Models Building}
\label{sec:buildingModels}
To build models of SmartApps under test, we have used a set of machine and deep learning algorithms including Logistic Regression, Naive Bayes, Decision Tree, Random Forest, K-Nearest Neighbors,  Support Vector Machine (SVM), Multilayer Perceptron (MLP), \hlc[highlight]{Convolutional Neural Network (CNN) and Long short-term memory (LSTM)}. 
The training sets to build the behavioral models of the SmartApps are prepared from the datasets presented in the section \ref{subsubsec:dataset}. Machine learning algorithms are trained on these datasets to build models. We use these models to identify vulnerable applications. The built model is a function predict(x), where x is a source code of a SmartApp. The function returns “vulnerable” if the SmartApp is vulnerable and “non-vulnerable”, otherwise. The evaluation metrics \hlc[highlight]{Area under the ROC Curve (AUC), F1 Score, Matthews correlation coefficient (MCC),} accuracy, precision and recall are used to evaluate the built models.

\subsubsection{Dataset \hlc[highlight]{Selection}}
\label{subsubsec:dataset}
We have used two datasets \hlc[highlight]{to evaluate the proposed techniques.} One dataset is prepared by Celik et al. by collecting SmartApps from marketplace, community, SmartThings forum and IoTBench \cite{Ref8}.
This dataset contains 217 SmartApps and we name it as $Corpus1$. 

\begin{figure}[!ht]
\centerline{\includegraphics[width=.66\textwidth]{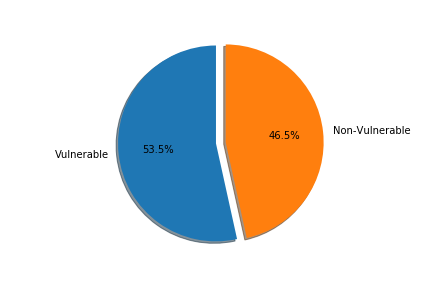}}
\caption{Ratio of Vulnerable and Non-Vulnerable SmartApps in $Corpus1$}
\label{fig:piechart_Vulnerable_Non-Vulnerable_originalDataset.png}
\end{figure}

From these 217 SmartApps, 101 are labelled as Non-Vulnerable, whereas, 116 are labelled as Vulnerable. The ratio of vulnerable and non-vulnerable SmartApps in the dataset is shown in Figure \ref{fig:piechart_Vulnerable_Non-Vulnerable_originalDataset.png}.

\begin{figure}[!ht]
\centerline{\includegraphics[width=.66\textwidth]{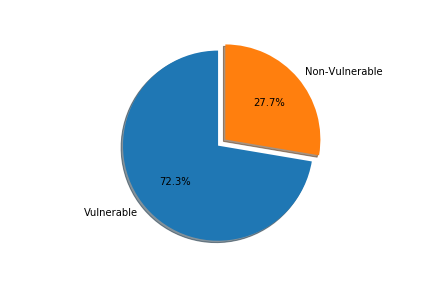}}
\caption{Ratio of Vulnerable and Non-Vulnerable SmartApps in $Corpus2$}
\label{fig:piechart_Vulnerable_Non-Vulnerable_mutatedDataset.png}
\end{figure}

$Corpus2$  was created by Parveen and Alalfi \cite{Ref17}
\hlc[highlight]{where they altered the order of statements in benign SmartApps to introduce vulnerable apps that leak sensitive information. Mutations created by them mainly focused on generating tainted flows vulnerabilities by changing the order of statements in the sequential code flow. The flow insensitive analyzers can not capture these vulnerabilities,} this case is presented in the background section, Listing \ref{lst:flow}.\hlc[highlight]{ Other mutants generated generated by them introduced changes to the conditional statements so it can produce vulnerabilities that path insensitive analyzers can not detect, this case is presented in the background section, Listing} \ref{lst:path}.\hlc[highlight]{ Finally the third category of mutants introduced by them changes to method calls so that vulnerabilities related to context are produced, this case is presented in the background section, Listing} \ref{lst:context}. 
$Corpus2$ consists of 1186 SmartApps, from these SmartApps 858 are Vulnerable and 328 are Non-Vulnerable. The ratio of these SmartApps is presented in Figure \ref{fig:piechart_Vulnerable_Non-Vulnerable_mutatedDataset.png}.

\begin{figure}[!ht]
\centering
\begin{subfigure}{.5\textwidth}
  \centering
  \includegraphics[width=1\linewidth]{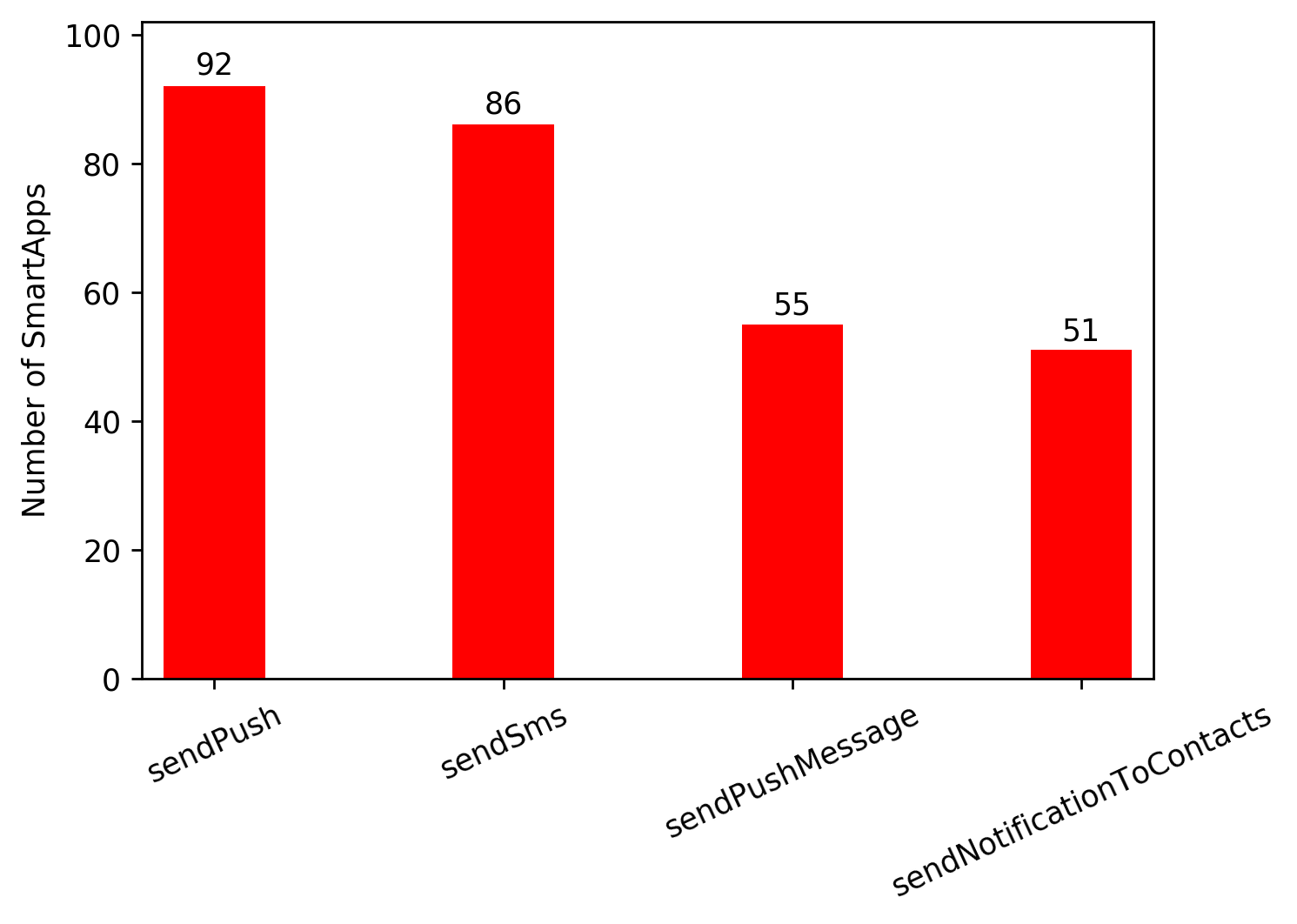}
  \caption{SmartApps in $Corpus1$}
  \label{fig:frequency_sinks_orig}
\end{subfigure}%
\begin{subfigure}{.5\textwidth}
  \centering
  \includegraphics[width=.98\linewidth]{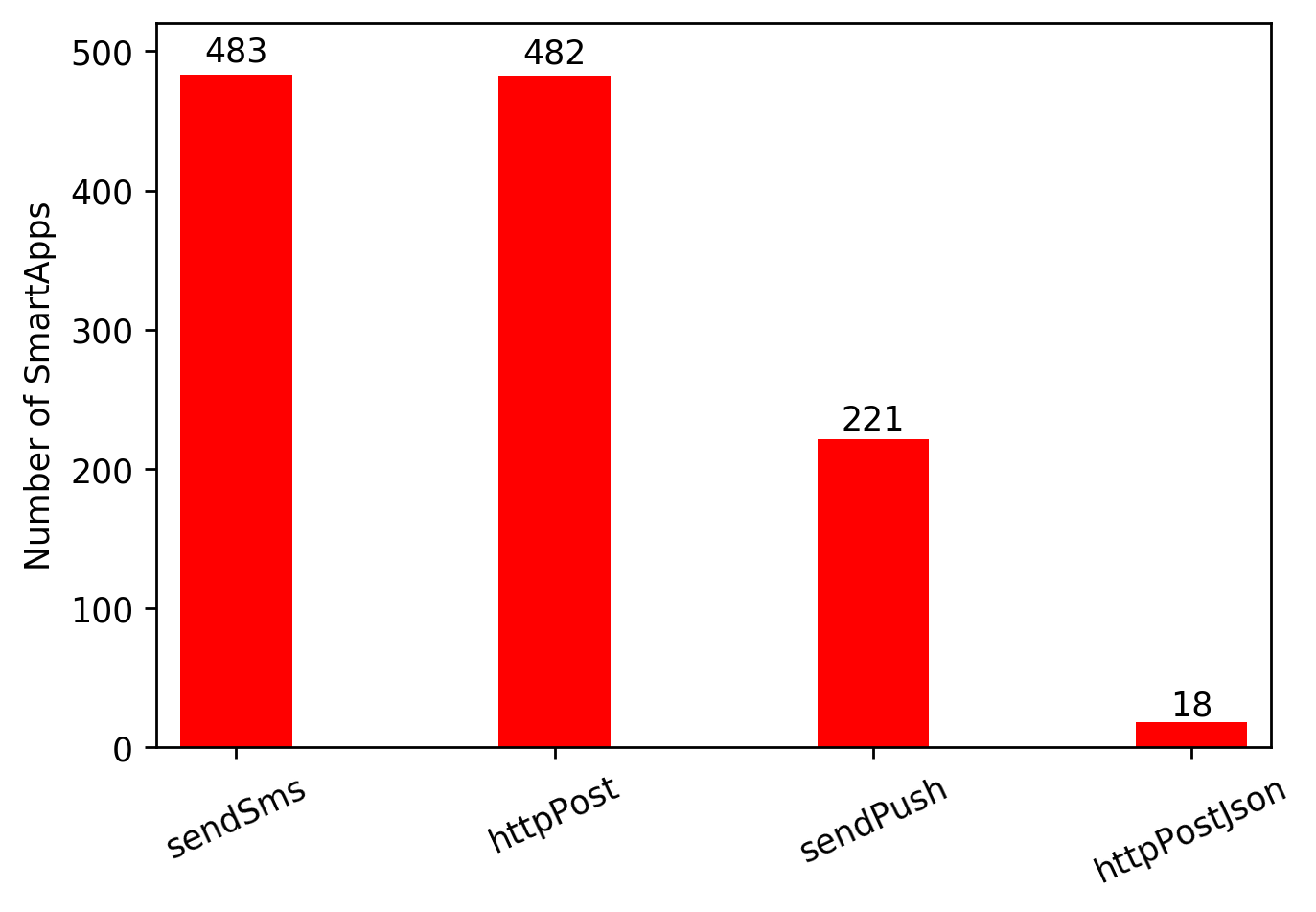}
  \caption{SmartApps in $Corpus2$}
  \label{fig:frequency_sinks_mut}
\end{subfigure}
\caption{Number of SmartApps with Specified Sinks in $Corpus1$ and $Corpus2$ Datasets}
\label{fig:how_many_apps_have_sinks_both_datasets}
\end{figure}

\hlc[highlight]{To understand both of the datasets, lets perform an exploratory analysis for the distribution of sinks in the datasets.}
Figure \ref{fig:how_many_apps_have_sinks_both_datasets} shows the number of SmartApps with any of the specific sinks for both datasets. $Corpus2$ has only 4 types of sinks and number of SmartApps with any of these are presented in the figure. Whereas, $Corpus1$ has 14 types of sinks and in the figure only the number of SmartApps with 4 most frequent sinks are presented. We can note that 92 SmartApps have $sendPush$ and 86 SmartApps have $sendSms$ sink. Whereas, for $Corpus2$ $sendSms$ and $httpPost$ are the most frequent sinks. The number of SmartApps with $sendSms$ sink is 483 and $httpPost$ is 482.

\begin{figure}[ht!]
\centerline{\includegraphics[width=.7\textwidth]{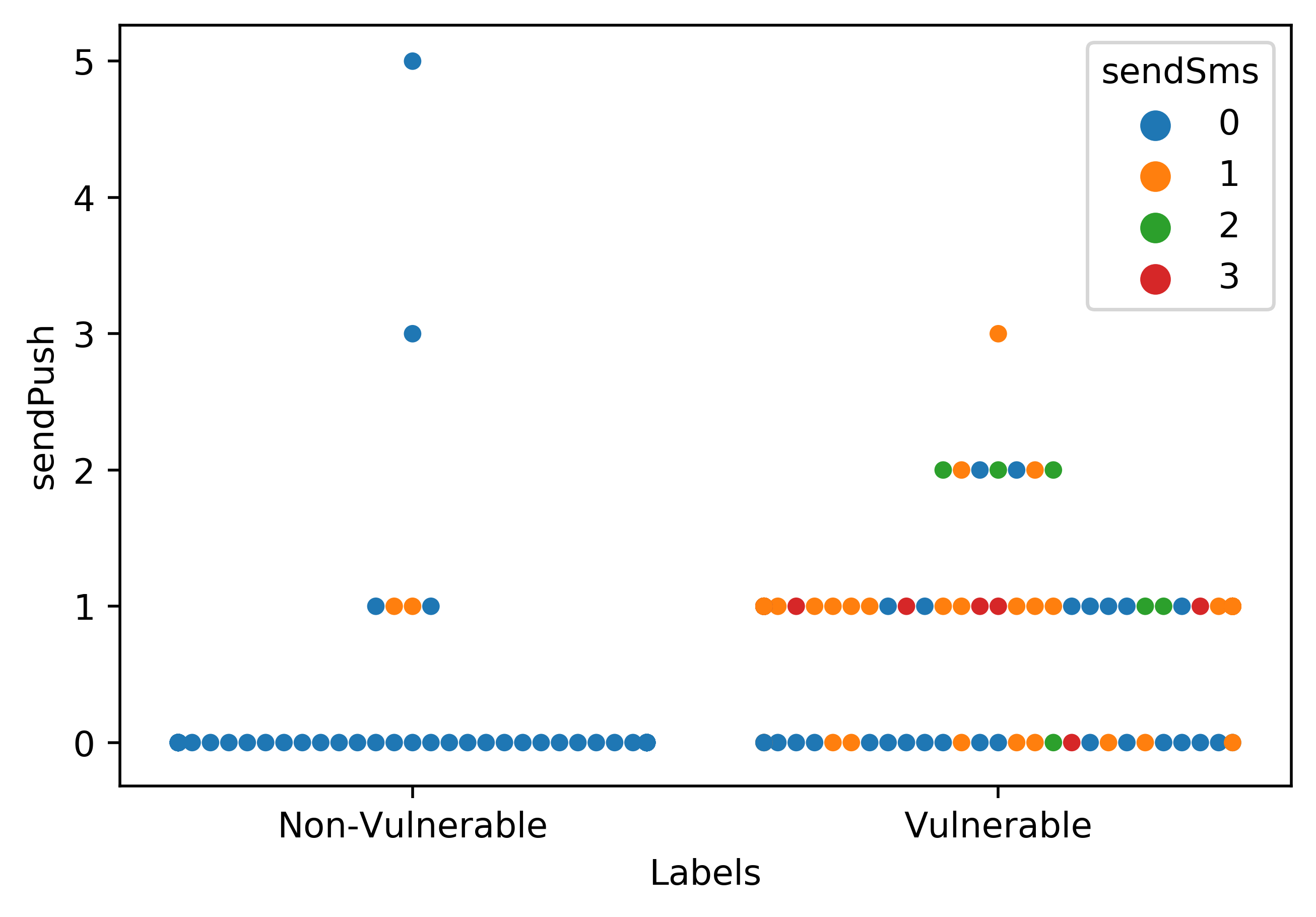}}
\caption{Frequency distribution of SendPush w.r.t sendSMS in all SmartApps of $Corpus1$}
\label{fig:sendPush_sendsmsOrig}
\end{figure}

In the Figure \ref{fig:sendPush_sendsmsOrig}, we have presented the frequency distribution of two most frequent sinks for  $Corpus1$ by showing a swarmplot. It can be noted that vulnerable SmartApps tend to have more instances of sinks. We have only 6 non-vulnerable SmartApps that have one or more instances of $sendPush$ and 2 SmartApps with any instance of $sendSMS$ sink. \hlc[highlight]{More figures about distribution of sinks for this dataset are provided in the Appendix} \ref{app:Third}. \hlc[highlight]{It can be noted that the frequency distribution of sink tokens across vulnerable and non-vulnerable SmartApps is different.}

\begin{figure}[ht!]
\centerline{\includegraphics[width=.7\textwidth]{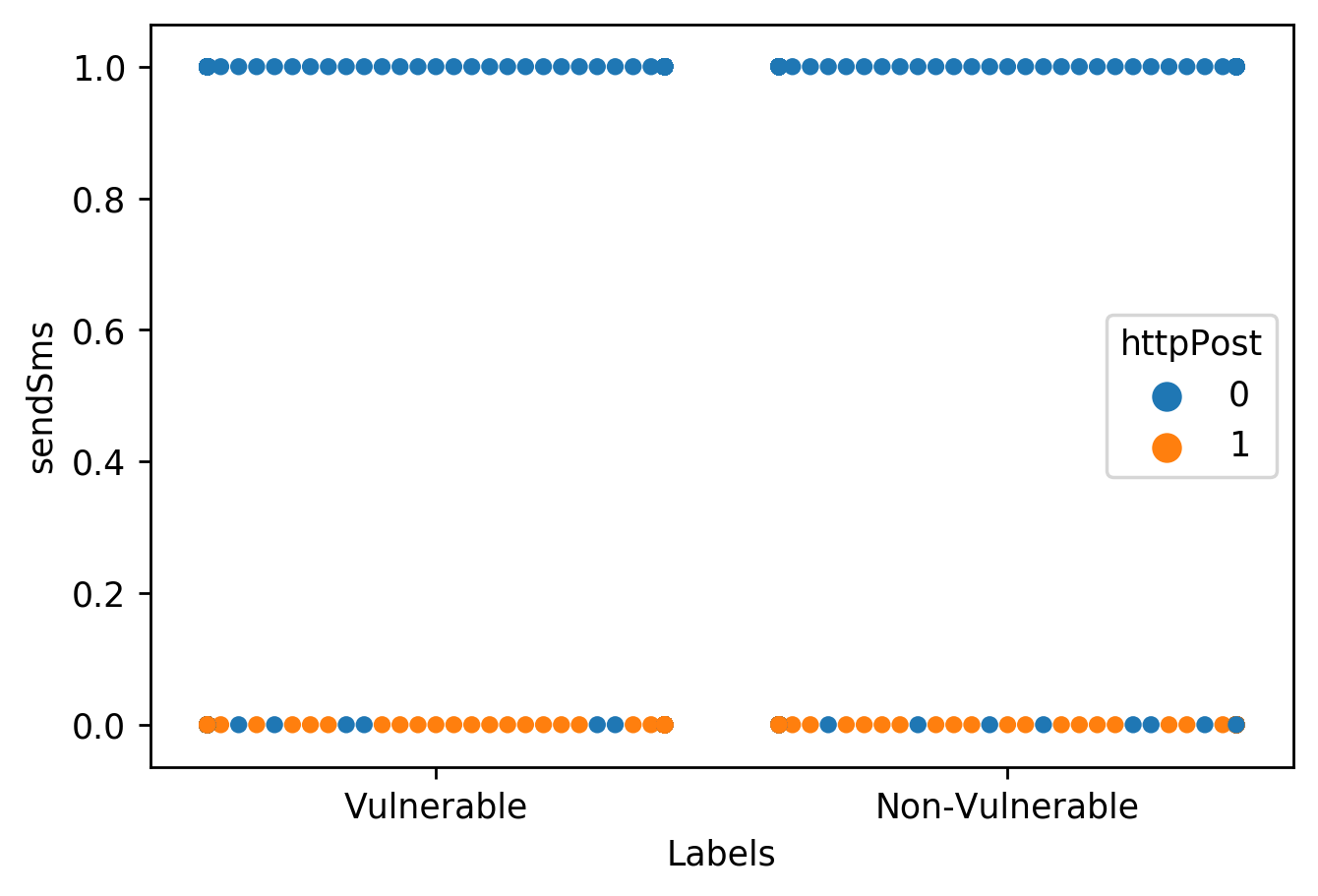}}
\caption{Frequency distribution of SendSMS w.r.t httpPost in all SmartApps of $Corpus2$}
\label{fig:sendSms_httpPost}
\end{figure}

The swarmplot for the SmartApps with most frequent sinks for $Corpus2$ is shown in the Figure \ref{fig:sendSms_httpPost}. 

\hlc[highlight]{It can be noted for two most frequent sinks of $Corpus2$ that the distribution of sinks across the vulnerable and non-vulnerable SmartApps is similar. Same is supported from the figures presented in Appendix} \ref{app:Third} \hlc[highlight]{for the distribution of sinks in $Corpus2$. The analysis of sinks is important and it can be concluded that presence of sinks is an important indicator for a SmartApp to be vulnerable. The vulnerable SmartApps may have more frequent instances of sinks. Similar analysis can be done to identify more tokens whose distribution across vulnerable and non-vulenrable SmartApps is different.}

\subsection{Models Testing}\label{sec:TestingBOW}
\hlc[highlight]{Learning a model with a machine learning algorithm is a complex task. Each instance of data contains knowledge and is useful for improving the model. Evaluation of a model helps to report its performance. A dataset that is previously not known to the model is required for the evaluation. The data should have neither been used for building the model nor for tuning features of the model. This can be done by dividing the historical data in two folds and use one fold for training the models with machine learning algorithms and the other for testing the models. This technique is called hold out cross validation which is an approximation of k-fold cross validation. More data points are kept in the training set for building models and relatively fewer data points are kept in the testing set.}

For training the machine learning algorithms to build models, the dataFrame prepared from set of vectors prepared in \hlc[highlight]{section} $\ref{subsec:token2Vec}$ and section \ref{sec:flow2Vec} is split in training and testing sets using stratified holdout cross validation. The algorithms are trained on the training set and then the test set is used to evaluate the built models. \hlc[highlight]{We have used the $train\_test\_split$ method from the $sklearn$ library} \cite{Ref21} which is an implementation of hold out cross validation. \hlc[highlight]{~

To compare the performance of machine learning algorithms we have ensured that the same training data is used to build the models for all machine learning algorithms and we have tested the built models on the same stratified hold out test data for all the algorithms. In order to find the best split in training and test data, we have tried various combinations of splitting the data starting from 51\% for training and 49\% for testing to 95\% for training and 5\% for testing. We ran the same experiment for all algorithms and found that all machine learning algorithms on the average perform well on the split of 70\% for training and 30\% for the testing. These sets are stratified on the labels of SmartApps which is achieved by setting the $stratify$ parameter to $True$ in the $train\_test\_split$ method.
}

\subsubsection{Evaluation Metrics}
\label{sec:EvaluationMetrics}
The built models are used to predict vulnerabilities in the SmartApps of the test set and the findings about the predictions are saved in a confusion matrix. All measures of a confusion matrix are shown in the Table \ref{tab:ConfusionMatrix}.

\begin{center}
\begin{table}[!ht]
\caption{\label{tab:ConfusionMatrix}Confusion Matrix}
\begin{tabular}{l|l|c|c|c}
\multicolumn{2}{c}{}&\multicolumn{2}{c}{Predicted Class}&\\
\cline{3-4}
\multicolumn{2}{c|}{}&Vulnerable&Non-Vulnerable&\multicolumn{1}{c}{Total}\\
\cline{2-4}
\multirow{2}{*}{Actual Class}& Vulnerable & $TP$ & $FN$ & $TP+FN$\\
\cline{2-4}
& Non-Vulnerable & $FP$ & $TN$ & $FP+TN$\\
\cline{2-4}
\multicolumn{1}{c}{} & \multicolumn{1}{c}{Total} & \multicolumn{1}{c}{$TP+FP$} & \multicolumn{    1}{c}{$FN+TN$} & \multicolumn{1}{c}{$N$}\\
\end{tabular}
\end{table}
\end{center}

A confusion matrix is a table which comprises of the following four measures: True Positives (TP), True Negatives (TN), False Positives (FP) and False Negatives (FN). 

\begin{description}
\item[TP:] When the actual class is vulnerable and the predicted class is also vulnerable.
\item[TN:] When the actual class is non-vulnerable and predicted class is also non-vulnerable.
\item[FP:] When actual class is non-vulnerable and predicted class is vulnerable.
\item[FN:] When actual class is vulnerable but predicted class in non-vulnerable.
\end{description}

To evaluate the learnt models, we have used the evaluation measures \hlc[highlight]{AUC, F1 Score, MCC,} and accuracy. Descriptions for all of \hlc[highlight]{these metrics} are provided in the following:

\textbf{Area Under The Curve (AUC)}
\hlc[highlight]{The AUC metric shows how good a model is at distinguishing between vulnerable and non-vulnerable applications. A good model has an AUC closer to 1, whereas, a poor model has an AUC near to 0.}

\textbf{Accuracy} is a ratio of correct predictions to the total predictions on the testing data. 
\[
Accuracy = \frac{TP+TN}{TP+TN+FP+FN}
\]





\textbf{F1 Score}
\hlc[highlight]{The $F_1$-Score is a weighted average of precision and recall and can be calculated by using the following formula.}
\[
F_1 = \frac{TP}{TP+ \frac{1}{2}(FP+FN)}
\]

\textbf{Matthews correlation coefficient (MCC)}
\hlc[highlight]{The MCC is high only if prediction is high in all confusion matrix categories. For this measure all classes have same importance, if you change a negative class to be positive, the MCC stays the same. It is calculated by using the following formula.}
\[
MCC = \frac{TP \times TN - FP \times FN}{\sqrt{(TP+ FP)(TP+ FN)(TN+ FP)(TN+ FN)}}
\]
\hlc[highlight]{The maximum value of MCC is 1 when both FP and FN are zero which represents a perfect positive correlation and it is -1 when both TP and TN are zero which represents a perfect negative correlation.}

\section{Evaluation}\label{sec:allEvaluation}

We have evaluated the proposed techniques by answering the following research questions.
\begin{description}

\item[RQ1:] \hlc[highlight]{How does automatically extracted features from the BoW of source code perform when the BoW representations of vulnerable and non-vulnerable applications are almost similar?}
\item[RQ2:] \hlc[highlight]{Whether a text mining approach is capable of extracting tainted flows features from SmartApps? and how such approach compares to static analysis techniques?
}
\item[RQ3:] \hlc[highlight]{What would be the impact of adding the tainted flows features automatically  extracted from text mining to automatically extracted features from the BoW?}
\end{description}

\hlc[highlight]{The questions are answered in the following sections.}

\subsection{Evaluation of ``BOW based Machine Learning"}
\label{sec:EvaluationBOW}
\hlc[highlight]{To address $RQ1$, this section presents an experiment using two datasets, the first dataset $Corpus1$ was captured from real world apps as described in section} \ref{subsubsec:dataset}. \hlc[highlight]{While the dataset in $Corpus2$ was produced from the benign subset of $Corpus1$ by injecting cases of information leakage using a mutation analysis framework. The goal of conducting the experiment in this section is to compare whether the BOW-based machine learning approach is capable of capturing the vulnerabilities introduced in $Corpus2$. As explained in section} \ref{subsec:informationLeakageVulnerability}, \hlc[highlight]{the changes introduced to the dataset were mainly by altering the order of statements in the code in order to produce the vulnerability that may leak sensitive information. The process of preparing numeric feature vectors for $Corpus1$ is initiated by preparing BoW from the source code of each SmartApp by using the technique described in section} \ref{tokenFrequencies}. The BoW of SmartApps are converted to feature vectors by using the tool Token2Vec presented in the \hlc[highlight]{section} \ref{subsec:token2Vec}. \hlc[highlight]{We have considered a token to be part of features if that is present in at least 4 SmartApps.}
The data prepared from feature vectors are split into training and testing sets that are stratified on the class of SmartApps by using the stratified hold out cross validation \hlc[highlight]{where the training data and the test data are 70\% and 30\% of the total number of SmartApps}.

\begin{figure}[!ht]
\centerline{\includegraphics[width=1\textwidth]{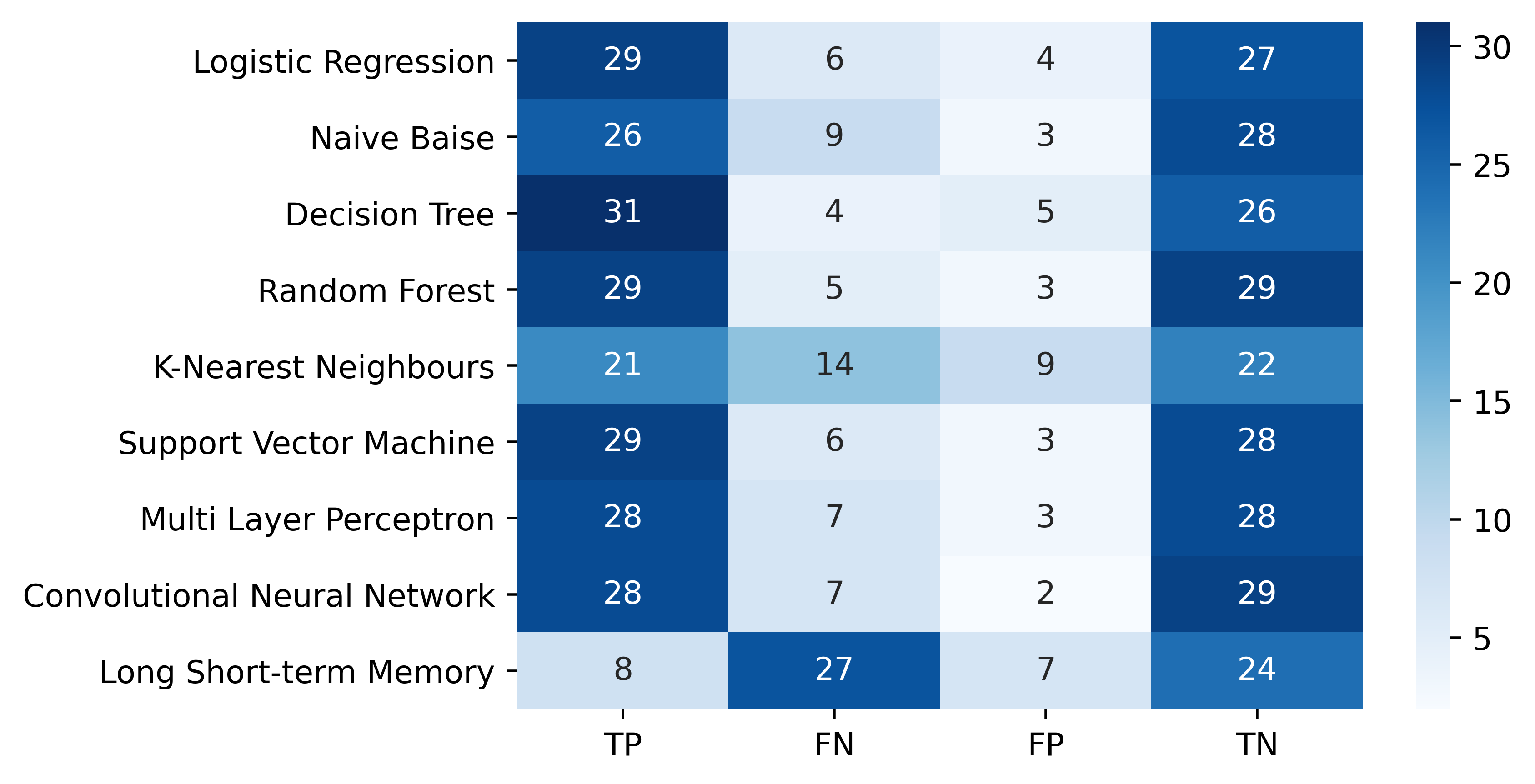}}
\caption{Confusion Matrices for $Corpus1$ Dataset}
\label{fig:CM_BOW}
\end{figure}

The set of machine learning algorithms listed in section \ref{sec:buildingModels} are tuned for this experiment. We have used the Python implementations of these algorithms available in the $sklearn$ library \cite{Ref21}. 
We have used the solver ($solver=`liblinear$') for Logistic Regression. For the Decision Tree, we have used entropy ($criterion=`entropy$'). We have used 112 and 120 estimators for Random Forest i.e. ($n\_estimators=112$) and ($n\_estimators=120$). For K-Nearest Neighbors, we have used K as 4. For Support Vector Machine, we have used a linear kernel i.e. ($kernel=`linear$'). 
\hlc[highlight]{We have used the Grid-search for hyperparameters tuning to identify and use optimal parameters.
We have used 12 and 8 hidden layers and ($solver=`lbfgs$') for the Multilayer Perceptron. 
For CNN, we have used Conv1D and set epochs to 20 and 10, and use relu as an activation with a 0.3 and 0.2 drop out rates. For LSTM, we have used the sequential model with 20 epochs, and relu and softmax as the activation function with 0.2 dropout rate.}

\hlc[highlight]{The prediction results for the set of classifiers on the $Corpus1$, in the form of confusion matrices, are shown in the Figure} \ref{fig:CM_BOW}.
From the confusion matrices, it can be observed that the Decision Tree algorithm has identified most number of true positive cases, whereas the random forest \hlc[highlight]{along Convolutional Neural Network has identified most number of true negative cases. The LSTM algorithm has most number of false negative cases and KNN has most number of false positive cases. The Decision Tree algorithm has identified least number of false negative cases and  Convolutional Neural Network algorithm has identified least number of false positive cases. From the confusion matrices, We can conclude that most of the machine learning algorithms performed well considering the true positive and true negative cases.}

\begin{figure}[!ht]
\centerline{\includegraphics[width=.7\textwidth]{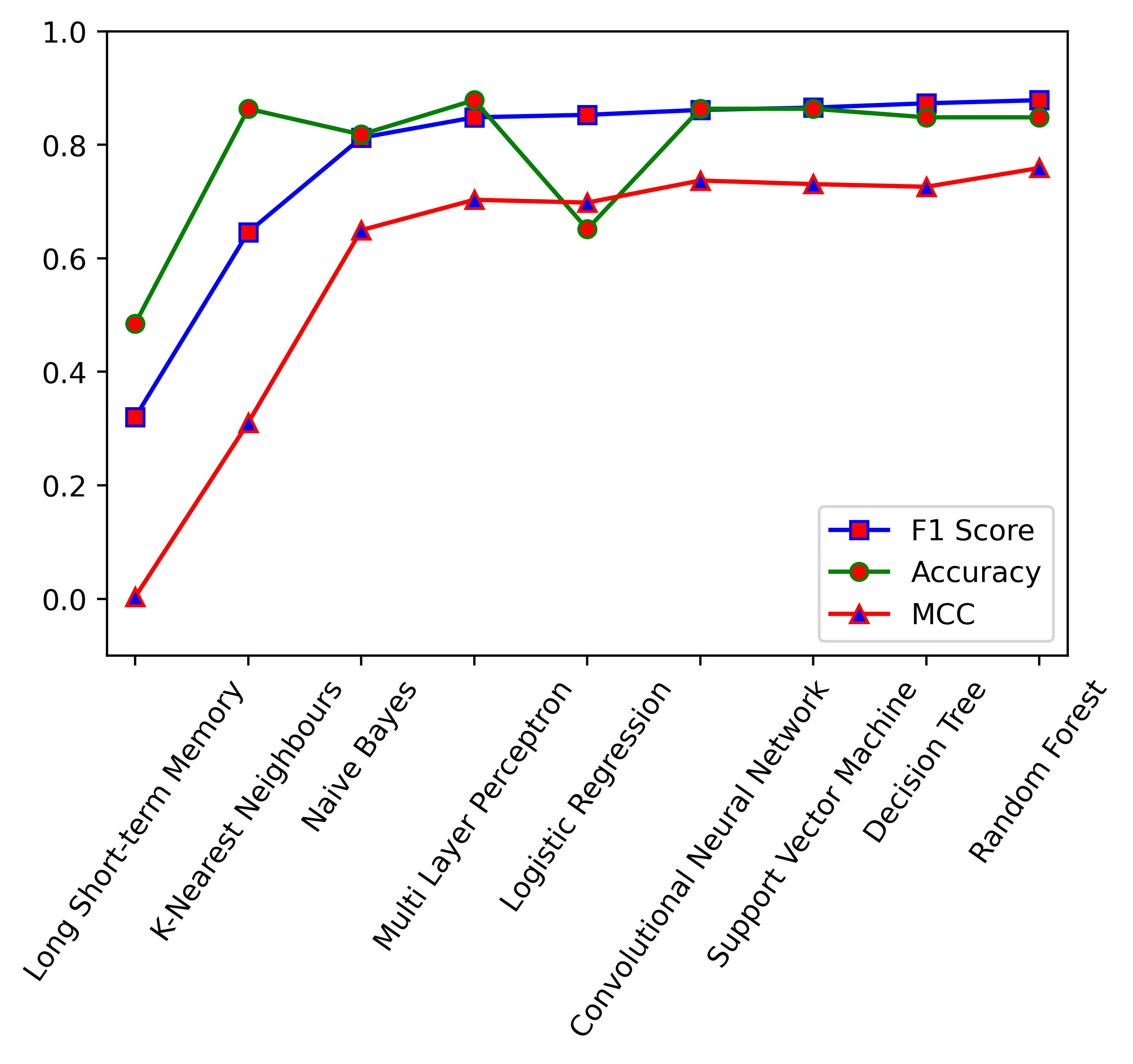}}
\caption{F1 Score, MCC and Accuracy for $Corpus1$ Dataset}
\label{fig:C1_metrics_BOW}
\end{figure}

\begin{figure}[!ht]
\centerline{\includegraphics[width=.7\textwidth]{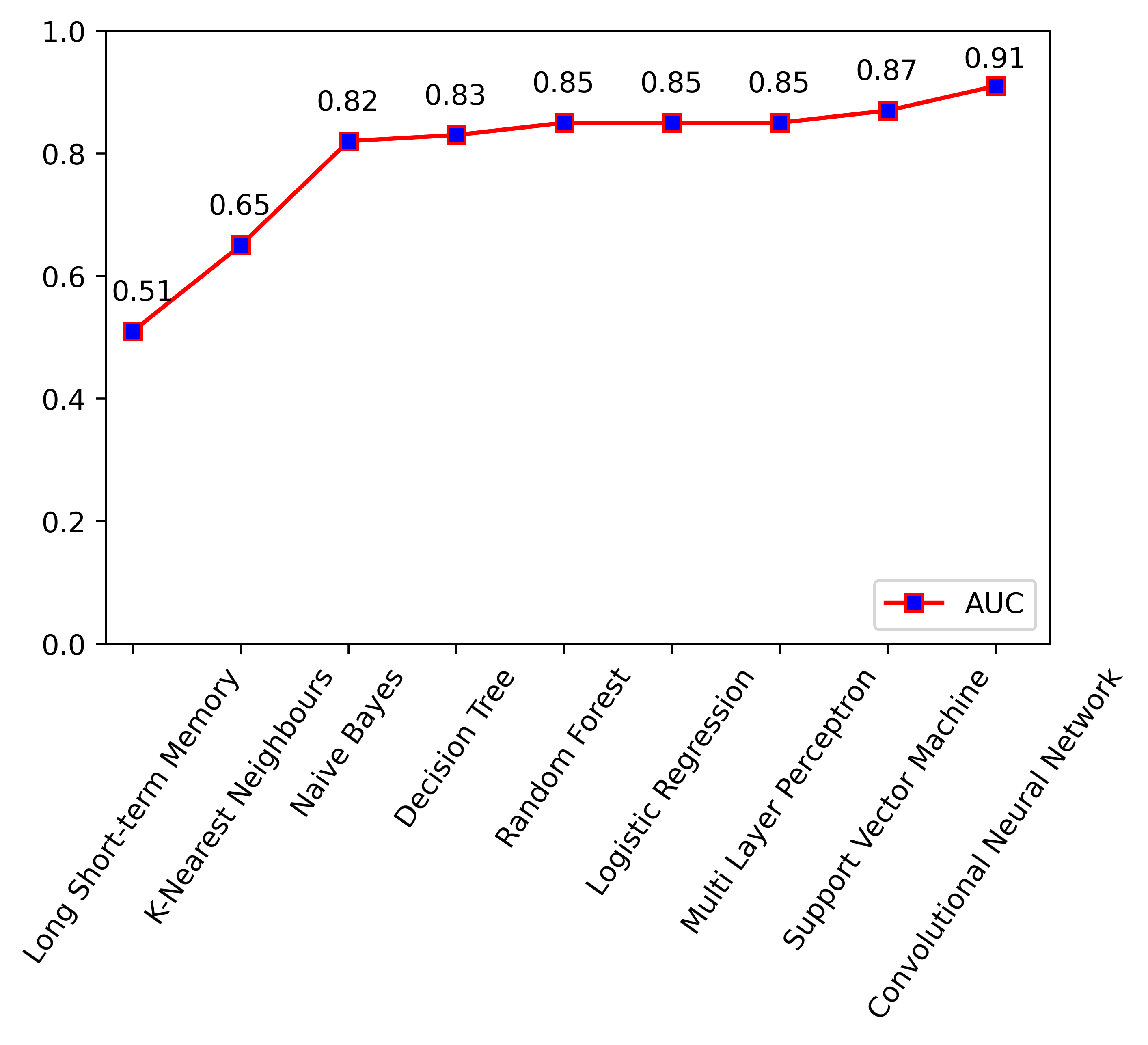}}
\caption{AUC for $Corpus1$ Dataset}
\label{fig:AUC_BOW_Corpus1}
\end{figure}

\hlc[highlight]{The F1 Score, MCC and Accuracy are shown in the Figure} \ref{fig:C1_metrics_BOW}.
\hlc[highlight]{The Random Forest algorithm performs the best considering the MCC and F1 Score, whereas, Multilayer perceptron performs the best considering the accuracy metric. The LSTM algorithm performs the worst considering all of the metrics. }
\hlc[highlight]{From Figure} \ref{fig:AUC_BOW_Corpus1} \hlc[highlight]{ and in terms of AUC, we can note that the 0.91 is AUC for the CNN. We can also observe that most of the algorithms have a reasonable performance with the $Corpus1$ dataset. The K-Nearest Neighbors and LSTM algorithms were not able to perform like the others. We were able to get these results with a little effort on hyperparameter tuning.} 

\hlc[highlight]{We believe that performance of the algorithms that have not performed well can be improved by feature engineering and hyperparameter tuning. The K-Nearest Neighbors algorithm considers all features as equally important to calculate the distances from K nearest neighbors to predict a target class. Increasing weights of important features and decreasing weights of less important features may improve the AUC of this algorithm.}
\hlc[highlight]{Considering all the metrics presented for this set of experiments, it can be concluded that BoW can be successfully used to detect vulnerable SmartApps for the $Corpus1$.}

\begin{figure}[!ht]
\centerline{\includegraphics[width=.75\textwidth]{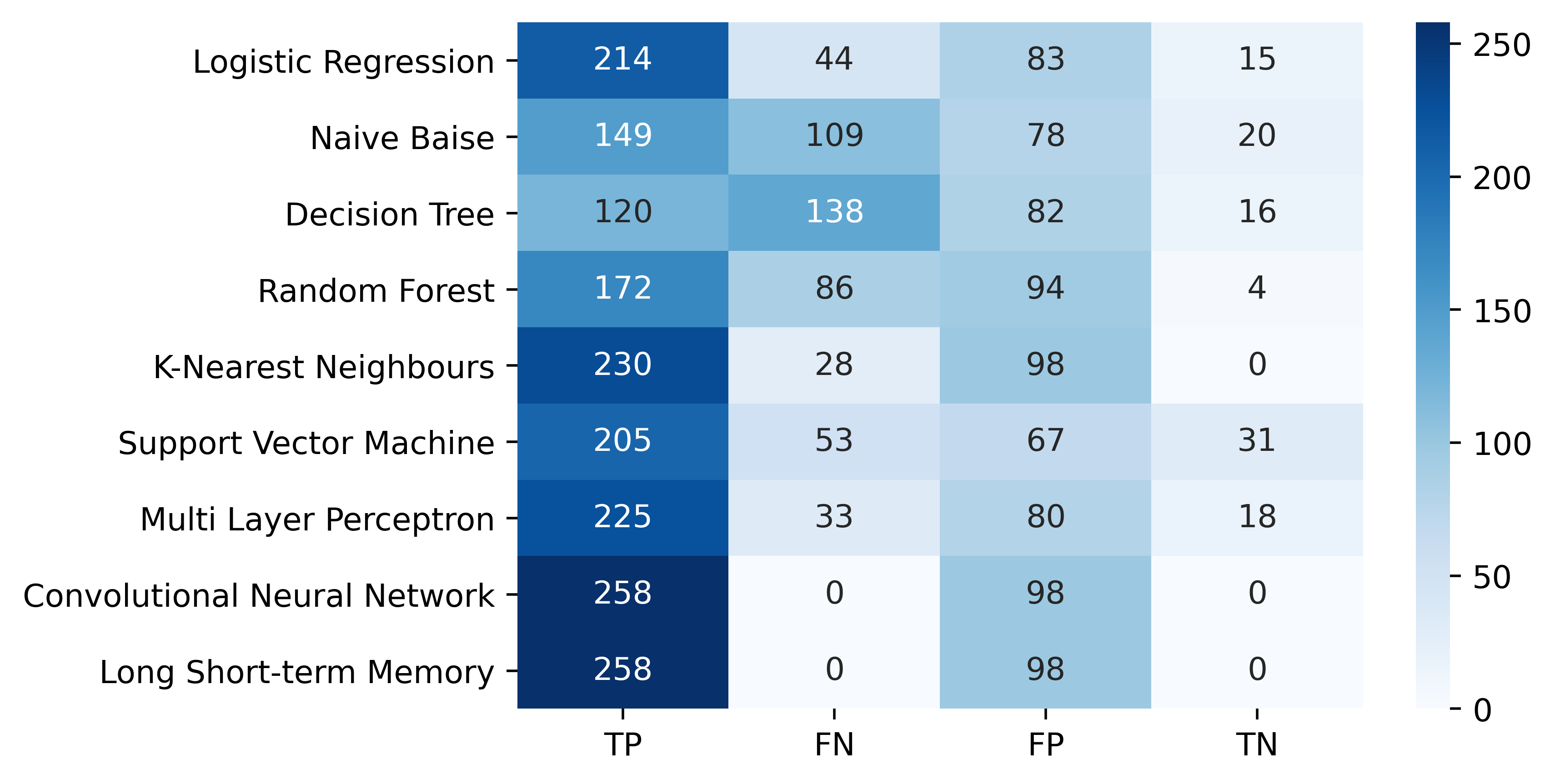}}
\caption{Confusion Matrices for $Corpus2$ Dataset}
\label{fig:CM_Mut_BOW}
\end{figure}

\hlc[highlight]{The second experiment is conducted on $Corpus2$. The BOW representation of vulnerable and non-vulnerable applications of this dataset end up with almost similar tokens}. 
We prepare the feature vectors data and divide the data into training and testing data exactly like the first experiment. We have used the same threshold like the first experiment to select a token to be a feature for building the models. 
The models built with the classifiers are tested on the testing data and the prediction results in the form of confusion matrices are presented in the Figure \ref{fig:CM_Mut_BOW}. 
From the confusion matrices, it can be noted that most of the classifiers are predicting the majority class. Support Vector Machine is the algorithm which is able to predict the most number of correct negative SmartApps.  

\begin{figure}[!ht]
\centerline{\includegraphics[width=.65\textwidth]{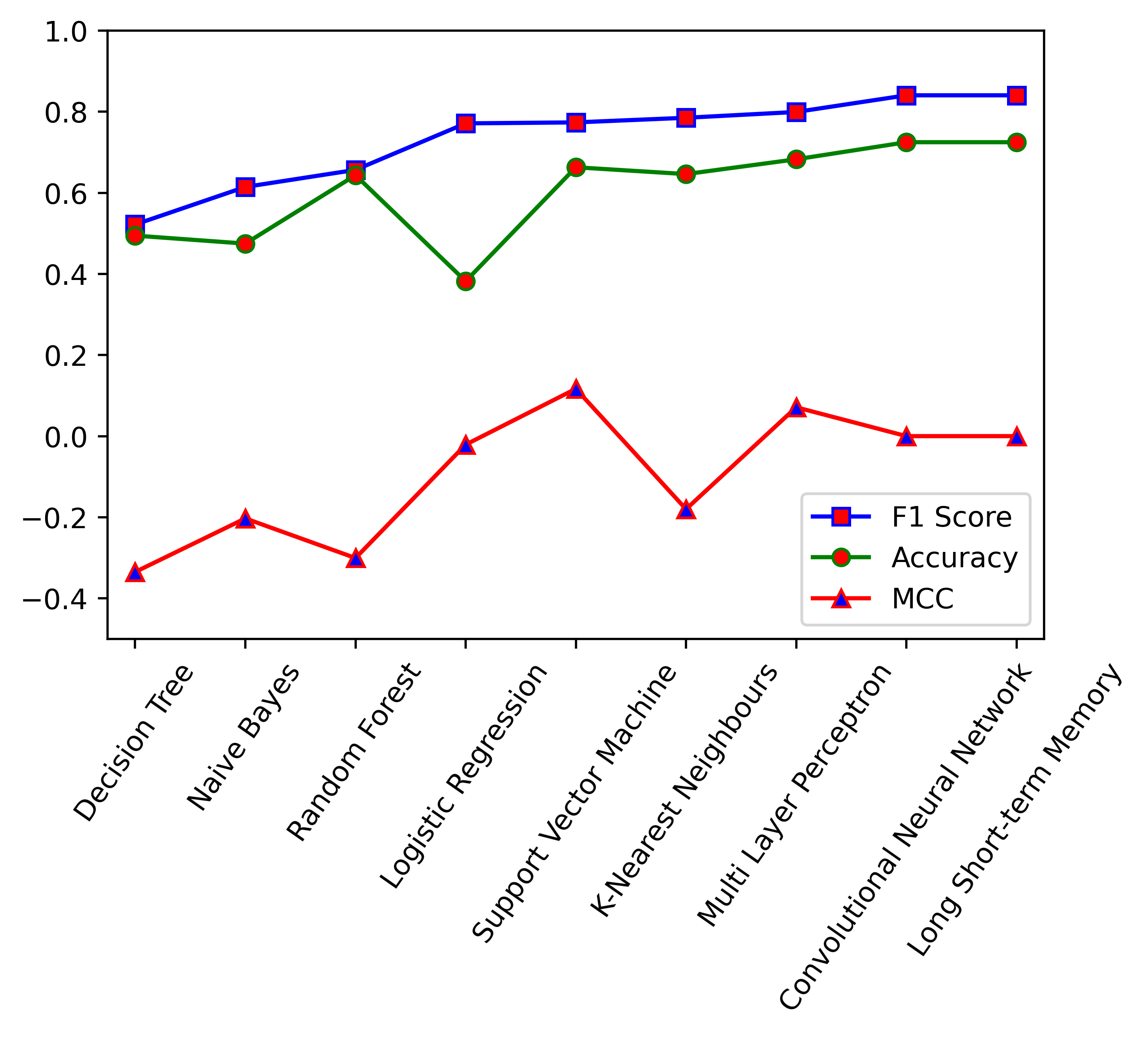}}
\caption{F1 Score, MCC and Accuracy for $Corpus2$ Dataset}
\label{fig:C2_metrics_BOW}
\end{figure}

\begin{figure}[!ht]
\centerline{\includegraphics[width=.8\textwidth]{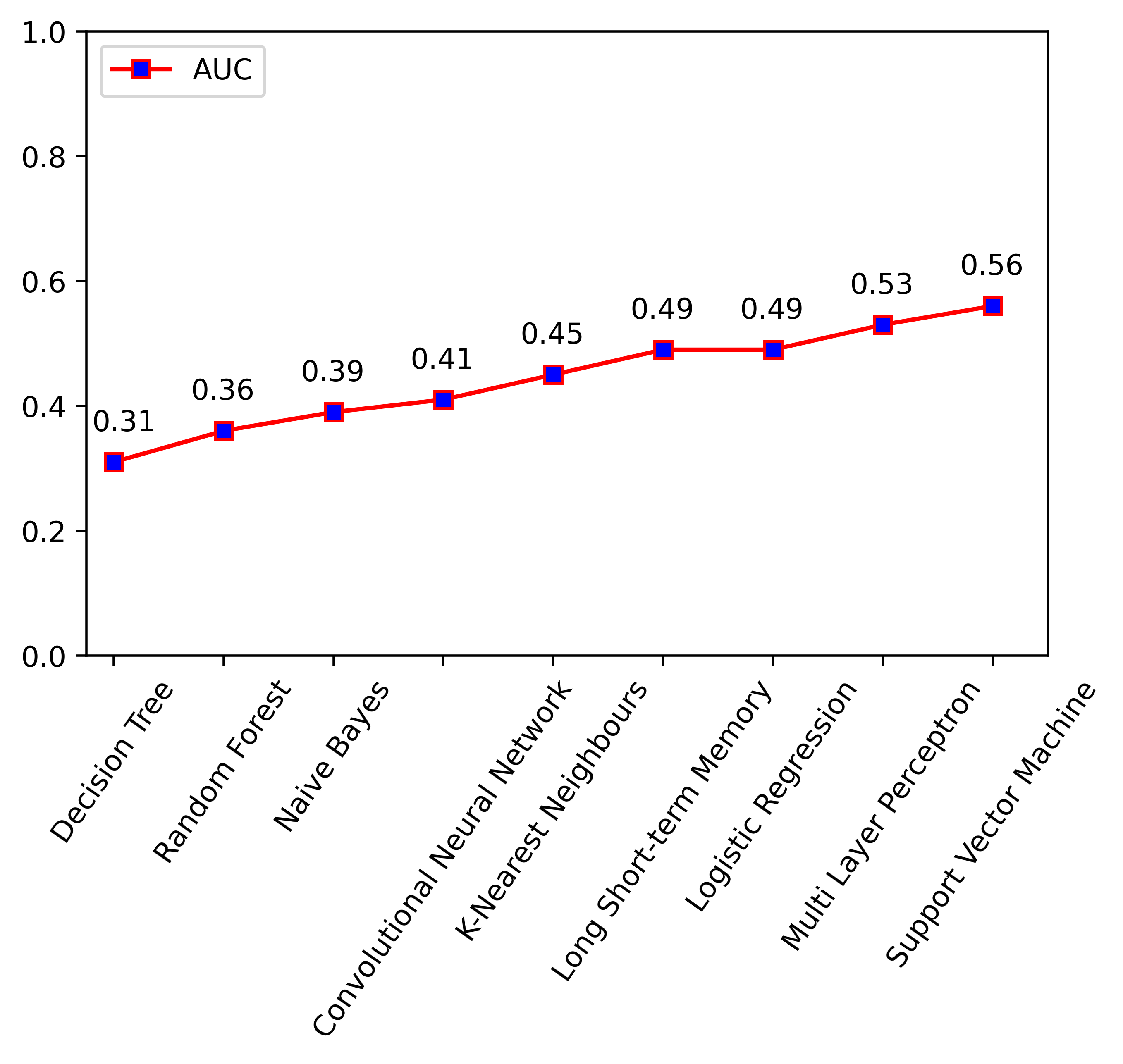}}
\caption{AUC for $Corpus2$ Dataset}
\label{fig:APR_Mut_BOW}
\end{figure}

\hlc[highlight]{For the $Corpus2$, F1 Score, MCC and Accuracy are shown in the Figure} \ref{fig:C2_metrics_BOW}. \hlc[highlight]{It can be noted that the LSTM algorithm performs the best considering all of these measures, but, this algorithm predicts only the majority class.}
\hlc[highlight]{It can be noted from Figure} \ref{fig:APR_Mut_BOW} \hlc[highlight]{that the best performer is  Support Vector Machine with 0.56 AUC. The AUC of other algorithms is also very low, it is 0.31 in one case. we can observe that the performance of the built models for all classifiers deteriorates for the $Corpus2$. This answers the $RQ1$, the automatically extracted features from the BoW of source code does not perform well when BoW representations of vulnerable and non-vulnerable SmartApps are almost similar}.
\hlc[highlight]{From the sink token analysis presented in section} \ref{subsubsec:dataset}, \hlc[highlight]{we can note that for $Corpus1$ frequency distribution of the sink tokens in vulnerable and non-vulnerable SmartApps are different, whereas, it is almost the same for $Corpus2$.} 

We have learned from these experiments that we can use \hlc[highlight]{BOW of a SmartApp to predict whether a SmartApp is vulnerable or non-vulnerable}. But, it can be problematic to identify the vulnerable SmartApps if the sequence of some lines of source code change an application from non-vulnerable to vulnerable. If distribution of all tokens for both vulnerable and non-vulnerable SmartApps are approximately the same then it becomes hard for a classification algorithm to distinguish between them only on the basis of features extracted from BoW. 
\hlc[highlight]{This result demonstrates the need for finding more features from the SmartApps that can help the machine learning algorithms to build correct behavioural models. For that reason, we have proposed to extract the tainted flows using text mining and use them as features.}

\subsection{Evaluation of Tainted Flows Identification \hlc[highlight]{Using} Text Mining}\label{sec:TaintedFlowsEvaluation}
\hlc[highlight]{In this section we answer research question $RQ2$.}
To report a SmartApp to be vulnerable, we need to find features which can help machine learning algorithms to distinguish it from non-vulnerable SmartApps. In section \ref{sec:featuePrep}, we  have presented a technique that uses BoW as features of a SmartApp. After evaluating the technique on $Corpus2$, we noted that the approach may fail when changing order of statements in a SmartApp  makes the app non-vulnerable, where initially it was vulnerable. 

Tainted flow analysis of SmartApps are required in such scenarios. Tools like SAINT proposed by Celik et al. \cite{Ref8}
perform static analysis to find such flows. They build some dependency graphs and then used algorithms to reduce the graphs to flows. 
We have proposed a text mining alternative which is less expensive as it requires less time to extract the flows. The technique focuses on tracking only the information that can end up in a tainted flow. We have evaluated the technique by implementing a tool, FlowsMiner, presented in section \ref{sec:flow2Vec}. The experiments with FlowsMiner were conducted on the datasets $Corpus1$ and $Corpus2$ presented in \hlc[highlight]{section} \ref{subsubsec:dataset}. The experiments were run on a laptop machine with Intel Core i5-7200 CPU @ 2.5GHz with a 8.00 GB RAM. The specifications of the machine are more or less similar to the machine used by Celik et al. \cite{Ref8}
\hlc[highlight]{for} their experiments. The results calculated by us are presented in the Table \ref{tab:flows}.

\begin{center}
\begin{table}[!ht]
\caption{\label{tab:flows}Duration to identify Flows in both $Corpus1$ and $Corpus2$ Datasets}
\begin{tabular}{ |c|c|c|c|c|c|c|c| } \hline

\textbf{Dataset} & \textbf{Sc\_Sn} & \textbf{eSc\_Sn} & \textbf{Sc\_eSn} & \textbf{eSc\_eSn} & \textbf{Sn\_C} & \textbf{eSn\_C} & \textbf{Duration}\\
\hline
\textbf{\emph{Corpus1}} & 159 & 35 & 2 & 61 & 145 & 50 & $\sim$17.5s\\ 
\hline
\textbf{\emph{Corpus2}} & 4 & 858 & 0 & 0 & 7 & 0 & $\sim$128s\\
\hline
\end{tabular}
\end{table}
\end{center}

It can be noted that the proposed technique is able to extract flows from all 217 SmartApps of $Corpus1$ in around 17.5 secondes whereas Celik et al. report that their static analysis tool requires 23$\pm$5 seconds for only one SmartApp of the dataset $Corpus1$ \cite{Ref8}.
The extraction of flows with our technique for complete data set takes less time than extracting flows for one SmartApp using the static analysis. The FlowsMiner is able to extract flows from 1196 SmartApps of dataset $Corpus2$ in around 128 seconds.
The proposed text mining technique implemented as FlowsMiner to identify the flows in SmartApps answers the research question $RQ2$ by providing an alternative to the expensive static analysis. The number of flows identified by the FlowsMiner are presented in the Table \ref{tab:flows} by dividing them into 6 categories. Where source to sink, extended source to sink, source to extended sink, extended source to extended sink, sink in a body of tainted conditional and extended sink in a body of a tainted conditional are denoted by $Sc\_Sn$, $eSc\_Sn$ ,	$Sc\_eSn$ , $eSc\_eSn$ , $Sn\_C$ and $eSn\_C$, respectively.


\subsection{Evaluation of Flows-aware Machine Learning}\label{sec:evaluationF2V}
\hlc[highlight]{To answer $RQ3$ and observe the impact of combining tainted flows features with the BoW features, we run the experiments on $Corpus1$ and $Corpus2$ datasets presented in section} \ref{subsubsec:dataset}. The features set is prepared by converting BoW to numeric vectors $\mathbb{V}$ as presented in \hlc[highlight]{section} \ref{subsec:token2Vec}.
\hlc[highlight]{The flows are extracted using FlowsMiner tool presented in section} \ref{sec:FlowsMiner}
\hlc[highlight]{and are converted to feature vectors $FLs$ using Flow2Vec tool presented in the section} \ref{sec:flow2Vec}. \hlc[highlight]{The vectors} $\mathbb{V}$ \hlc[highlight]{and $FLs$ are combined together as} $\mathbb{V \cup}$ \hlc[highlight]{$FLs$}. The output of this step, numeric feature vectors for all SmartApps are used for training the machine learning algorithms to build the behavioural models of the SmartApps. The numeric feature vector datasets prepared for $Corpus1$ are split into stratified training and testing samples on class variables of the SmartApps with 70\% and 30\% samples, respectively. 

\begin{figure}[!ht]
\centerline{\includegraphics[width=.9\textwidth]{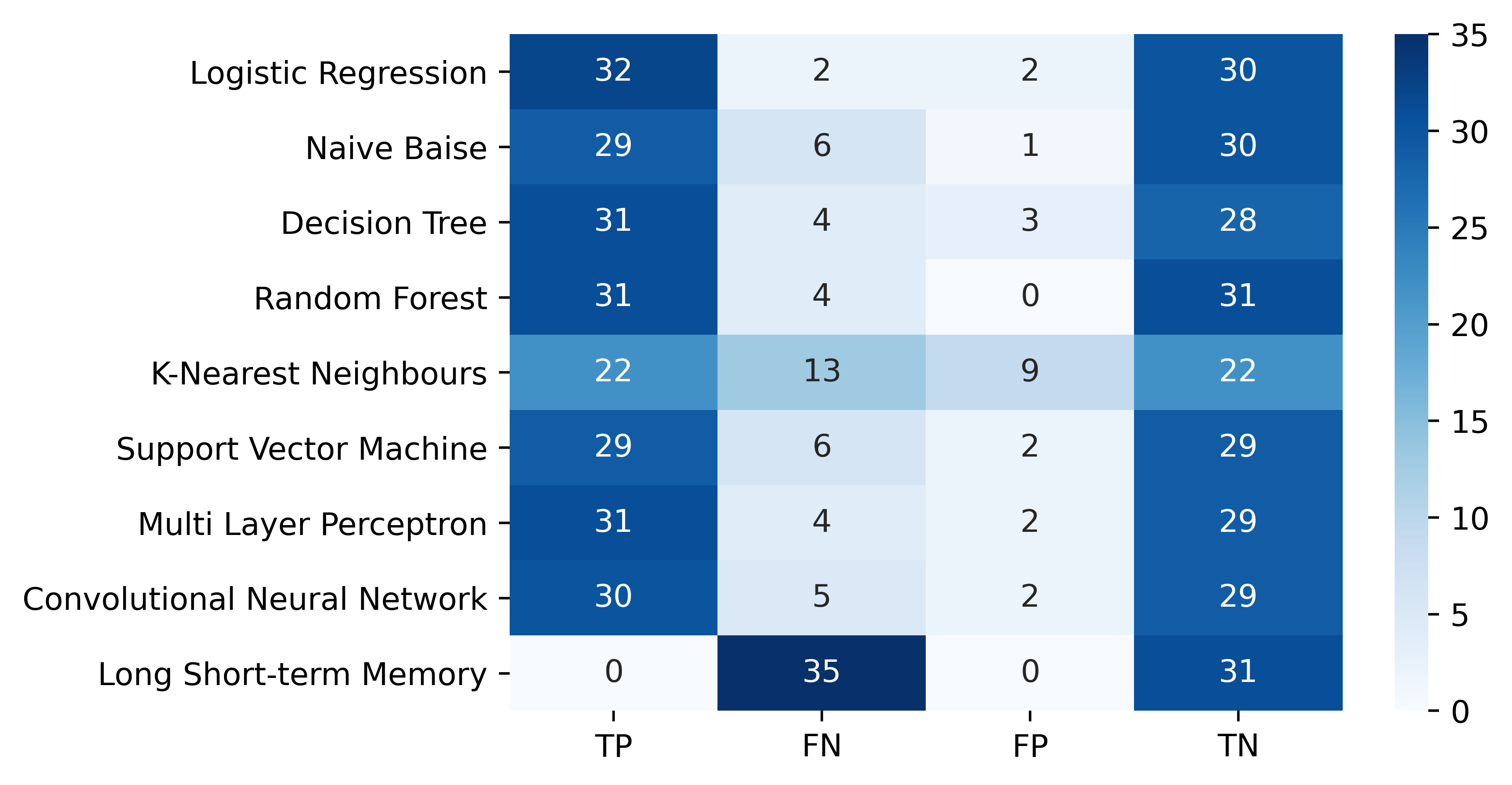}}
\caption{Confusion Matrices for $Corpus1$ Dataset}
\label{fig:CM_BOW_Flows}
\end{figure}

The set of machine learning algorithms are tuned to get best performance results for this experiment. We have used the Python implementations of these algorithms available in the $sklearn$ library \cite{Ref21}. 
By hyperparameters tuning, we found that for this experiment the Logistic Regression requires ($solver=`liblinear$') solver. For the Decision Tree, we have used entropy ($criterion=`entropy$'). We have used 112 and 120 estimators for Random Forest i.e. ($n\_estimators=112$) and ($n\_estimators=120$). For K-Nearest Neighbors, we have used K as 5. For Support Vector Machine, we have used a linear kernel i.e. ($kernel=`linear$'). We have used 4 and 9 hidden layers, and ($solver=`lbfgs$') for the Multilayer Perceptron. For CNN, we have used Conv1D and set epochs to 20 and 13, and use relu as an activation with a 0.2 and 0.2 drop out rates. For LSTM, we have used the sequential model with 20 epochs and softmax as the activation function with 0.2 dropout rate. 

The confusion matrices of this experiment are presented in the Figure \ref{fig:CM_BOW_Flows}.  
\hlc[highlight]{From the confusion matrices, it can be noted that the Random Forest has the most number of true positive and true negative cases. The LSTM is predicting only the negative class.}

\begin{figure}[!ht]
\centerline{\includegraphics[width=.65\textwidth]{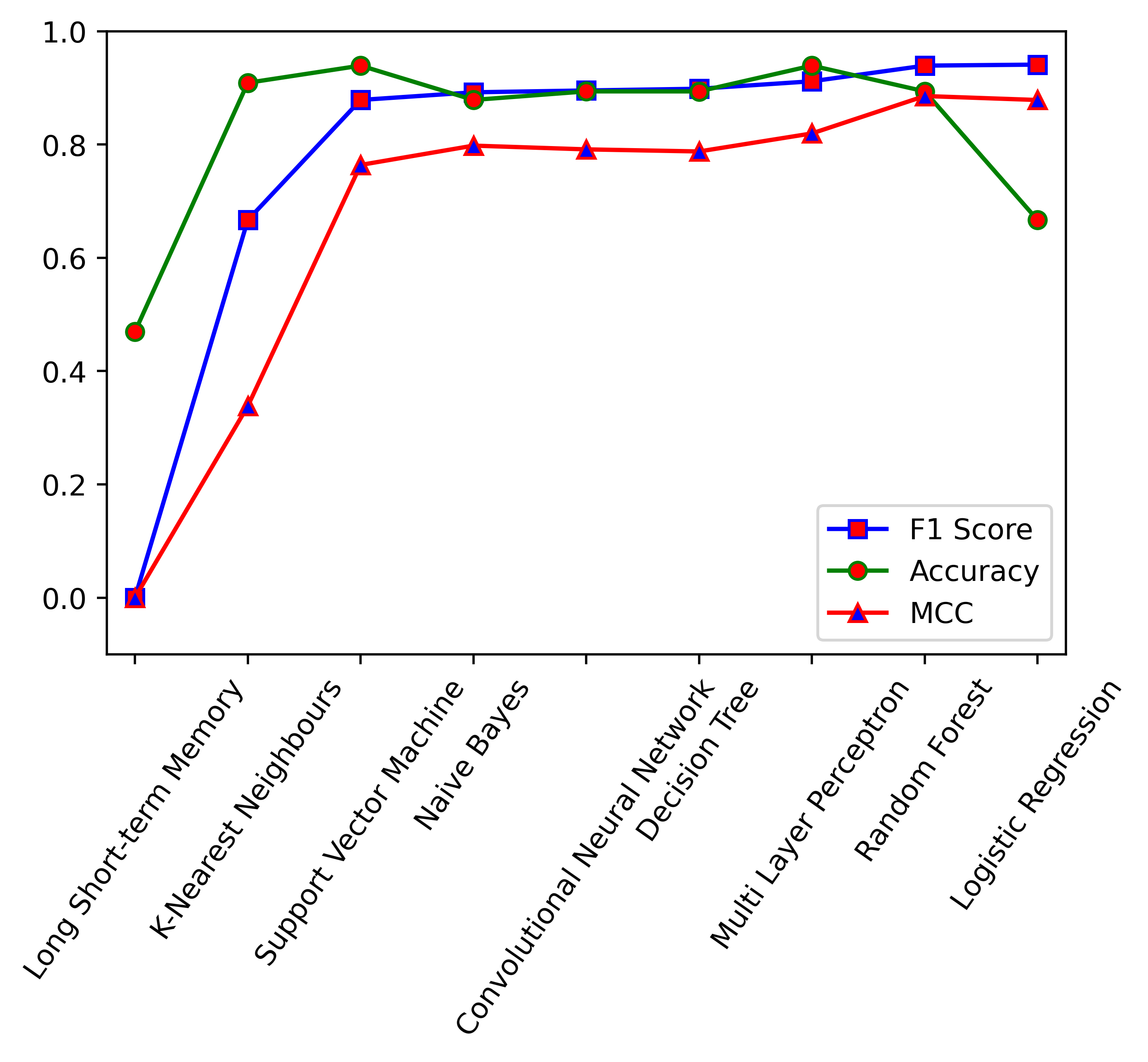}}
\caption{F1 Score, MCC and Accuracy for $Corpus1$ Dataset}
\label{fig:C1_Flows}
\end{figure}

\hlc[highlight]{From Figure} \ref{fig:C1_Flows}, \hlc[highlight]{it can be noted that the Logistic Regression algorithm performs the best considering the F1 Score and MCC, whereas, SVM performs the best considering the Accuracy metric.}

\begin{figure}[!ht]
\centerline{\includegraphics[width=.65\textwidth]{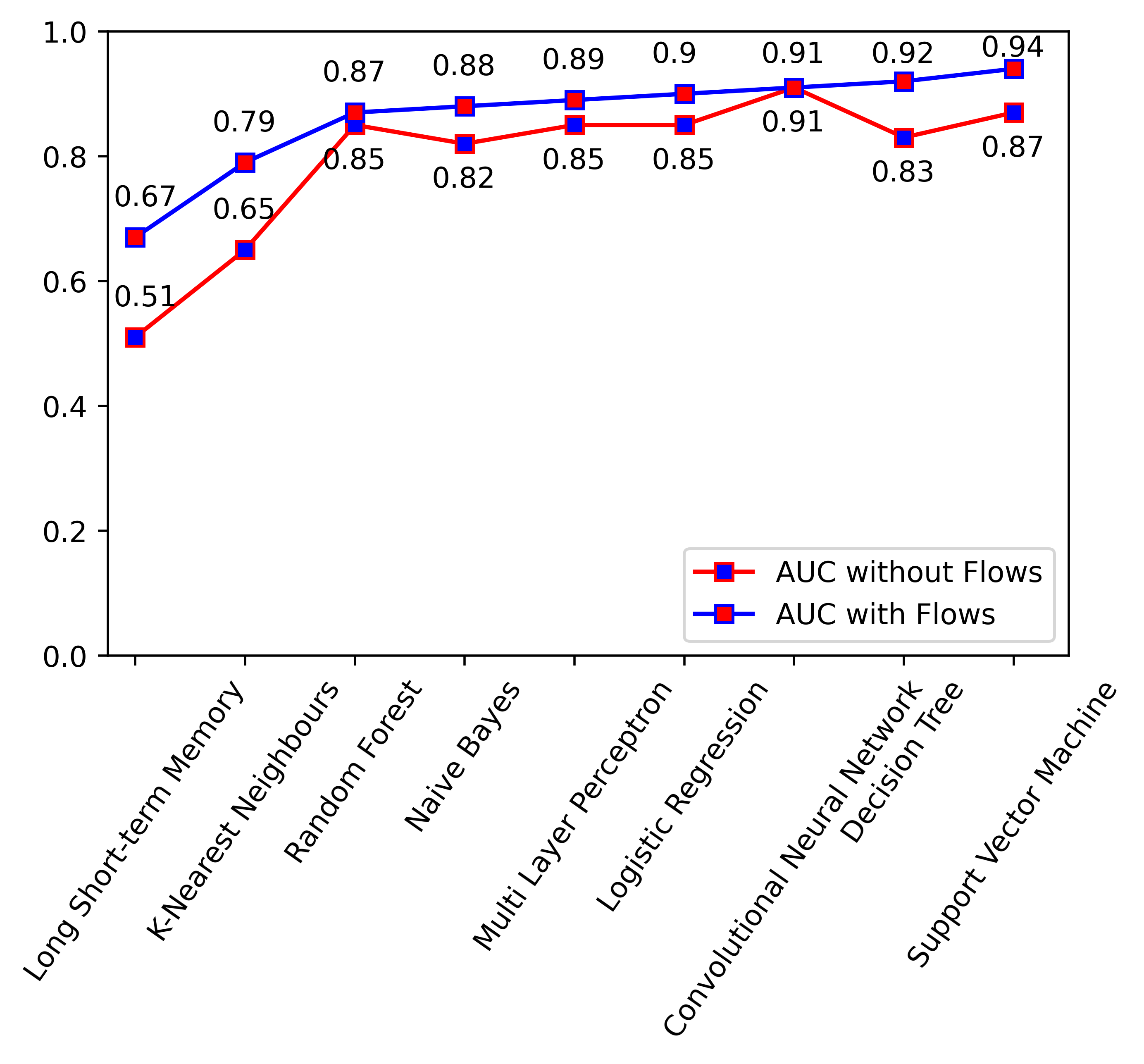}}
\caption{AUC for $Corpus1$ Dataset with/without flows}
\label{fig:Both_ACU_Line_BOW_Flows}
\end{figure}

\hlc[highlight]{The AUC calculated for all classifiers show that the Logistic Regression has the highest AUC of 0.94 as shown in Figure} \ref{fig:Both_ACU_Line_BOW_Flows}. \hlc[highlight]{As such, we can observe that the other algorithms also have high AUC with an exception of the LSTM.}
\hlc[highlight]{The AUC comparison for the set of machine learning algorithms with the experiments conducted without considering tainted flows are presented in Figure} \ref{fig:Both_ACU_Line_BOW_Flows}.
\hlc[highlight]{It can be noted that if the tainted flows are used for building and testing the models, the AUC for all machine learning algorithms improves. This answers research question \textbf{RQ3}}. We conclude that: the identification of tainted flows in the SmartApps and their usage for building models improves the performance of machine learning algorithms to identify vulnerable SmartApps.

\begin{figure}[!ht]
\centerline{\includegraphics[width=0.8\textwidth]{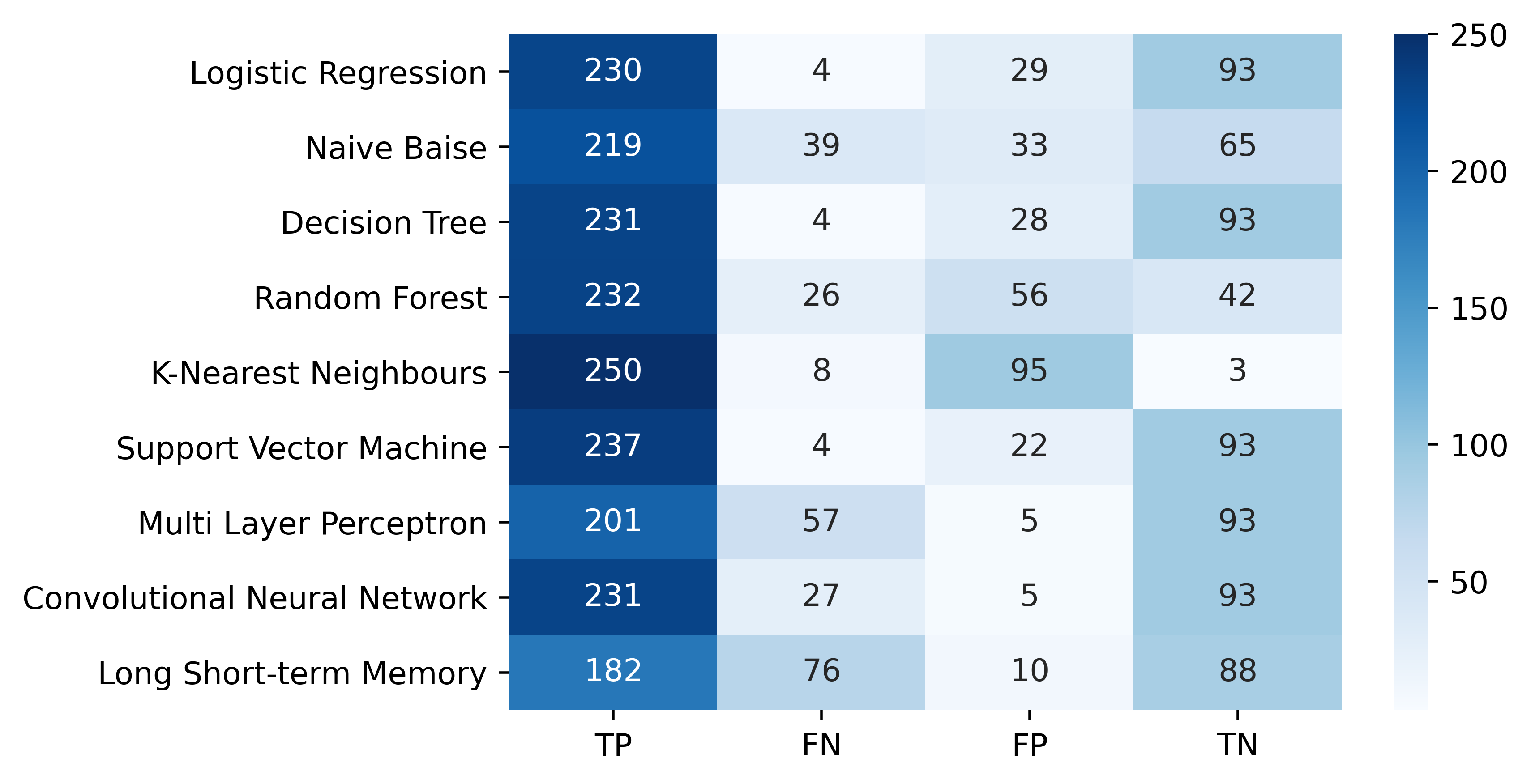}}
\caption{Confusion Matrices for $Corpus2$ Dataset}
\label{fig:CM_Mut_BOW_Flows}
\end{figure}

\hlc[highlight]{To answer the same research question on a different dataset, we have conducted our experiments on $Corpus2$, where authors of the dataset have introduced some mutations to convert non-vulnerable SmartApps to vulnerable. The changes were mainly focused on introducing  structural changes to the program such as statements reordering in the sequential program flow, the conditional flow, and contextual flow, i.e., method calls. Changes in some cases would add few statements to the original program but those statements belongs to the same vocabulary of the original code, as such, the BOW representation for both original code and mutants of the SmartApps have almost similar tokens.} 
\begin{figure}[b!]
\centerline{\includegraphics[width=.8\textwidth]{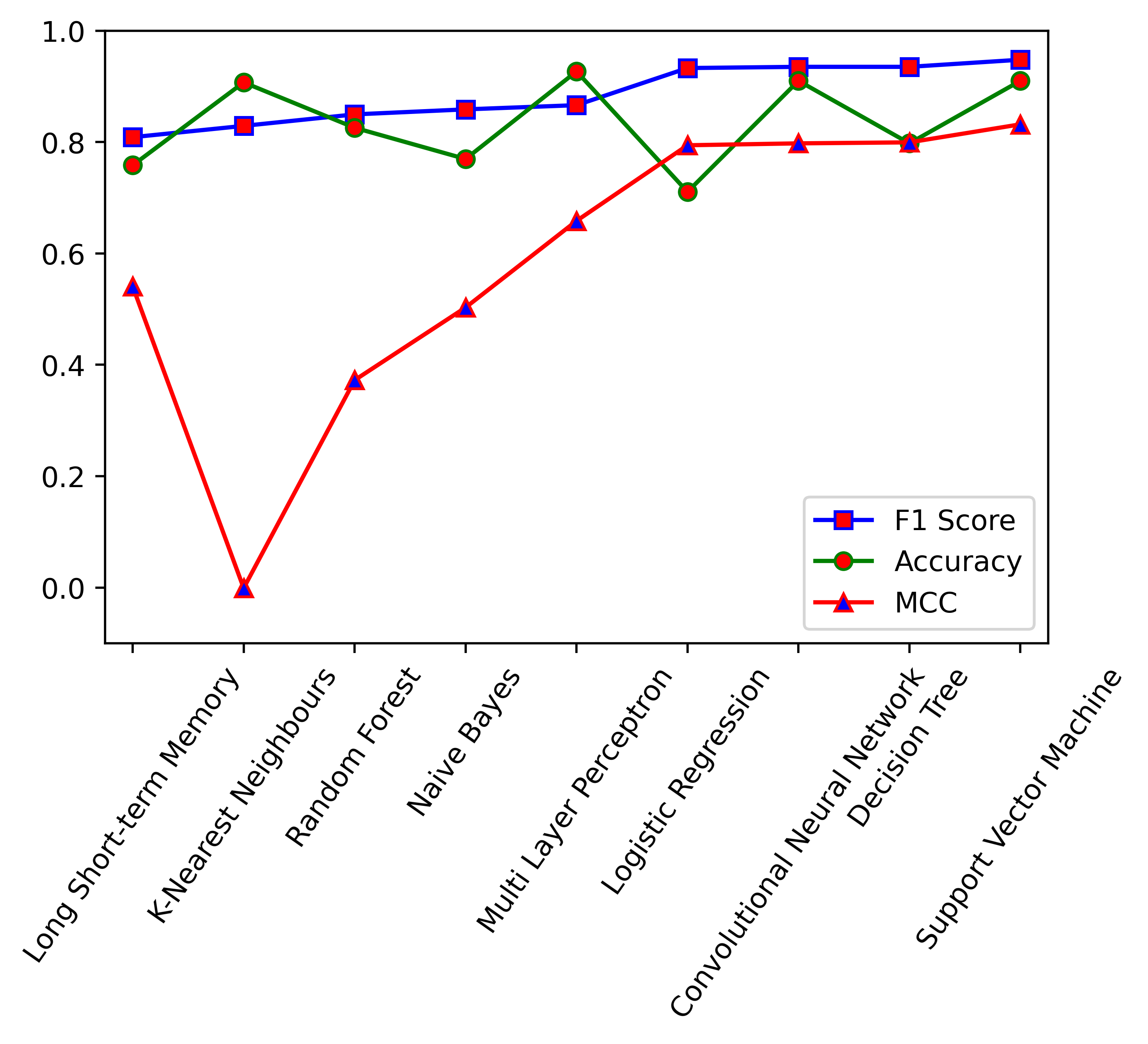}}
\caption{F1, MCC and Accuracy for $Corpus2$ Dataset}
\label{fig:C2_Flows}
\end{figure}
The feature vectors are prepared exactly like we prepared the feature vectors in our first experiment presented at the \hlc[highlight]{beginning} of this section. Once the feature vectors are prepared they are split into training and testing samples which are stratified on the class labels of the SmartApps. The size of the training set is 70\% and size of the test set is 30\% of the size of actual $Corpus2$ dataset.
The results by training the set of machine learning algorithms and then testing them on test sets are presented in the confusions matrices presented in Figure \ref{fig:CM_Mut_BOW_Flows}. It can be noted that the most number of true positive cases are identified by the K-Nearest Neighbors algorithm. We can also note that the same algorithm identifies most number of false positive cases. 
\begin{figure}[t!]
\centerline{\includegraphics[width=.80\textwidth]{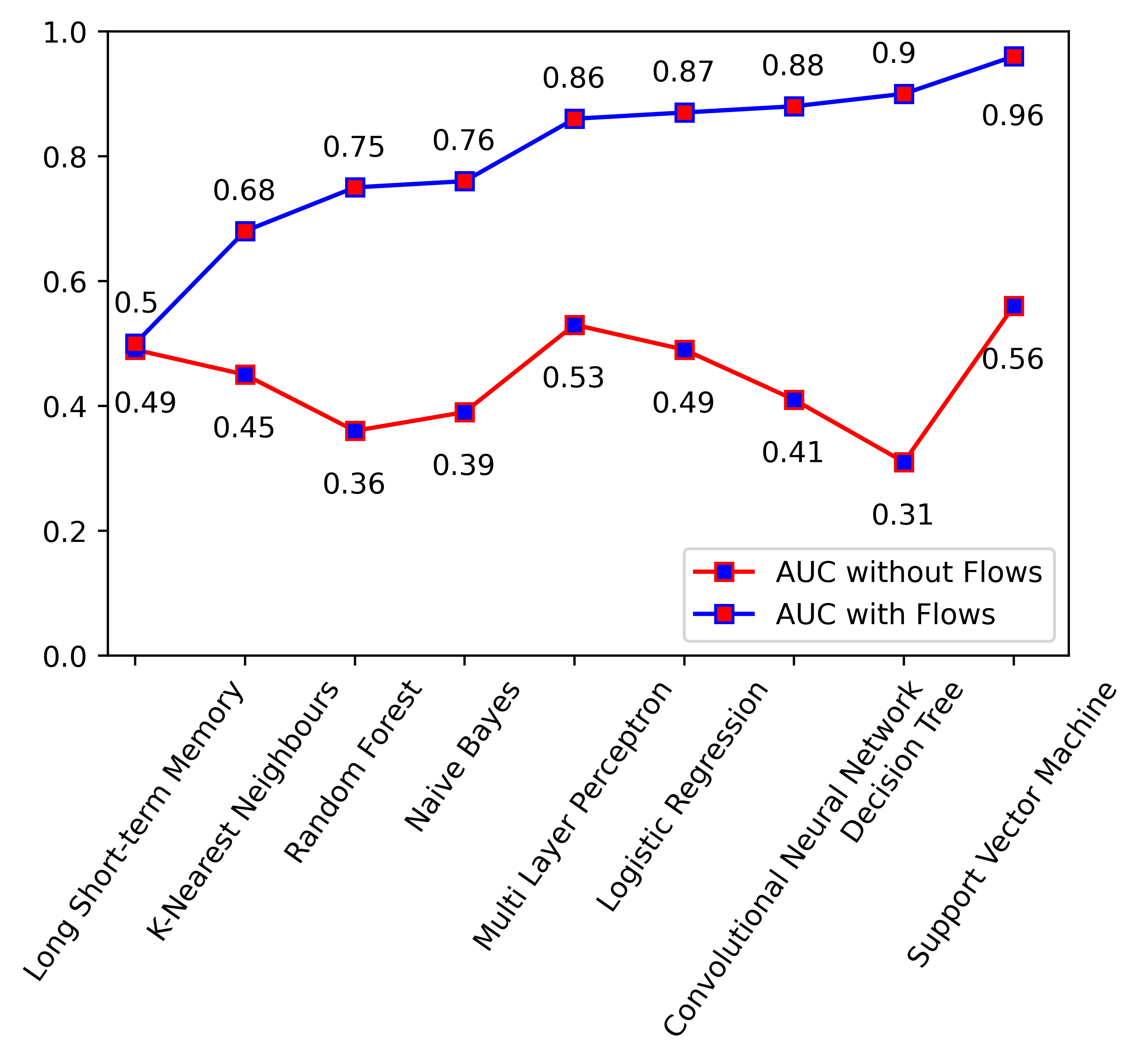}}
\caption{AUC for $Corpus2$ Dataset with/without flows}
\label{fig:Both_ACU_Line_Mut_BOW_Flows}
\end{figure}

It can be noted that the algorithm is predicting the majority class. This is because $Corpus2$ is not a balanced dataset and the algorithm is predicting by considering neighbors of the test data. In $Corpus2$, the vulnerable application are 3 times more than the non-vulnerable applications which can be noted from the \hlc[highlight]{section} \ref{subsubsec:dataset}.

We can also note that there are five algorithms that are predicting the same number of true negative cases and the algorithm from these five with high  number of true positive cases is Support Vector Machine. Which makes the Support Vector Machine, best predictor along CNN for this dataset.
\hlc[highlight]{The Support Vector Machine Algorithm has highest F1 Score and MCC and MLP has highest Accuracy, as presented in Figure} \ref{fig:C2_Flows}.
\hlc[highlight]{The comparison for the AUC of the machine learning algorithm's
by using the flows features and without using them is presented in Figure}\ref{fig:Both_ACU_Line_Mut_BOW_Flows}. \hlc[highlight]{It can be noted that we have a significant improvement on AUC for all machine learning algorithms. For the previous experiment the best AUC was 0.56. For this experiment, AUC for all algorithms is improved. There are two algorithms that have more than 0.9 AUC, where CNN algorithm has the highest, 0.96. We can conclude that inclusion of the tainted flows features have significantly improved the performance of the machine learning algorithms for this dataset. This answers our research question \textbf{RQ3} and we conclude that the identification of tainted flows is important and helps the algorithms to distinguish better between vulnerable and non-vulnerable SmartApps. We can also note from the performance measures that the gain with inclusion of flows features for} $Corpus2$ is more clear than $Corpus1$.

\section{Discussion}
\hlc[highlight]{This research focuses on identifying vulnerable SmartApps using machine learning algorithms. Feature selection is an important step for training the machine learning algorithms. Manual features selection is a complex and time consuming task. We opt to use an automated process of features selection. To do so, a BOW is prepared from the source code of SmartApps and Token2Vec is used to convert these BOW to feature vectors. This process automatically prepares the set of features. To check the reliability of this technique, we use the research question $RQ1$ to conduct experiments on $Corpus1$ and $Corpus2$. 
We found that this technique works well for $Corpus1$ but fails for $Corpus2$. This is due to the nature of $Corpus2$, where different versions of a SmartApp are prepared by injecting tainted flows in benign SmartApps using a mutation analysis framework. The framework produces around 18 different versions (mutants) for each SmartApp from the base benign Dataset. The BOW representation of vulnerable and non-vulnerable versions of such SmartApps, in many cases, is almost the same. For that reason, machine learning techniques that rely on BOW for feature preparation fail to differentiate between vulnerable and non-vulnerable versions of SmartApps which is evident from the experiments presented in section} \ref{sec:EvaluationBOW}. 

\hlc[highlight]{In such a situation, tainted flows features can play an important role. Therefore, we decided to use tainted flows for distinguishing the vulnerable SmartApps from the non-vulnerable ones like Avdiienko et al.} \cite{Ref5}\hlc[highlight]{. We found different tools from the literature that can identify flows from source code of SmartApps} \cite{Ref5,Ref8}. \hlc[highlight]{However, all of them require static analysis of the source code. The static analysis of SmartApps is expensive and time consuming. To investigate this problem, we have devised the research question $RQ2$. To answer this question, we have proposed a text mining approach 
which was able to identify the tainted flows cheaply. Contrary to static code analysis, this technique uses text mining approach to identify and remember only relevant part of statements that may result in a tainted flow. We have implemented this technique to develop a tool called FlowsMiner. The experiments presented in} section \ref{sec:TaintedFlowsEvaluation} \hlc[highlight]{show that the proposed approach requires far less time than the static analysis counterpart.}

\hlc[highlight]{We then aimed to examine the use of flows features to differentiate between vulnerbale and non-vulnerable SmartApps. This gives rise to our research question $RQ3$. To answer this research question, the flows identified from FlowsMiner are converted to feature vectors by using the Flow2Vec. We have combined both types of features and are able to improve the AUC for all of the considered machine learning algorithms. The best case AUC for the $Corpus1$ and $Corpus2$ data sets improve by 0.03 and 0.4, respectively.}

\hlc[highlight]{

As discussed above, we conducted our experiments for the evaluation of the proposed techniques on two datasets. The dataset named $Corpus1$ has an approximately balanced number of examples for each class label, whereas $Corpus2$ examples are imbalanced. We want to split the dataset into train and test sets in a way that preserves the same proportions of examples for each class as observed in the actual dataset. We want to ensure that the training and testing sets represent the actual dataset. However, splitting the data into training and testing by randomly selecting data points is not a good choice as it can end up in placing more representation of one class in a subset and less representation in the other subset as compared to the actual data which will end up with an estimate of the model error in high variance. As such, we have opted for splitting the data with stratification because stratification ensures that each subset of the data represents the whole data. All the splits in training and testing were stratified on the class variable to have the training and test sets as best representatives of the actual data.

In our experiments we want to achieve confidence and stable estimates about the performance of the built models. In a try to find a method that is a good compromise between complexity and accurate performance estimates, we found that the stratified holdout cross validation is the best fit. We tried various validation methods including repeatedly running experiments with stratified 10-fold cross validation but we always ended up approximately with similar gain in performance estimates.  We were not getting an advantage for using computationally more expensive techniques since both validation methods are consistent with each other and with our findings that the added flows features improved the prediction accuracy, therefore, we decided to stick with a relatively less expensive validation technique, stratified holdout cross validation.} 

\hlc[highlight]{While the focus of this research is to devise a technique to identify vulnerable SmartApps, the proposed technique is universal and can be used to identify vulnerable applications developed in other programming languages. However, it would require some tuning for adapting the proposed technique which includes sets of sensitive sources and sensitive sinks in addition to adaptation for the extraction of the tainted flows. 
An immediate future work would extend our research to evaluate the proposed techniques on applications developed in other programming languages and frameworks.}

\hlc[highlight]{Our approach operates at App level, meaning that the approach predicts whether the app as a whole is vulnerable or not. The granularity of the analysis, however,  is done at method level, since we extract vulnerable flows features at sequence flow, conditional flow, and method scope/context flow. One could wonder whether our approach can operate at another granularity level, such as class level, (Motivation: If an app is vulnerable, it would be more helpful for a developer if a tool/technique tells them which class is vulnerable rather then saying that the App (whole) is vulnerable), however, this is not possible for SmartApps, since developers can not create their own classes in SmartApps. SmartApps are developed in a sandbox environment, and for performance and security reasons, developers are limited to a specific subset of Groovy language.  while this seems to be a significant restriction, however, with the various methods available to developers, the need to create their own classes is rarely needed} \cite{Ref58}.

\hlc[highlight]{However, for our approach, and since our solution can be generalized to applications in other languages that may allow/have multiple classes, we think our approach can operate at class level. In order to extract features for applications that have classes, we need to extract features in the form of tainted flows and source code token frequencies for every class and then merge them to serve as features of that application. A class can consist of one method or multiple methods. We need to extract features in the form of tainted flows and source code token frequencies for every method and then merge them to serve as features of the class. If the feature set of an application is prepared and is provided to a model built with our approach and is predicted as vulnerable. A developer might be interested in identifying a vulnerable class, if an application has only one class, the answer is straightforward. In the case of multiple classes, a feature set for each class can be provided to the model one by one to figure out the vulnerable classes and the same can be done for the methods. The process depends on the developer's decision on  which level of granularity they want to operate. All they have to do is prepare features set for that level.}

\hlc[highlight]{The proposed technique works fine for the SmartApps whose BOW tokens or their frequencies can be used to distinguish the vulnerable SmartApps from the non-vulnerable ones. If the SmartApps have information leakage vulnerability then the tainted flows can be used to distinguish the vulnerable SmartApps. This technique should work fine where the tainted flows are identifiable by the FlowsMiner. In addition, it can be adapted to handle tainted flows results obtained from taint flows static analysis tools. 

If we have vulnerabilities in the applications other than the information leakage vulnerability, such as: Cross site scripting (XSS), SQL Injection, Command Execution, Remote File Inclusion, File System Access, or Malicious Evaluation. The FlowsMiner can be adapted to detect these vulnerabilities. All it needs is the set of sources and sinks presented in Table} \ref{tab:VULTYPES} \hlc[highlight]{and adaptations to track flows for a different language. The FlowsMiner tool uses text mining to identify the flows, these adaptations should be straightforward for most of the languages.} 

\begin{table}[!t]
\centerline{\includegraphics[width=.99\textwidth]{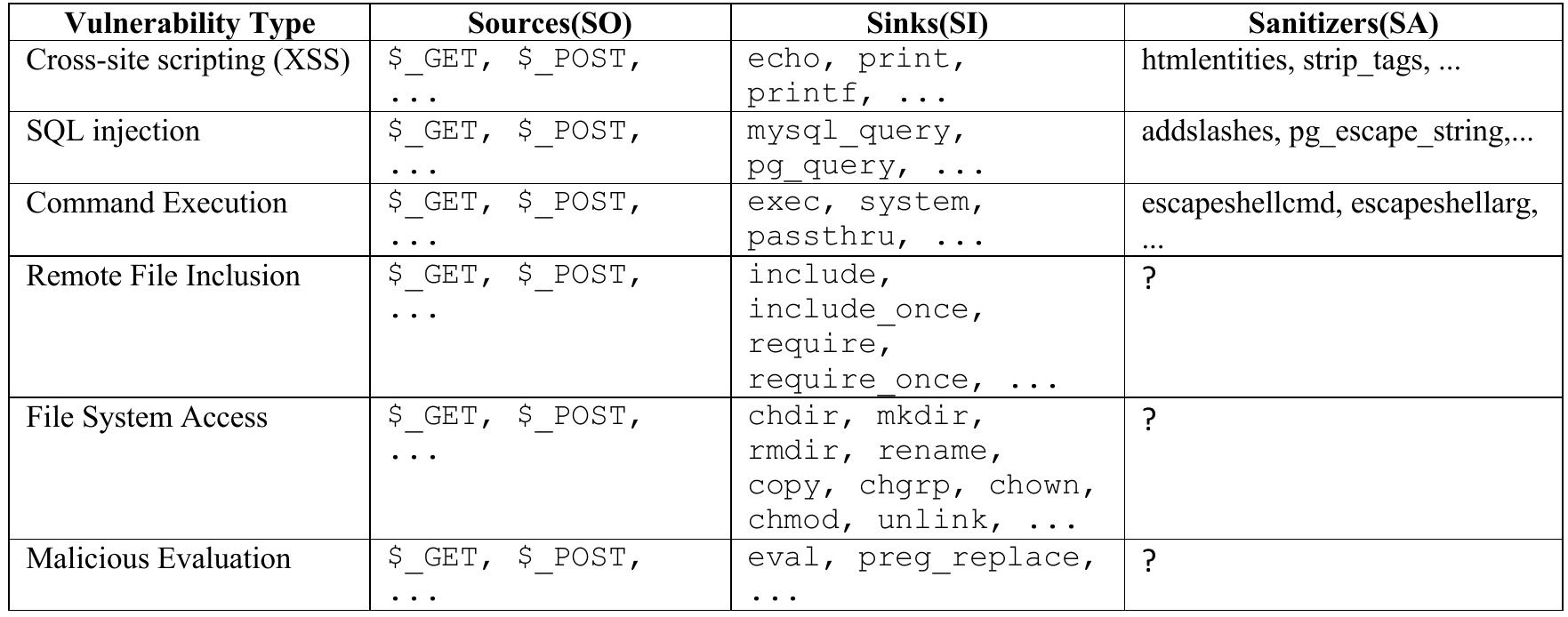}}
\caption{Examples of vulnerabilities caused by tainted flows,\\ \footnotesize{(?) refers to the absence of specific sanitization function for those vulnerabilities, other preventive techniques such as while list is required}}
\label{tab:VULTYPES}
\end{table}

\subsection{Threats to validity}
\hlc[highlight]{The potential limitations along with mitigation strategies of this empirical study are discussed in this section. Imbalanced datasets can have an impact on \textbf{internal and construct validity}. Due to fewer data points for the minority class, the algorithms can generate suboptimal models. Balanced datasets can be generated by randomly eliminating instances of the majority class from the dataset or by synthesizing new examples from the minority class by using the synthetic minority oversampling technique (SMOTE) but it does not always end up in performance gain}  \cite{Ref57}.

\hlc[highlight]{We have not applied any data balancing technique and have used stratified holdout cross validation to preserve the same class distribution in the training and test sets. Another possible threat for the construct validity is about labeling the SmartApps. We have used the SAINT and Taint-Things tools to label the SmartApps for Corpus1. Static analysis tools may be imprecise and result in incorrect labeling in some cases, as such, we have manually reviewed the apps to confirm the assigned labels. We are confident about the labels of $Corpus2$, as the vulnerable apps were created using a mutation framework and the framework developers have provided the SmartApps along with the labels. 

\textbf{Threat to external validity} can be related to the generalizability of our proposed technique. In this paper, our dataset is based on Groovy language and its usage in SmartApps. We have used datasets named as $Corpus1$ and $Corpus2$ for estimating the performance of the proposed technique. Adapting the proposed technique for another dataset developed in a different programming language may end up with different results. We have not tested the proposed techniques on various projects developed in other programming languages. To be more confident about our findings, it would be interesting to test the proposed techniques on larger datasets and projects developed in other programming languages. Although, we believe that the proposed approach is generalizable for the applications developed in different programming languages. Theoretically, the proposed technique will be more reliable once tested in other settings with applications developed in different programming languages. Another threat to the validity could be about the choice of selected machine learning algorithms. We have selected the algorithms that have been frequently used in other studies} \cite{Ref4}.

\hlc[highlight]{For the \textbf{conclusion validity}, it can be noted that we have used well-established metrics including AUC, F1-score and MCC for the performance estimates. We have compared the proposed technique with the base Bag of Words (BoW) approach to show the gain. There is no bias in the performance estimates due to the opted stratified hold out cross validation because the gain for the proposed technique for our experiments is similar to repeated stratified k-fold cross validation. We have provided the datasets used for the experiments on Github and the implementation details to address the validity threat about the reliability and reproducibility of our findings. If the technique is applied along with the provided settings, the same results will be obtained.}

\section{Related Work}\label{sec:RW}
This research presents an approach for identification of vulnerable IoT applications using machine and deep learning techniques. We have focused on the IoT applications developed in SmartThings. In the absence of literature available on the topic, we explored the closest relevant field that is identifying vulnerable android and web applications using machine learning algorithms \cite{Ref46,Ref47}.
Most of the techniques extract a set of features from web, android or other types of applications and then use the features to train machine learning algorithms to identify vulnerable applications \cite{Ref1,Ref4,Ref7,Ref48,Ref49}. 

In this section, we have reviewed the literature to explore how researchers have addressed similar problems. \hlc[highlight]{We start by reviewing the techniques to extract features using static analysis, then by extracting features using hybrid analysis and text mining.}

\subsection{Identifying Features using Static Analysis}\label{sec:FeaturesusingStaticAnalysis}
Zhu et al. build an android malware detection tool called DeepFlow \cite{Ref11}.
They have used FLOWDROID \cite{Ref12} to extract data flows from sensitive sources to sensitive sinks in android applications. They have applied the SUSI technique for features transformation from method to categories level. They choose 3000 benign apps and 8000 malicious apps for running their experiments. Once the dataset is ready they have applied deep learning techniques for building models to identify malicious applications. 
Similar techniques are applied to the web applications.
Medeiros et al. propose a technique to automatically detect web vulnerabilities using machine learning \cite{Ref10}.
They have implemented their approach in a tool called DEKANT. They have used Hidden Markov Models (HMM) to learn models. A HMM is composed of set of symbols or tokens called vocabulary, set of states, probabilities set that includes initial probabilities, transition probabilities and emission probabilities. The model built by them keeps a track of the order of code slices named as observation. The model learns the state of observations with the intent to expose hidden state. Then the states are checked to be vulnerable or non-vulnerable. 
Methods of a web application can also be used to infer a model to report vulnerabilities in that application. Shar et Tan took a set of web apps and extracted sanitization methods, divided them into different categories and used them to train machine learning models to predict SQL injection and cross site scripting vulnerabilities \cite{Ref2}.

Deciding about features to be extracted from software applications and then  selecting features for training a suitable model is a time consuming task. Zhao et al. introduced a tool named Fest \cite{Ref6} 
to extract meaningful features from android applications and evaluated which set of features are more effective in  predicting malicious applications.  
Avdiienko et al. proposed to distinguish between malicious and benign android applications on the basis of different flows \cite{Ref5}.
The name of their tool is MUDFLOW which uses FLOWDROID \cite{Ref12},
a static analysis tool to identify flows in the android applications. They compared the flow of sources to sinks for malicious applications with benign applications. They claim that their approach is good for prediction of any application which is having a novel malware. They used FlowDroid for the extraction of flows from different applications and then categorized those flows according to SUSI categories. Then they trained models on attributes of benign applications and compared test applications to be malicious or benign.
\hlc[highlight]{We can find other contributions that use flows for identification of vulnerable applications. SUI et al. have used static analysis to identify context-sensitive and alias-aware data flows} \cite{Ref53}. 
\hlc[highlight]{They name their technique as Flow2Vec. They consider all flows regardless of if they are tainted or not. They compile a program to low-level intermediate representation (LLVM-IR) and build interprocedural value-flow graph (IVFG) on top of that using Andersens pointer analysis} \cite{Ref55}. \hlc[highlight]{Then they prepare a matrix from IVFG. They formulate the flow reachability by using the matrix multiplication and the outcome is decomposed to source and target vectors for every node of IVFG.}
\hlc[highlight]{Kim et al. have designed a specification language for describing flow patterns of software vulnerabilities} \cite{Ref52}. \hlc[highlight]{They have implemented a vulnerability flow detector that matches the predefined syntactic patterns with source codes. Their static analysis solution constructs flow graphs to identify flows that are indicators of vulnerabilities.} 
\hlc[highlight]{The set of features used by each of the research papers discussed in this section is presented in the Table} \ref{tab:summary2}.

\subsection{Identifying Features using Hybrid Analysis}\label{sec:FeaturesusingHybridAnalysis}
Shar et al. enhanced their previously presented approach \cite{Ref2} by using hybrid attributes comprising of both static and dynamic, instead of just static attributes \cite{Ref3}. After applying different machine learning techniques they found that the results were better with hybrid attributes. The features set used by them is provided in the Table \ref{tab:summary2}.

\subsection{Identifying Features using Text Mining}\label{sec:FeaturesusingTextMining}
Scandariato et al. identify vulnerabilities in android software components using text mining of source code \cite{Ref4}.
They report a software component to be vulnerable, if it has any warnings. They convert source code of a software component to BoW and each word is considered as a feature that they use for predicting the warnings and vulnerabilities. The set of features that is used for building the predictor models is not fixed because it depends on the vocabulary of the source code. They have used Weka implementations of Decision Trees, K-Nearest Neighbours, Na\"{i}ve Bayse, Random Forest and Support Vector Machine to conduct the experiments. 
Walden et al. provide 223 vulnerabilities that they identified in three open source web applications written in PHP \cite{Ref1}.
They have prepared a dataset comprising of software metrics including lines of code, number of functions, maximum nesting complexity, cyclomatic complexity, fan in and fan out to compare two vulnerabilities identification techniques. One based on dataset of software metrics and other based on the mined text BoW representation. They report that the text mining technique performed better than the software metrics.

We can save time spent on feature engineering by adapting techniques like text mining which can automatically prepare set of features. 
A summary of the features used for identifying vulnerable applications by different researchers discussed in the previous sections is presented in the Table \ref{tab:summary2}.
After exploring the related work presented in this section, we were not able to find a set of features that can be used to detect vulnerable SmartApps. We opted to extract all possible features and used all of them to train the machine learning algorithms for building the behavioural models like we have done in \cite{Ref20}.

\begin{table}[t!]
\centerline{\includegraphics[width=.99\textwidth]{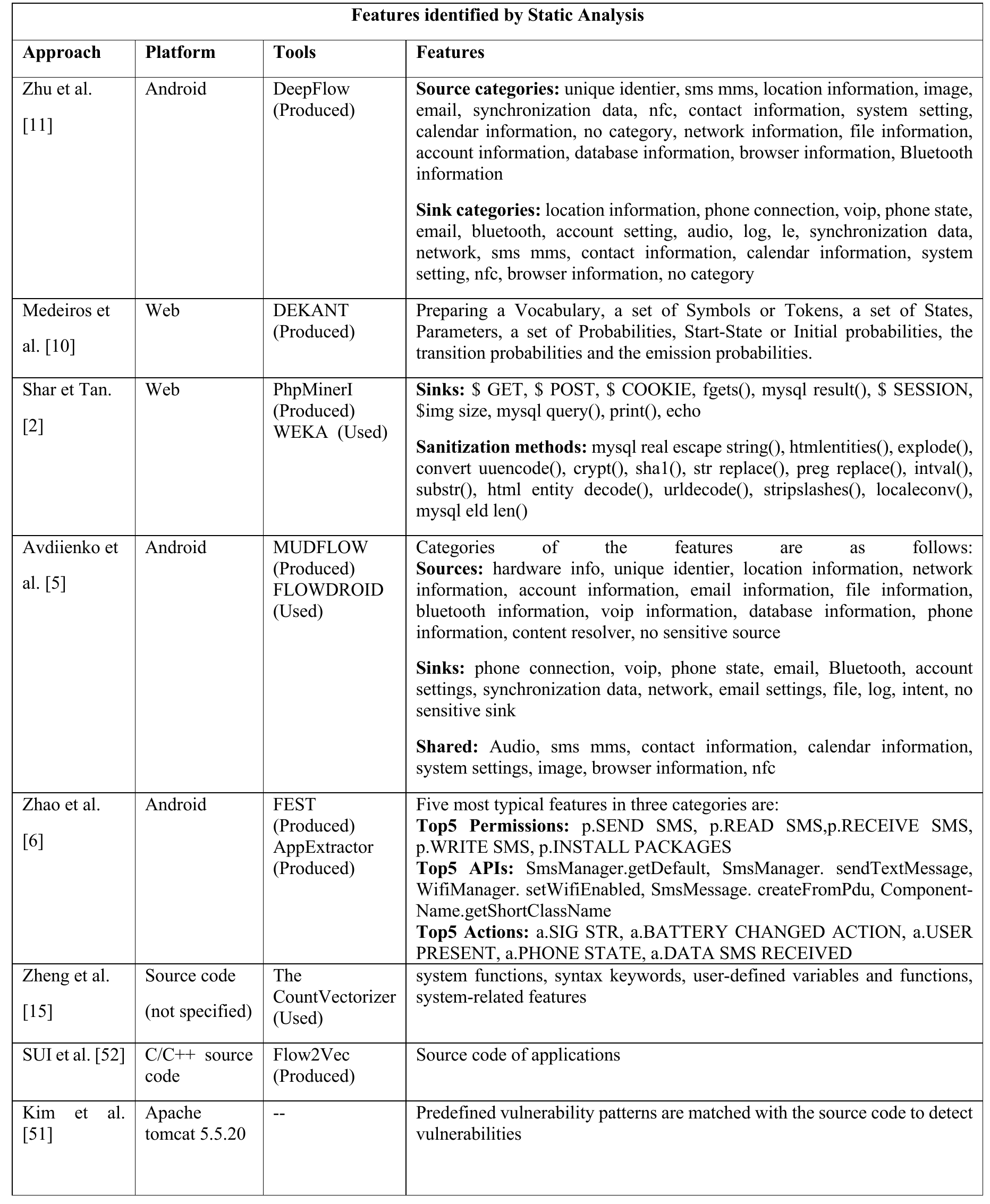}}
\label{tab:summary1}
\end{table}

\begin{table}[t!]
\includegraphics[width=.99\textwidth]{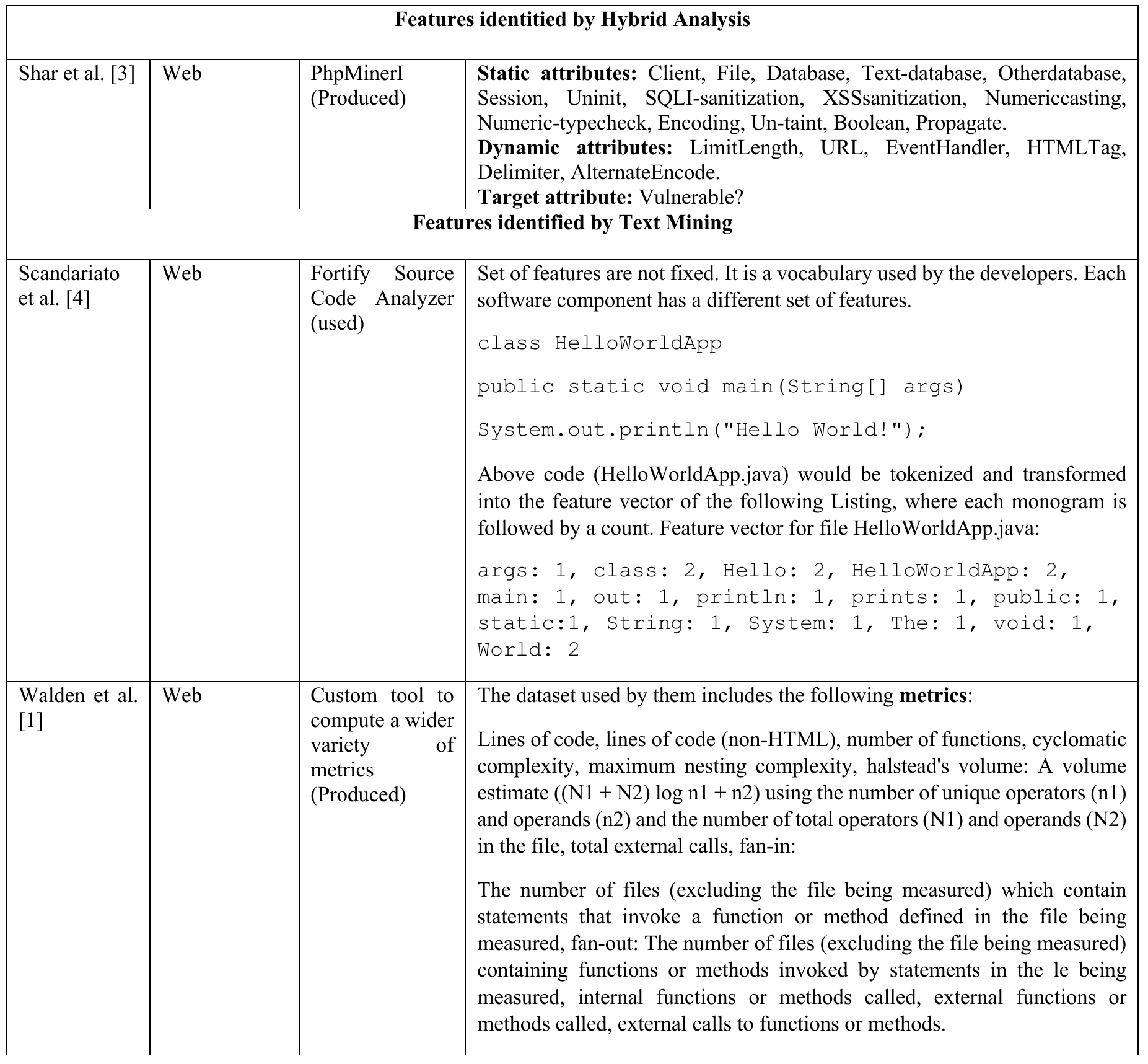}
\caption{Features used by researchers for vulnerability identification}
\label{tab:summary2}
\end{table}
When the performance of models is not satisfactory, we can select subset of these features by using domain knowledge, adjusting weights of features, feature selection or feature elimination techniques. We have used text mining to prepare the features set for our research. 

\section{Conclusion and Future Work}\label{sec:con}
\hlc[highlight]{We have proposed an automated technique to identify vulnerable SmartApps using machine learning algorithms and that by training them on features automatically extracted by the tools developed by us, Toekn2Vec and Flow2Vec}. Our research contributions not only flag a SmartApp to be vulnerable, but they can also report the exact location of vulnerabilities. \hlc[highlight]{A high level summary of our research contributions is presented as follows:}
\begin{itemize}
    \item  \hlc[highlight]{We have identified that the features extracted from the BoW are not always appropriate.}
    \item \hlc[highlight]{We have proposed a text mining approach to identify tainted flows in SmartApps. The approach is less expensive than  static analysis because it only considers relevant parts of the code to identify the tainted flows. It only keeps track of the flows which are tainted. We have implemented this technique as a tool called FlowsMiner.}
    \item \hlc[highlight]{We have developed two tools named as Token2Vec and Flow2Vec to prepare feature vectors from BoW and Flows.}
    \item \hlc[highlight]{We show that considering the features extracted from tainted flows in addition to features extracted from BoW of source code improve the AUC of machine learning algorithms.}
\end{itemize}
As for future work, there are numerous possibilities that could be considered for going forward to contribute more on the subject. Different feature engineering techniques can be used to run more experiments. Balancing the datasets for both vulnerable and non-vulnerable classes will reduce the chances for any algorithm to predict a majority class. Normalisation of features will improve the performance of the machine learning algorithms that are susceptible to different ranges of values for features. Dimensionality reduction techniques may also help to improve performance of some machine learning algorithms. 

Another possibility is to extend this study to other types of software like android applications. We have a lot of labelled android applications available in a variety of repositories. This will give us an opportunity to evaluate the proposed techniques on more datasets and other languages.

\pagebreak
\appendix 

\section{Visual Analysis of sinks in Corpus1 and Coprpus2 datasets}\label{app:Third}
\begin{subappendices}
\subsection{Figures for Multiple sinks in Corpus 1 and 2}\label{sec:sinkss}

\begin{figure}[ht!]
\centerline{\includegraphics[width=.7\textwidth]{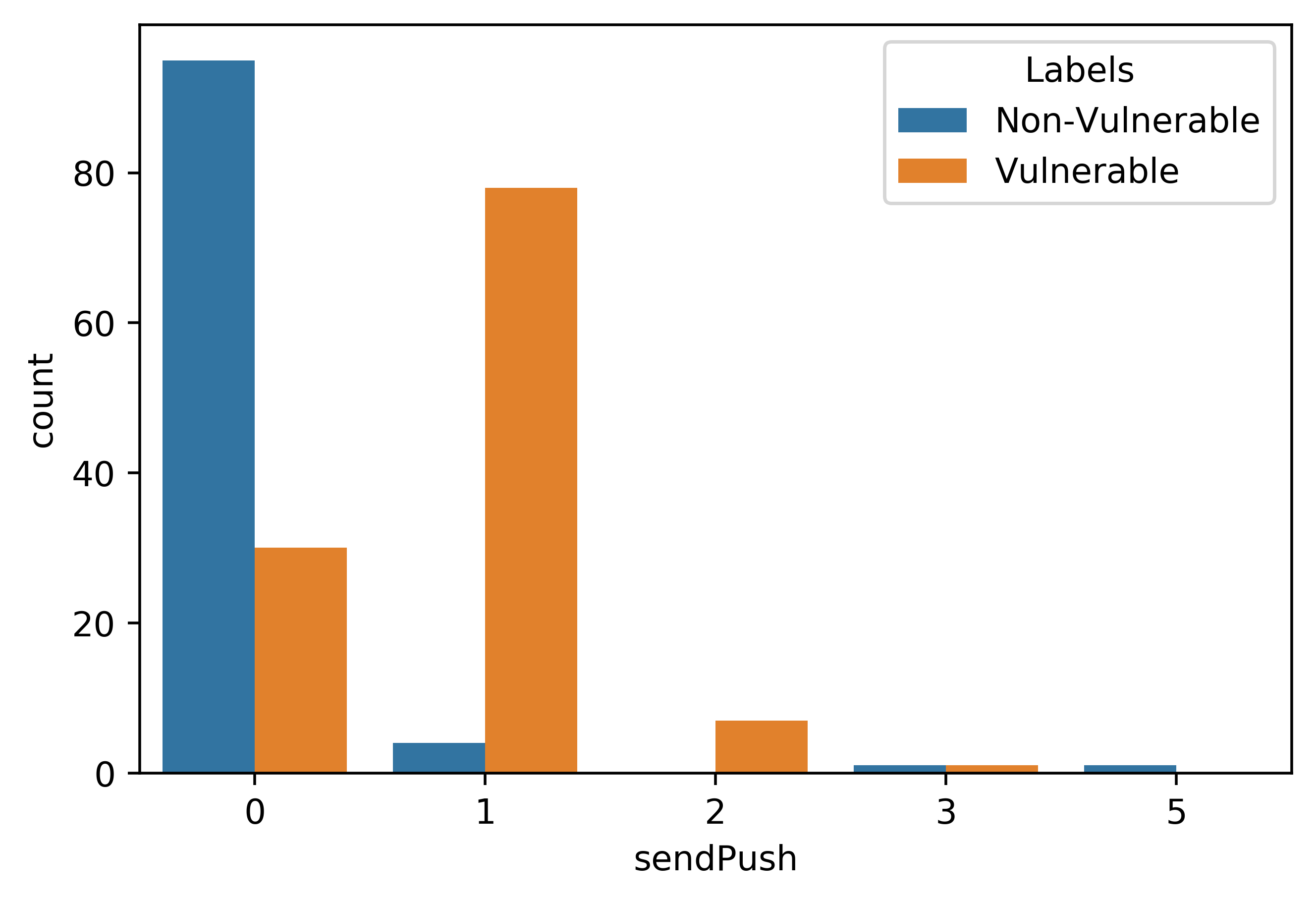}}
\caption{SendPush Frequency in all apps for Corpus 1}
\label{fig:sendPush}
\end{figure}

\begin{figure}[ht!]
\centerline{\includegraphics[width=.7\textwidth]{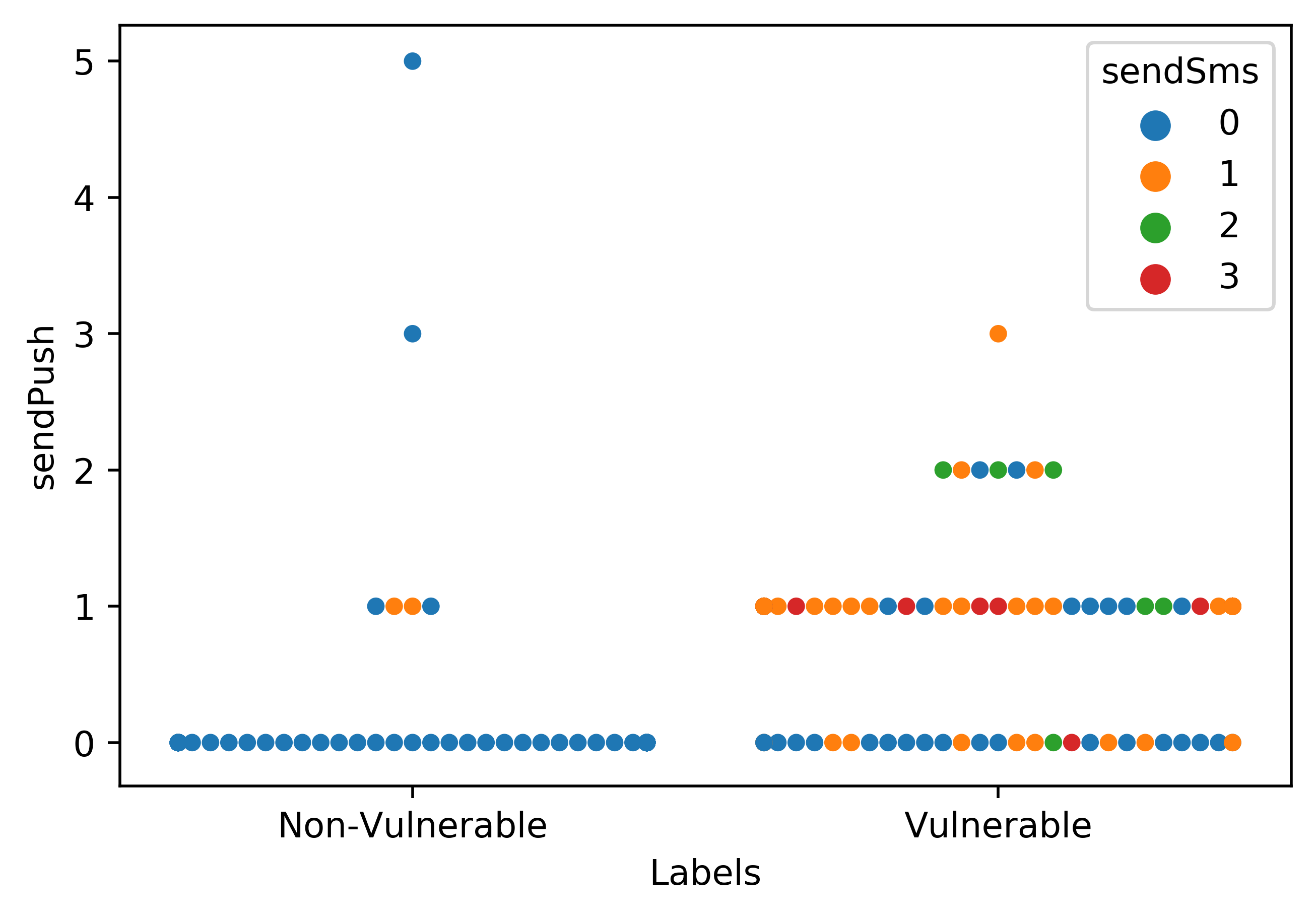}}
\caption{Occurrence of SendPush w.r.t sendSMS in all apps for Corpus 1}
\label{fig:sendPush_sendsms}
\end{figure}

\begin{figure}[ht!]
\centerline{\includegraphics[width=.7\textwidth]{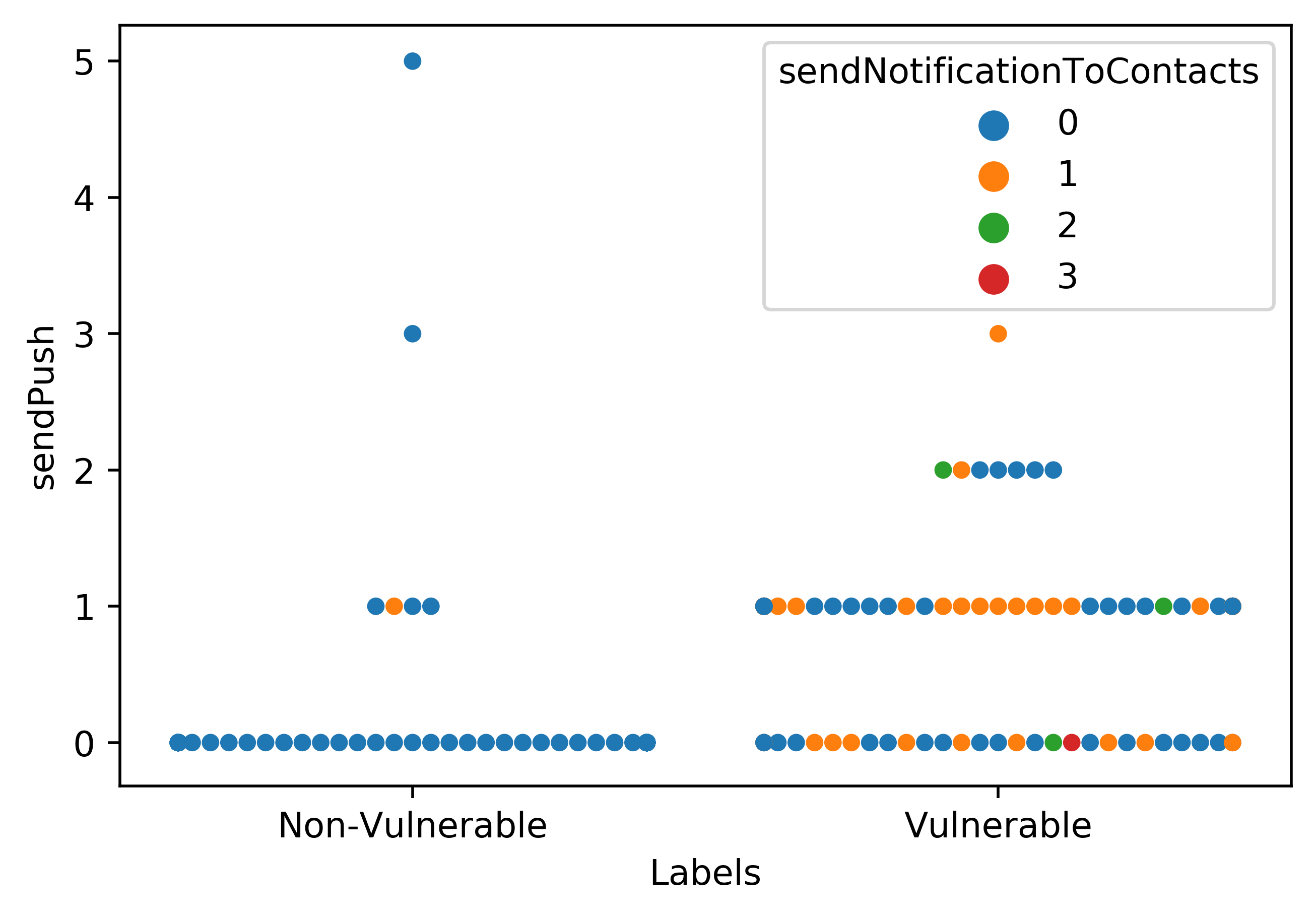}}
\caption{Occurrence of SendPush w.r.t sendNotificationToContacts in all apps for Corpus 1}
\label{fig:sendPush_senNotificationtocont}
\end{figure}

\begin{figure}[ht!]
\centerline{\includegraphics[width=.7\textwidth]{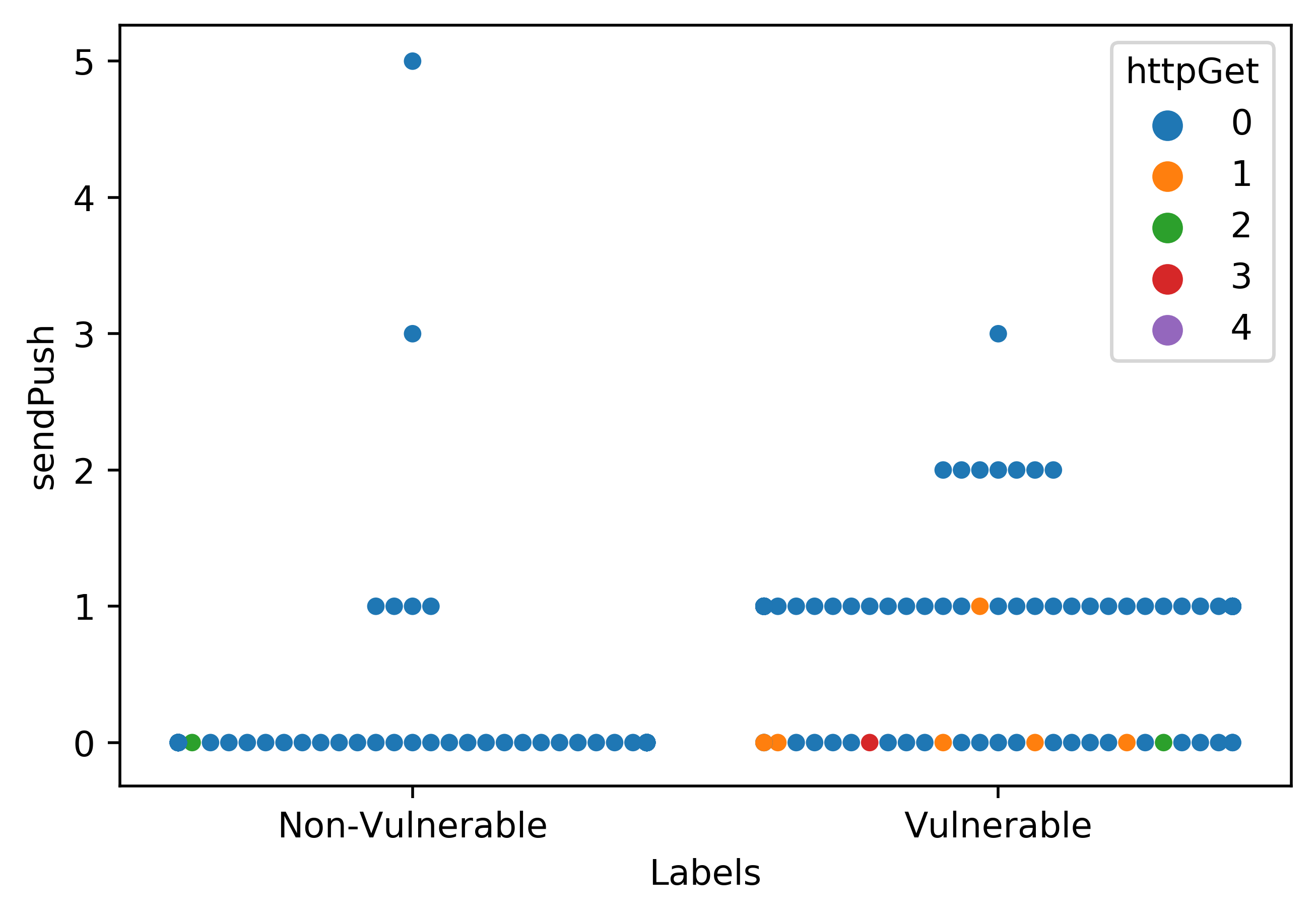}}
\caption{Occurrence of SendPush w.r.t httpGet in all apps for Corpus 1}
\label{fig:sendPush_httpget}
\end{figure}

\begin{figure}[ht!]
\centerline{\includegraphics[width=.7\textwidth]{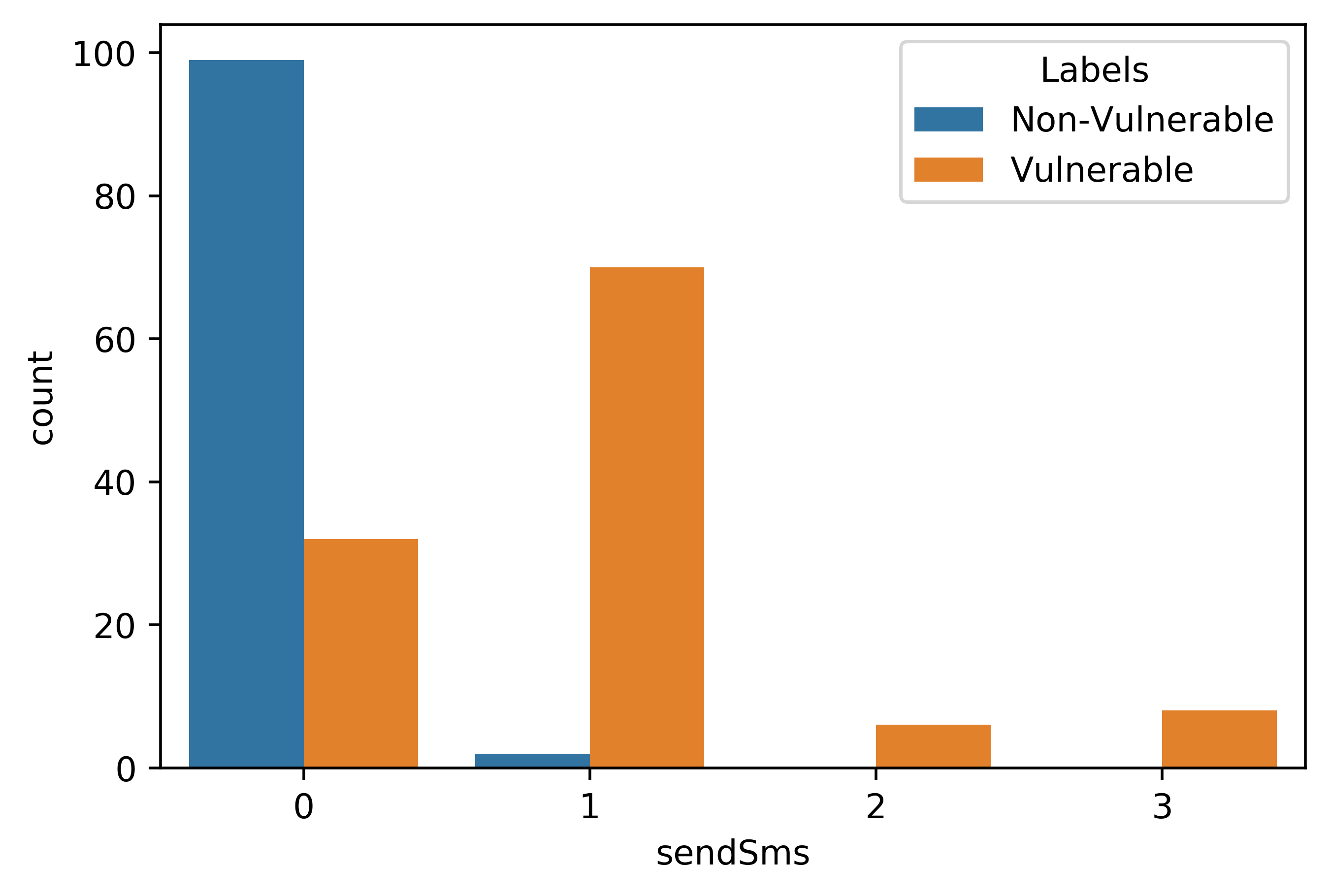}}
\caption{sendSMS Frequency in all apps for Corpus 1}
\label{fig:sendsmsh}
\end{figure}

\begin{figure}[ht!]
\centerline{\includegraphics[width=.7\textwidth]{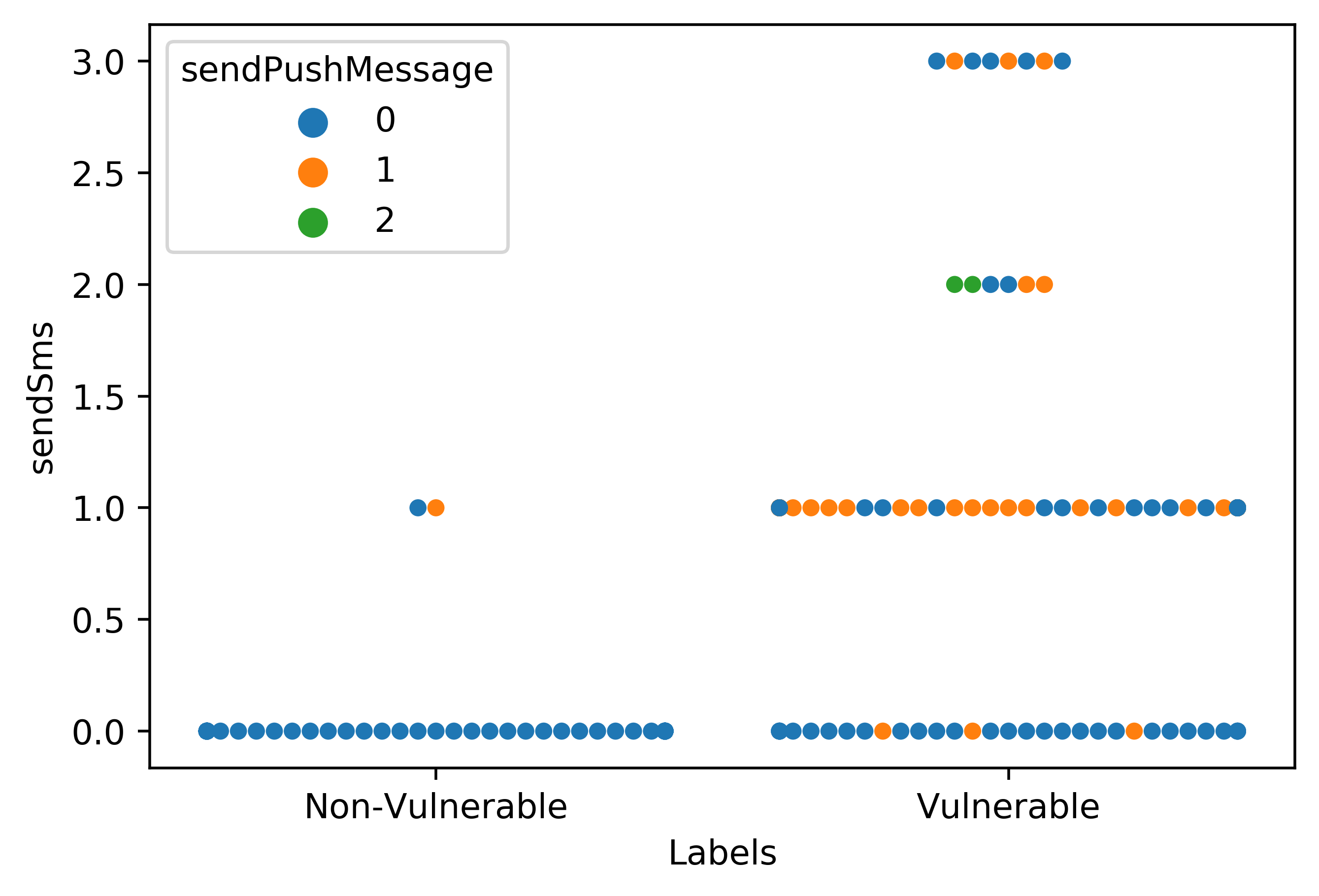}}
\caption{Occurrence of sendSMS w.r.t sendPushMessage in all apps for Corpus 1}
\label{fig:sendsms_sendpushmsg}
\end{figure}

\begin{figure}[ht!]
\centerline{\includegraphics[width=.7\textwidth]{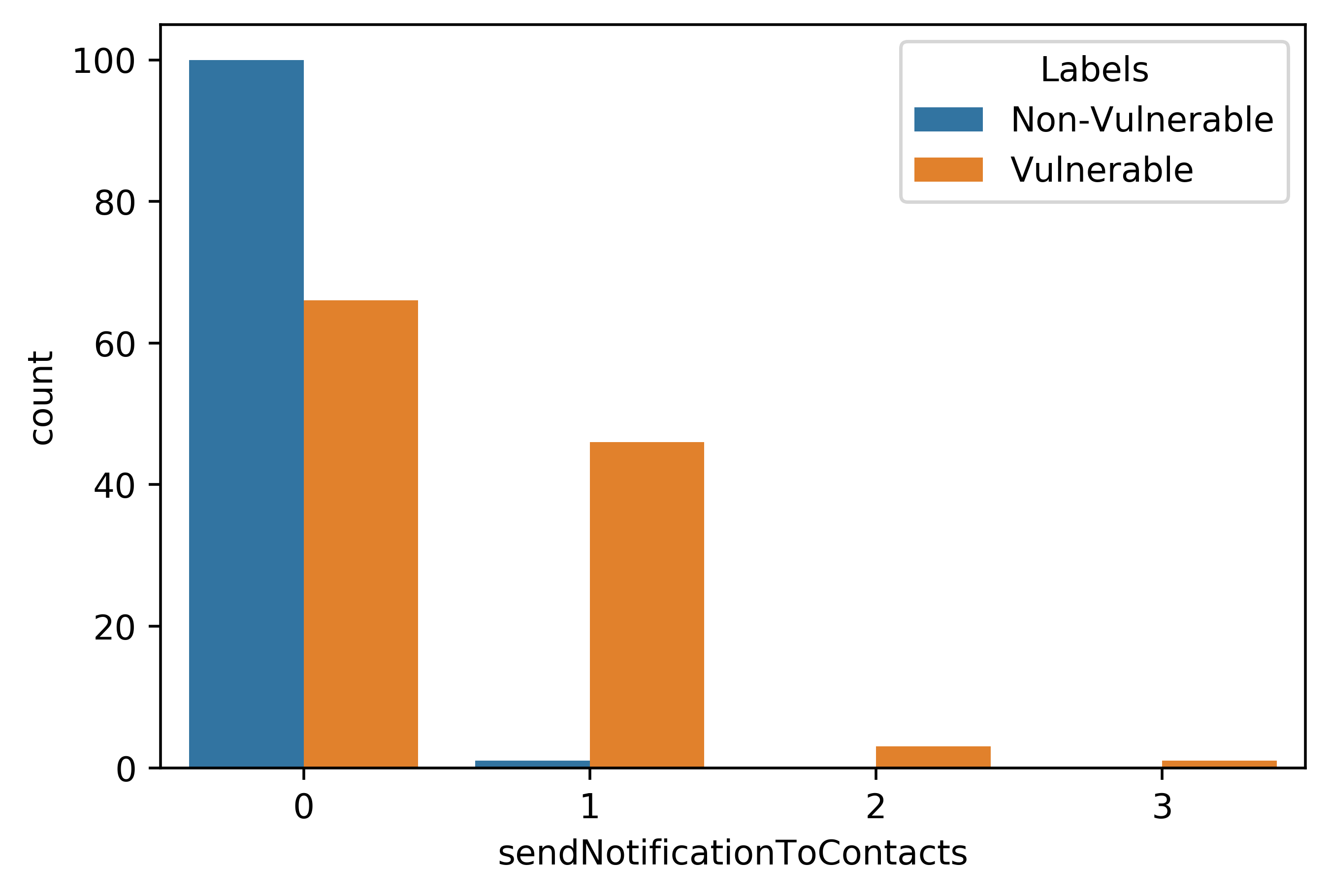}}
\caption{sendNotificationToContacts Frequency in all apps for Corpus 1}
\label{fig:sendNotificationtocont}
\end{figure}

\begin{figure}[ht!]
\centerline{\includegraphics[width=.7\textwidth]{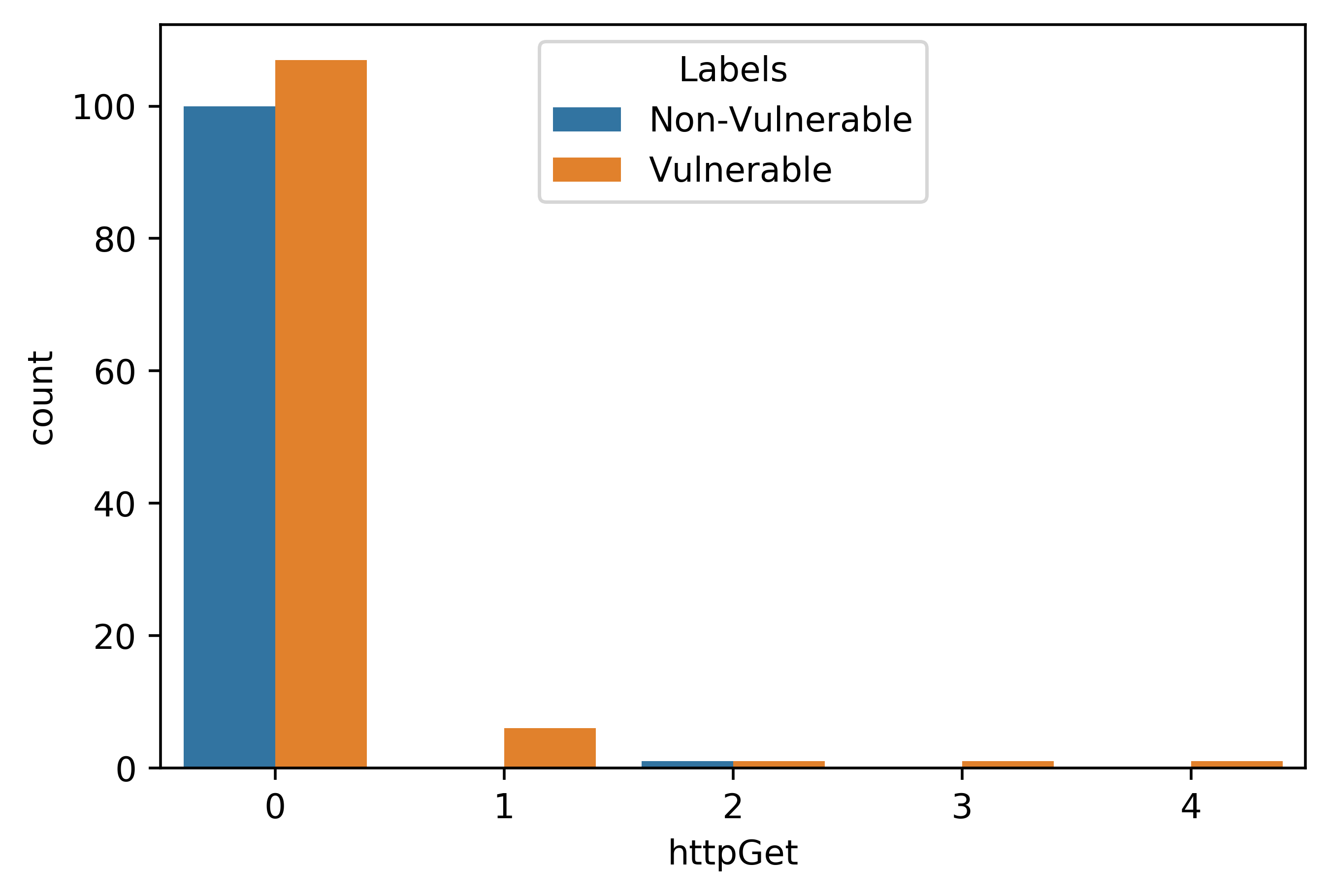}}
\caption{httpGet Frequency in all apps for Corpus 1}
\label{fig:httpget}
\end{figure}

\begin{figure}[ht!]
\centerline{\includegraphics[width=.7\textwidth]{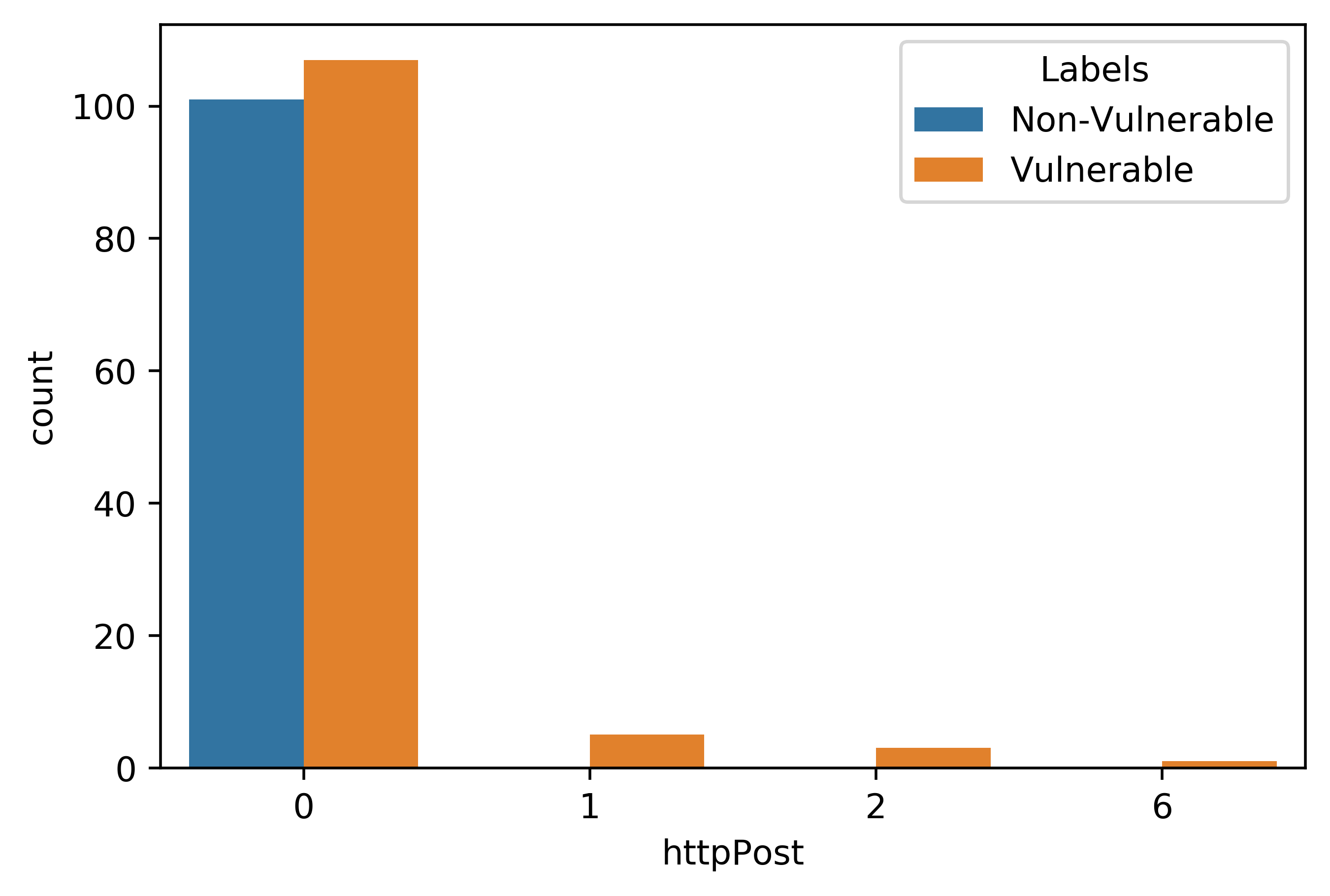}}
\caption{httpPost Frequency in all apps for Corpus 1}
\label{fig:httppost}
\end{figure}

\begin{figure}[ht!]
\centerline{\includegraphics[width=.7\textwidth]{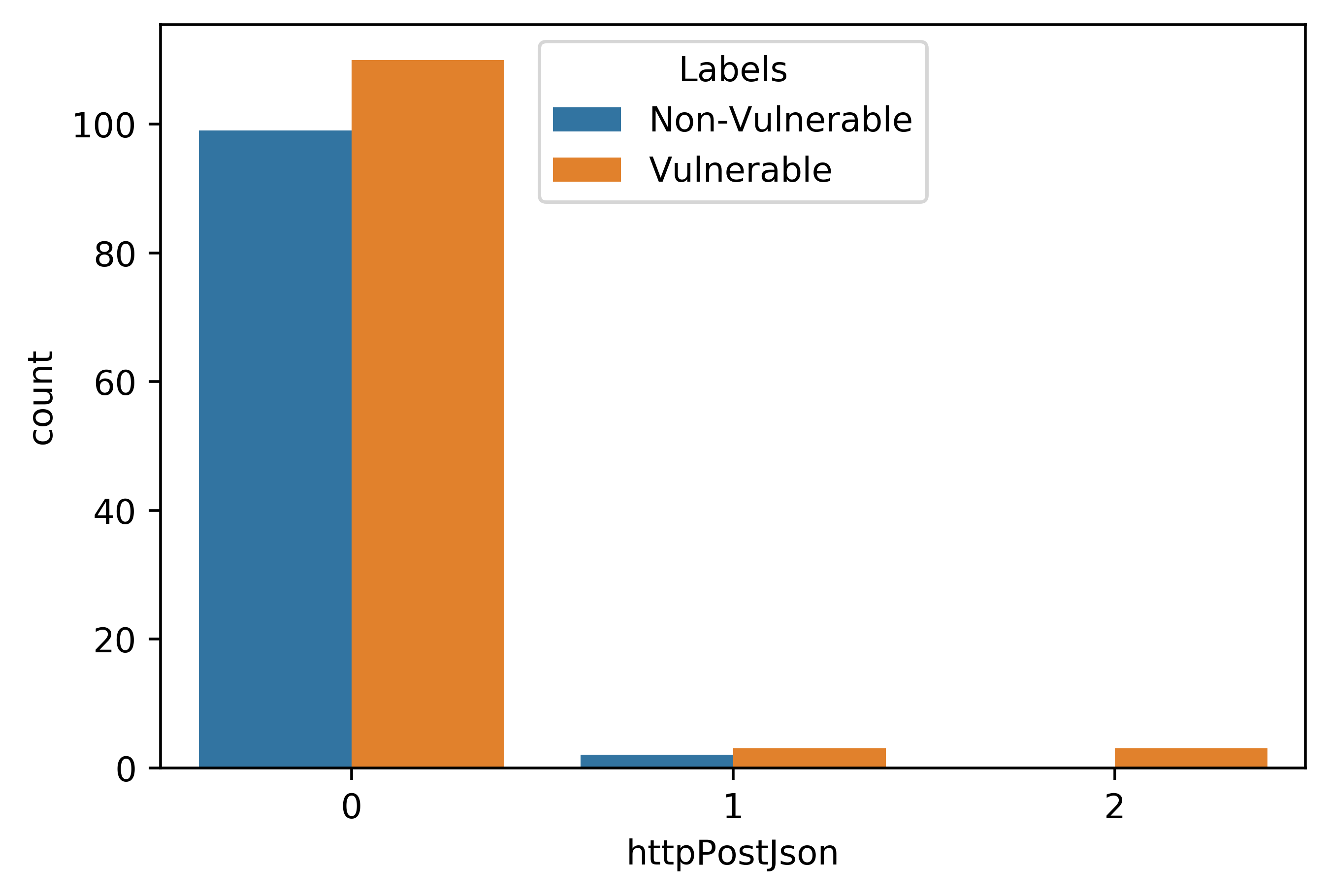}}
\caption{httpPostJson Frequency in all apps for Corpus 1}
\label{fig:httppostjson}
\end{figure}

\begin{figure}[ht!]
\centerline{\includegraphics[width=.7\textwidth]{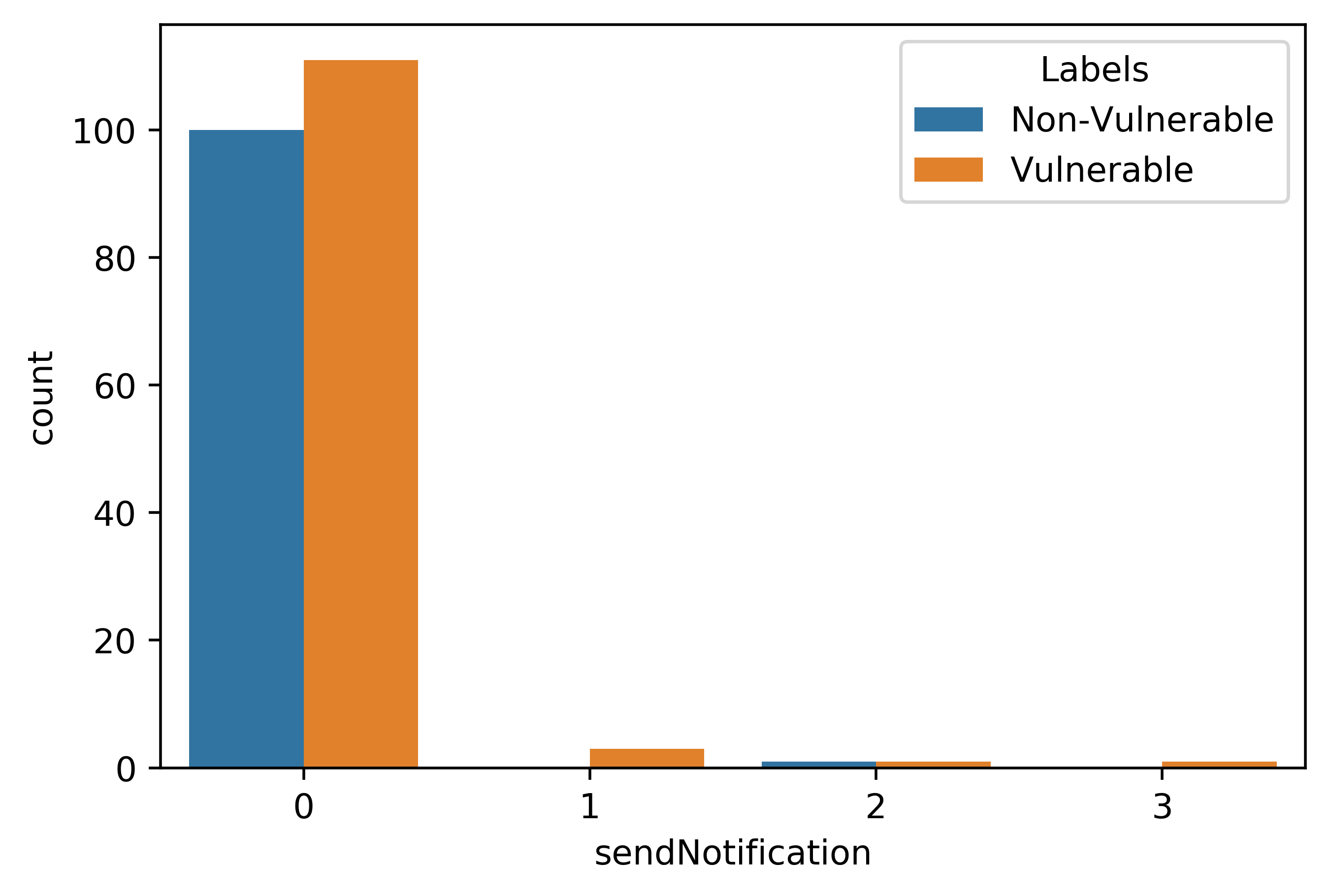}}
\caption{sendNotification Frequency in all apps for Corpus 1}
\label{fig:sendnotification}
\end{figure}

\begin{figure}[ht!]
\centerline{\includegraphics[width=.7\textwidth]{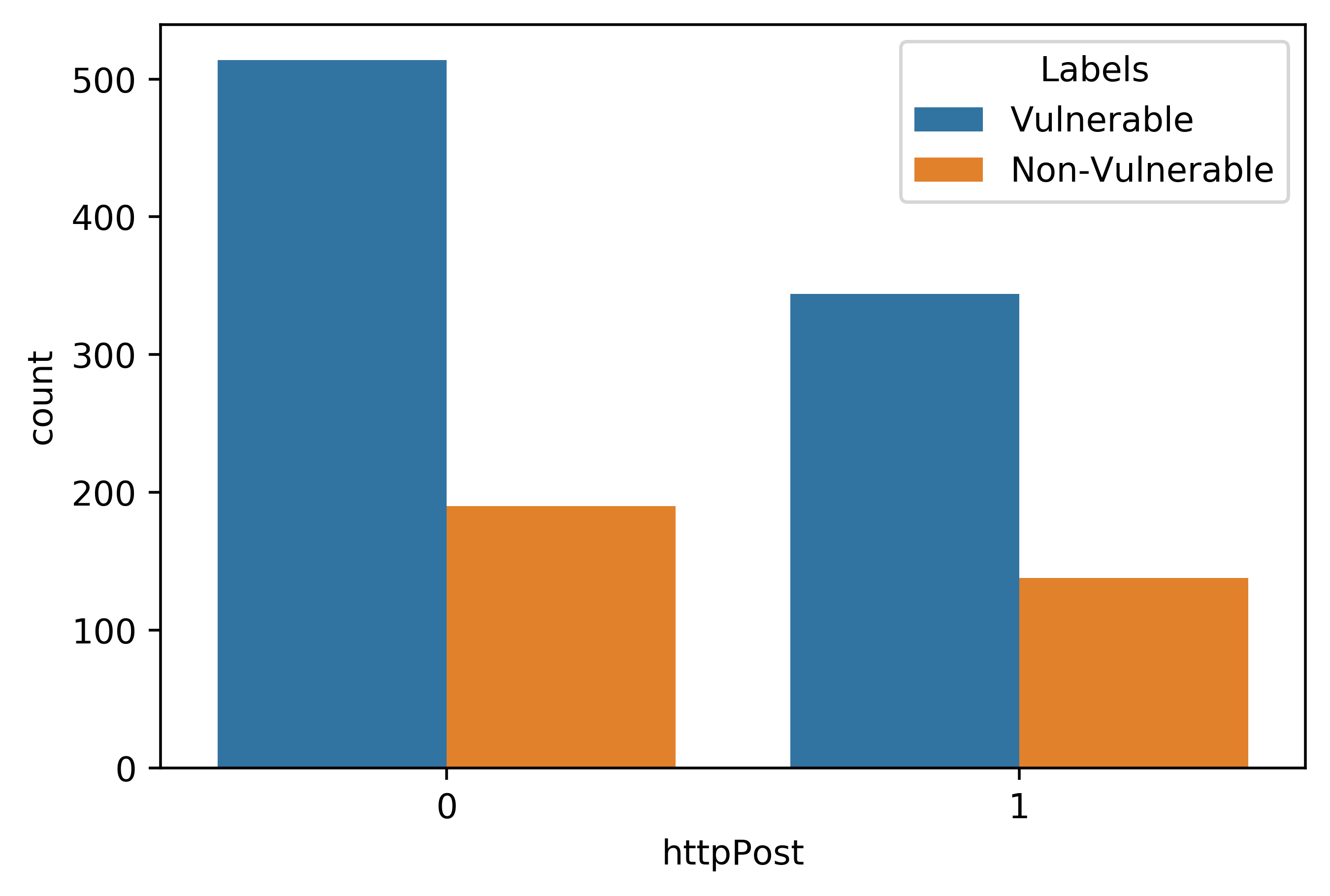}}
\caption{httpPost Frequency in all apps for Corpus 2}
\label{fig:shttppost_mut}
\end{figure}

\begin{figure}[ht!]
\centerline{\includegraphics[width=.7\textwidth]{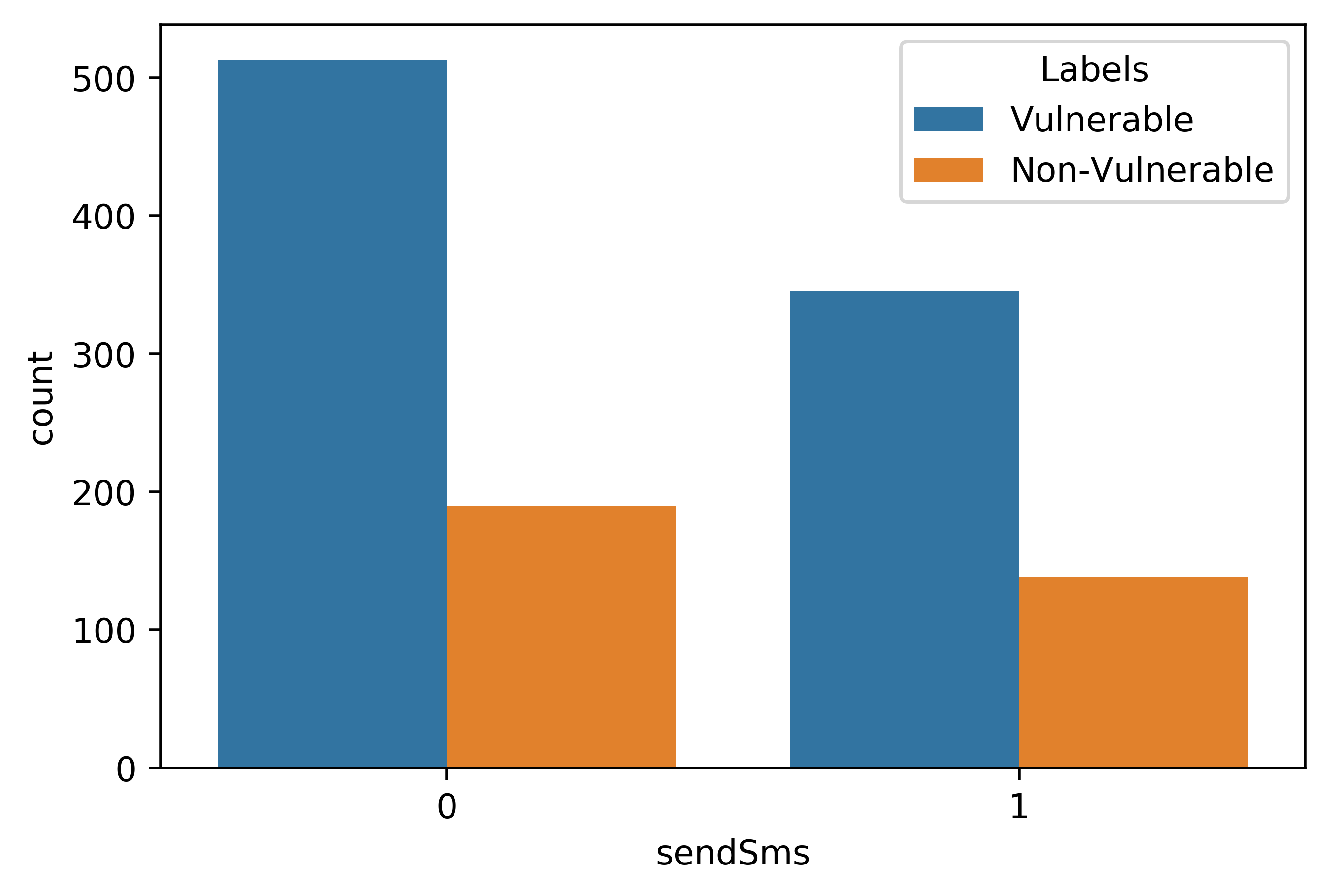}}
\caption{sendSMS Frequency in all apps for Corpus 2}
\label{fig:sendsms_mut}
\end{figure}

\begin{figure}[ht!]
\centerline{\includegraphics[width=.7\textwidth]{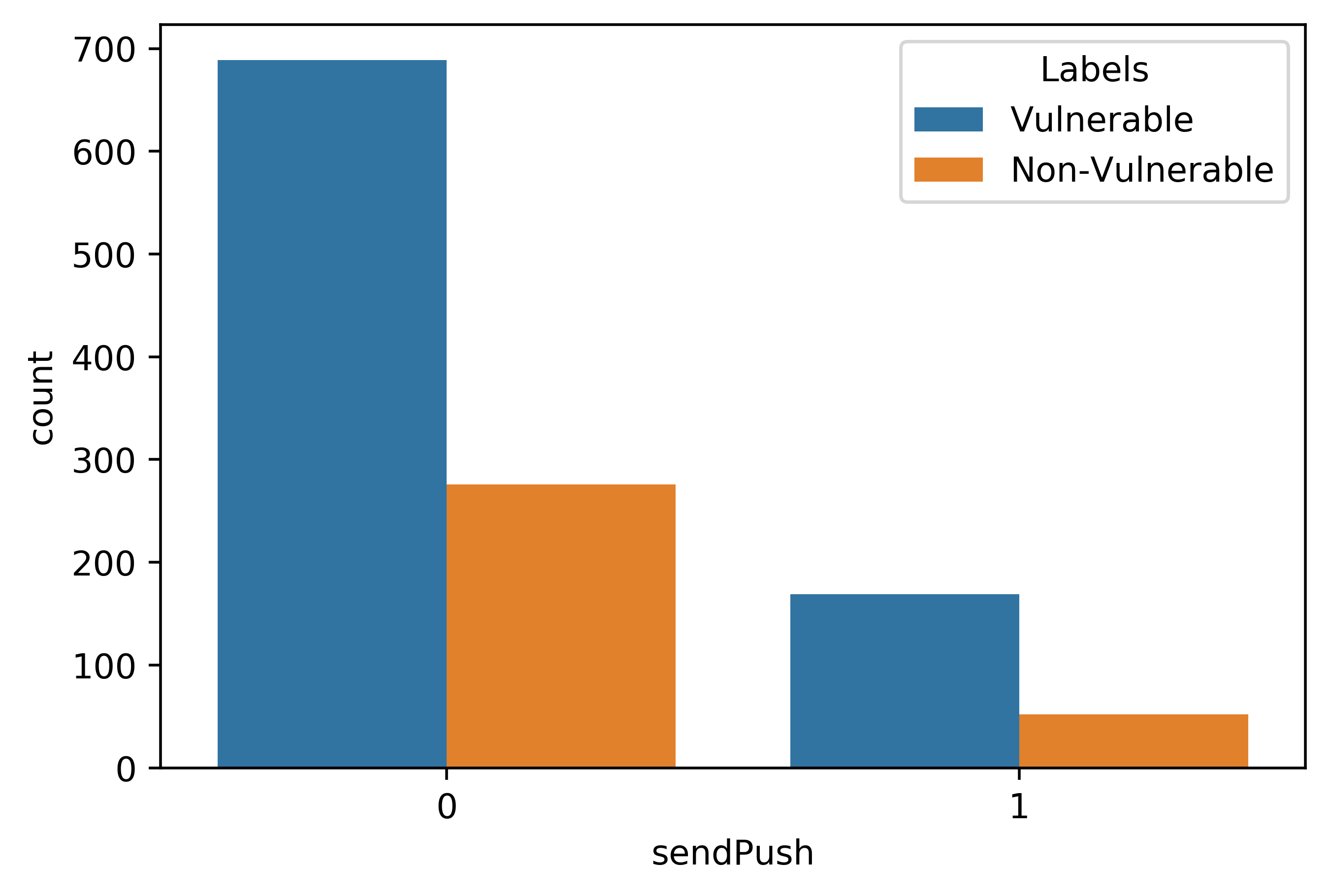}}
\caption{sendPush Frequency in all apps for Corpus 2}
\label{fig:sendpush_mut}
\end{figure}

\begin{figure}[ht!]
\centerline{\includegraphics[width=.7\textwidth]{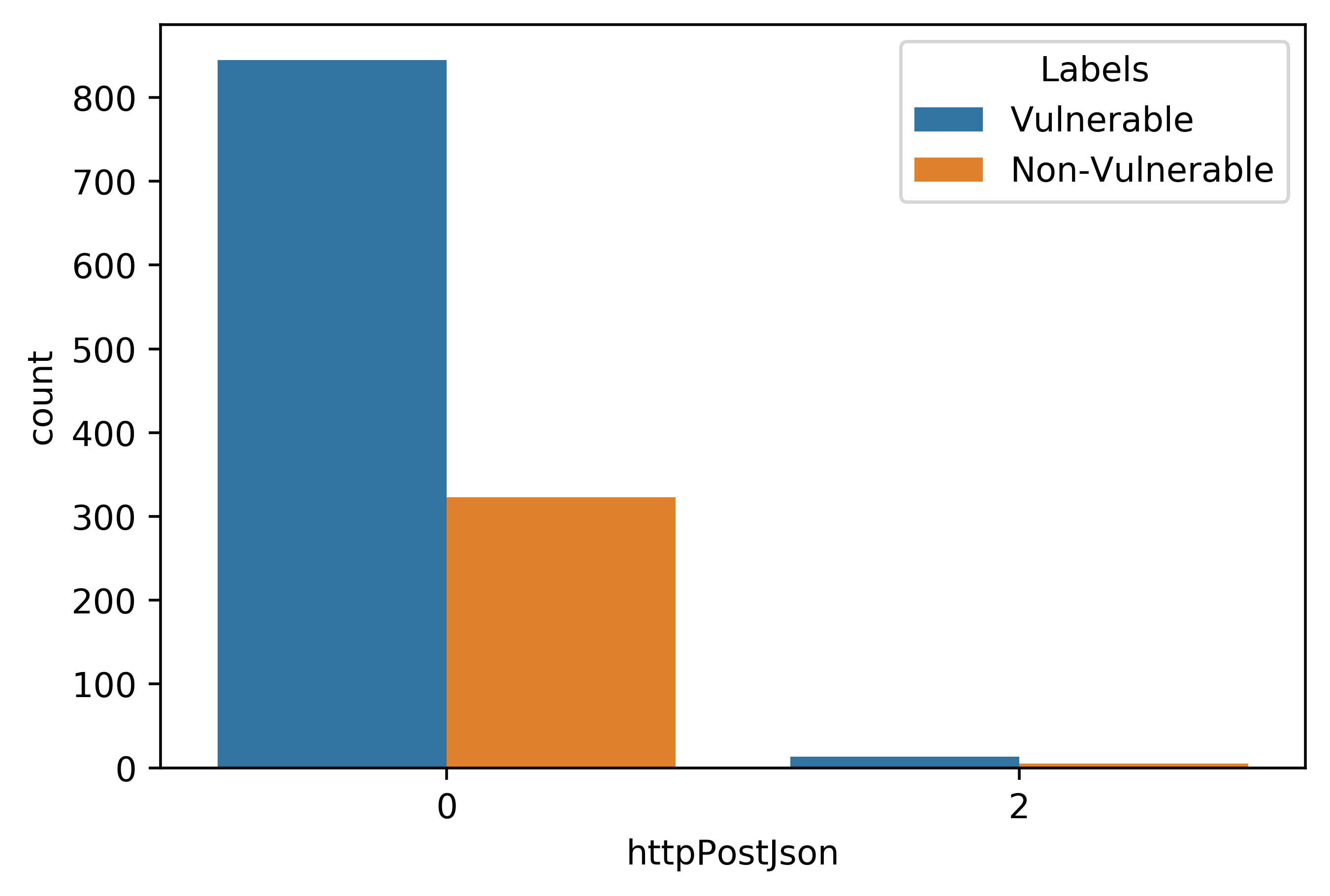}}
\caption{httpPostJson Frequency in all apps for Corpus 2}
\label{fig:postjson_mut}
\end{figure}

\begin{figure}[ht!]
\centerline{\includegraphics[width=.7\textwidth]{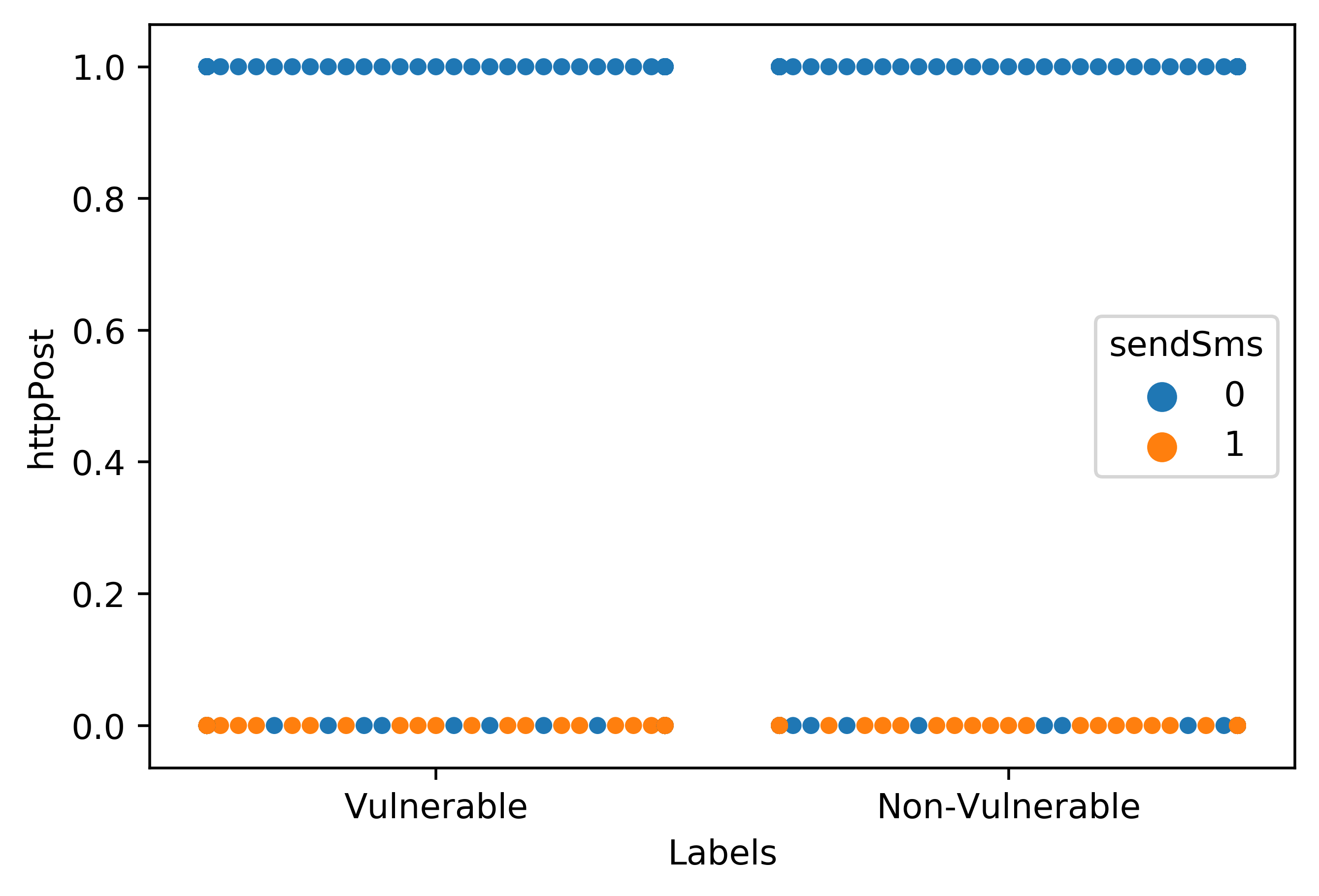}}
\caption{Occurrence of httpPost w.r.t sendSMS in all apps for Corpus 2}
\label{fig:postsms_mut}
\end{figure}

\begin{figure}[ht!]
\centerline{\includegraphics[width=.7\textwidth]{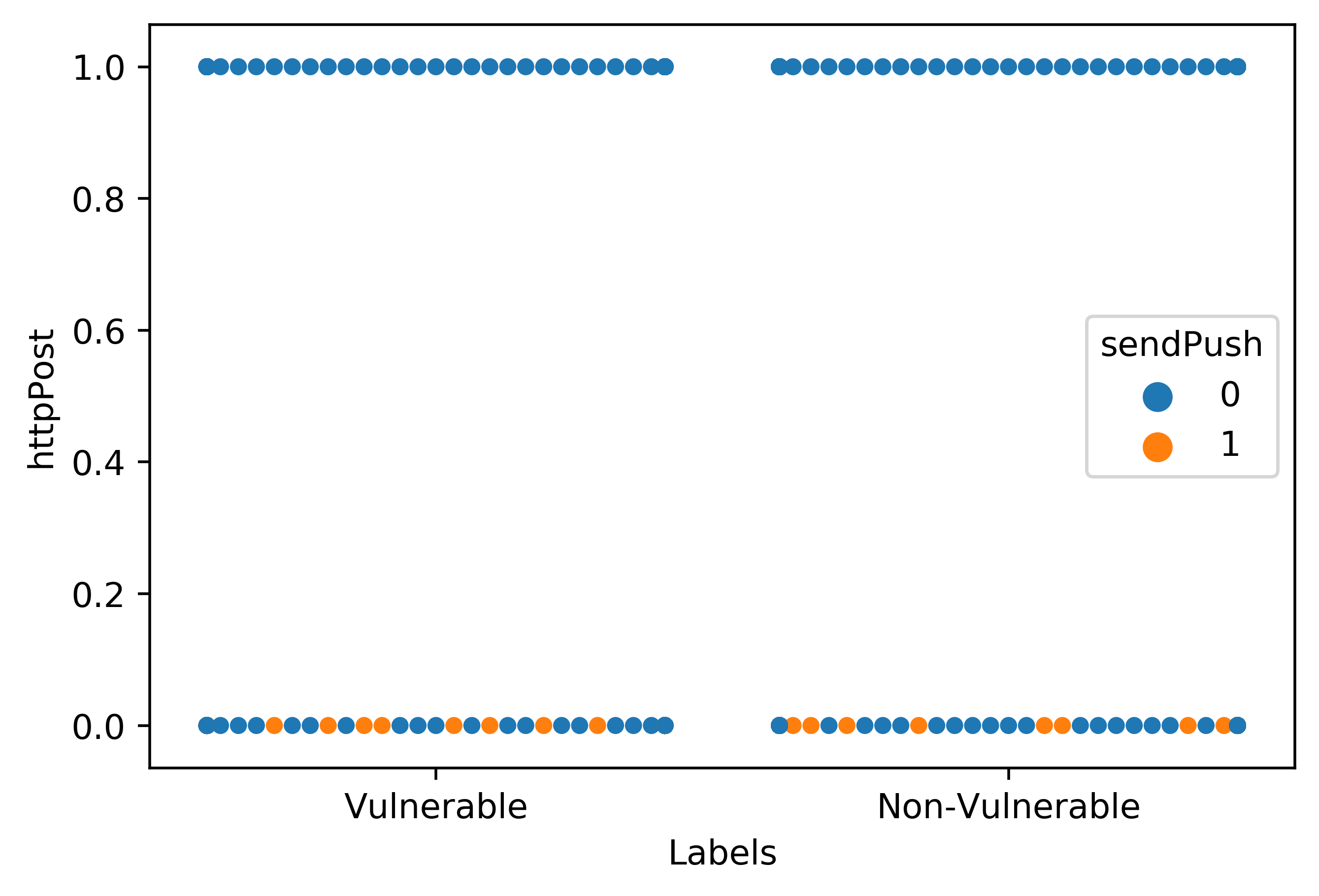}}
\caption{Occurrence of httpPost w.r.t sendPush in all apps for Corpus 2}
\label{fig:postpush_mut}
\end{figure}
\end{subappendices}

\end{document}